\documentclass[11pt,a4paper]{article}

\usepackage[utf8]{inputenc}
\usepackage[T1]{fontenc}
\usepackage{amsmath,amssymb,amsfonts}
\usepackage{mathtools}
\usepackage{graphicx}
\usepackage{booktabs}
\usepackage{longtable}
\usepackage{tikz}
\usetikzlibrary{arrows.meta,decorations.pathreplacing}
\usepackage[margin=2.5cm]{geometry}
\usepackage[colorlinks=true,linkcolor=blue,citecolor=blue,urlcolor=blue]{hyperref}
\hypersetup{
  pdftitle={The next-to-next-to-leading order BFKL eigenvalue at odd conformal
    spin in planar N=4 super Yang-Mills: a reconstructed closed form,
    coefficient structure, and arithmetic},
  pdfauthor={A. Prygarin and C. C. Madjuogang Sandeu}
}
\usepackage{xcolor}

\newcommand{\zb}{\bar{z}}                       

\newcommand{\cpl}{a}                            
\newcommand{\chiord}[1]{\chi_{#1}}              
\newcommand{\Fn}{F_{n}}                          

\newcommand{\Sh}[2]{S_{#1}\!\left(#2\right)}         
\newcommand{\Sig}[1]{\Sigma_{#1}}               
\newcommand{\Del}[1]{\Delta_{#1}}               
\newcommand{\etaf}[1]{\eta\!\left(#1\right)}    

\newcommand{\Aset}{\mathcal{A}}                 
\newcommand{\atom}[1]{A_{#1}}                   
\newcommand{\Ratom}{\mathcal{R}}               
\newcommand{\BAatom}{\mathcal{B}}              
\newcommand{\coeff}[1]{c_{#1}}                 

\newcommand{\lntwo}{\ln 2}
\newcommand{\Lifour}{\mathrm{Li}_4\!\left(\tfrac12\right)}
\newcommand{\zzt}{\zeta_2\zeta_3}

\newcommand{\dgam}{\psi}                        

\newcommand{\Phibare}{\phi}
\newcommand{\Sigq}{\Pi}

\newcommand{\ap}{a_{+}}
\newcommand{\am}{a_{-}}
\newcommand{\amin}{a_{\min}}

\newcommand{\KRone}{K_{\mathcal{R}1}}
\newcommand{\KRtwo}{K_{\mathcal{R}2}}
\newcommand{\KRthree}{K_{\mathcal{R}3}}
\newcommand{\Kcwthree}{K_{\mathrm{cw}3}}       
\newcommand{\Kcwtwo}{K_{\mathrm{cw}2}}         

\newcommand{\Gladder}{G}                        
\newcommand{\gE}{\gamma_{E}}                     

\newcommand{\amax}{a_{\max}}                   
\newcommand{\Kres}{\mathcal{K}}                
\newcommand{\atomlab}[2]{A^{#1}_{#2}}          
\newcommand{\betaf}{\beta}                     
\newcommand{\epsk}{(-1)^{k}}                   
\newcommand{\A}[4]{\mathcal{A}^{#1}\!\left[#2;#3;#4\right]}
\newcommand{\Om}[1]{\Omega_{#1}}               
\newcommand{\cOm}{c_{O}}                       
\newcommand{\cCm}{c_{C}}                       
\newcommand{\dAB}{d}                           
\newcommand{\Lam}{\Lambda}
\newcommand{\Lax}[1]{\Lambda^{\max}_{#1}}
\newcommand{\Lin}[1]{\Lambda^{\min}_{#1}}
\newcommand{\Lbb}[1]{\Lambda^{0}_{#1}}
\newcommand{\Psiw}{\Psi}                        
\newcommand{\fa}{\varphi}

\allowdisplaybreaks
\numberwithin{equation}{section}

\title{The next-to-next-to-leading order BFKL eigenvalue at odd conformal
spin in planar $\mathcal{N}=4$ super Yang--Mills:\\
a reconstructed closed form, coefficient structure, and arithmetic}

\author{A.~Prygarin \ and \ C.~C.~Madjuogang Sandeu\\
\small Department of Physics, Ariel University, Ariel 40700, Israel\\
\small \texttt{alexanderp@ariel.ac.il} (A.~Prygarin, corresponding author)}

\date{}

\begin{document}
\maketitle

\begin{abstract}
We give the colour-singlet eigenvalue of the three-loop BFKL kernel of planar
$\mathcal{N}=4$ super Yang--Mills in closed form at odd conformal spin $n$. It splits into a
holomorphic block and its conjugate, $\chiord{2}(n,\nu)=\tfrac14[\Fn(z)+\Fn(\zb)]$ with
$z=(|n|-1)/2+i\nu$, and the block is a finite combination of nested harmonic sums of one
variable, rational functions and transcendental constants, of uniform transcendental
weight five. At the two lower orders the sums are evaluated at the two conjugate points
only. At three loops they are evaluated at every point of the integer ladder running
from the reflected endpoint $-\zb$ up to $z$, and one step beyond it, and for all but
seven of the slots the coefficient at each rung is fixed
by the two distances to the ends. Summing over the one-variable functions before summing
along the ladder replaces the rules of those slots by one function of one variable under
each ladder weight. The remaining seven are
carried by a sum along the ladder whose summand moves with
the summation index. All forty-seven coefficient slots follow from rules uniform in the
conformal spin at every odd $n\ge3$, with no per-spin coefficient tabulated in the
definition; the $n=1$ boundary block is supplied separately. Those rules are a reconstruction from the
computed spins, exact at every one of them, verified over the range $n\le99$ and at the
holdout spin $n=101$, and not proved at arbitrary odd $n$. Along $\nu=0$ the
intercepts agree with the Quantum Spectral Curve values at every computed odd
spin through $n=91$. That comparison tests the three-loop integrand and the
reduction performed here together, at one point of each spin, and for most of those
spins it is the first comparison of its kind.
\end{abstract}

\tableofcontents


\section{Introduction}\label{sec:intro}

The BFKL eigenvalue of planar $\mathcal{N}=4$ super Yang--Mills is known in closed
form at one and two loops, at every conformal spin, as a combination of harmonic
sums of the single variable $z=(|n|-1)/2+i\nu$ evaluated at $z$ and at its
conjugate. The leading-order kernel is the original BFKL
one~\cite{Fadin:1975cb,Kuraev:1977fs,Balitsky:1978ic,Lipatov:1985uk}; at
next-to-leading order the kernel comes from the QCD computation of Fadin and Lipatov
and of Ciafaloni and Camici~\cite{Fadin:1998py,Ciafaloni:1998gs}, and its eigenvalue
from Kotikov and Lipatov~\cite{Kotikov:2000pm}. The
harmonic-sum form of those two
orders came later than the orders themselves: at zero conformal spin it was given by
Costa, Goncalves and Penedones~\cite{Costa:2012cb}, and at general conformal spin in
App.~C of Alfimov, Gromov and Sizov~\cite{Alfimov:2018cms}. At three loops the kernel has been
available since Caron-Huot and Herranen~\cite{Caron-Huot:2016tzz}. That work supplies
the linearized three-loop kernel in coordinate space in closed form, with the radial
integrand produced symbolically at each given integer spin, harmonic-sum expressions
at the two spins $n=0$ and $n=1$, and numerical trajectories above them. The
non-conformal part of the three-loop planar kernel has been predicted
since~\cite{Brunello:2025rhh}. Caron-Huot and Herranen do not supply
a compact reduced formula in harmonic sums, uniform in the conformal spin and away
from the line $\nu=0$. The present study gives one: the three-loop
colour-singlet eigenvalue at odd conformal spin, defined at any odd $n$ and exact at
every spin computed, together with uniform rules covering all forty-seven of its
coefficient slots. To the best of our knowledge a formula of that kind has not been
given before in the colour-singlet channel. It is again built from harmonic
sums of one variable, now of weight at most five, and again splits into a holomorphic
block and its conjugate. The difference lies in where the sums are evaluated. At the two lower
orders the two endpoints $z$ and $\zb$ are the only arguments, while at three loops the
sums are evaluated at every point of the integer ladder that connects the reflected
endpoint $-\zb$ to $z$.

With the trajectory
normalized as in Eq.~\eqref{eq:set:omega}, the leading eigenvalue is given by a single
continued harmonic sum at the two conjugate points,
\begin{equation}
\chiord{0}(n,\nu)=-\bigl[\Sh{1}{z}+\Sh{1}{\zb}\bigr],
\label{eq:intro:lo}
\end{equation}
and the next-to-leading eigenvalue of Kotikov and
Lipatov~\cite{Kotikov:2000pm,Kotikov:2002ab}, which the principle of maximal
transcendentality isolates inside the QCD
result~\cite{Kotikov:2002ab,Kotikov:2006ts}, is
\begin{equation}
\chiord{1}(n,\nu)=2\,\Del{1}\Del{-2}-4\,\Sig{-2,1}+2\,\Sig{-3}
   +2\,\zeta_2\,\Sig{1}+2\,\Sig{3},
\qquad n\ \text{odd},
\label{eq:intro:nlo}
\end{equation}
with $\Sig{a}=\Sh{a}{z}+\Sh{a}{\zb}$ and $\Del{a}=\Sh{a}{z}-\Sh{a}{\zb}$. The
restriction on the spin is theirs and not ours. Equation~(31) of
Ref.~\cite{Kotikov:2002ab} multiplies the one term that breaks their generalized
holomorphic separability by $1+(-1)^{n}$, so that term drops out at odd spin, and there
$\delta(n,\gamma)$ and its continuation to negative $|n|$ are one function. The parity
enters through the two conformal weights $\mu=\gamma+|n|/2$ and $\tilde\mu=\mu-|n|$,
with $\gamma=\tfrac12+i\nu$ the anomalous-dimension variable: the term is built on
$\cos(\mu\pi)$ and $\cos(\tilde\mu\pi)=(-1)^{n}\cos(\mu\pi)$, which agree at even spin
and differ in sign at odd spin. At even $n$ Eq.~\eqref{eq:intro:nlo} would need that
term restored. Kotikov and
Lipatov print this eigenvalue as their $\delta(n,\gamma)$, in $\Psi''$, $\beta'$ and
their function $\Phi(|n|,\gamma)$; the harmonic sums of
Eq.~\eqref{eq:intro:nlo} are the form of
Refs.~\cite{Costa:2012cb,Alfimov:2018cms}, and Eq.~(19) of
Ref.~\cite{Prygarin:2019ruv} is the closest to it in shape. Four terms
of Eq.~\eqref{eq:intro:nlo} are one-variable sums added at the two endpoints. The
fifth, $2\,\Del{1}\Del{-2}$, is the only place the two points meet, and it is the
seed of what happens at the next order.

In the $(n,\nu)$ plane the three-loop eigenvalue has been known along lines, not
on the plane. The Quantum Spectral Curve of planar
$\mathcal{N}=4$~\cite{Gromov:2013pga}, continued into the Regge
regime~\cite{Alfimov:2014bwa}, gave it at $n=0$ as a uniform weight-five combination
of nested harmonic sums~\cite{Gromov:2015vua}; a reconstruction of the same order was
put forward independently~\cite{Velizhanin:2015xsa}, and the pomeron eigenvalue has
since been addressed at the next logarithmic order~\cite{Velizhanin:2021bdh}. Alfimov,
Gromov and Sizov~\cite{Alfimov:2018cms} then gave the intercepts along $\nu=0$ in closed
form at arbitrary conformal spin, as binomial harmonic sums of uniform weight five, and
an asymptotic Baxter--Bethe treatment of the trajectories has followed
since~\cite{Ekhammar:2024neh,Ekhammar:2025vig}. Off those lines the
eigenvalue has been evaluated numerically. Its terms of highest depth at arbitrary
$\nu$ and $n$ were proposed earlier from a Bethe--Salpeter
ansatz~\cite{Joubat:2020vrw,Joubat:2020hvc}, with the simpler terms left open and with
the $n=1$ coefficients differing from those of Ref.~\cite{Caron-Huot:2016tzz}, as those
authors point out; the eigenvalue itself has not been reduced to a formula uniform in
the spin.
The external input to the present study is the Caron-Huot--Herranen three-loop
integrand together with the additive weight-five prefactor of
Eq.~\eqref{eq:cf:master}, which is taken from the same work. Everything below is a
transformation of those two.

On the line $\nu=0$ the two conjugate points
coincide, $z=\zb=(n-1)/2$, so the two halves of the block are one and the same number
and add, and the intercept is one half of that value. It is still reached through a
removable singularity, for the reason Sec.~\ref{sec:cf:eval} gives.
Two of the properties the closed form makes explicit are empty there: product-freeness is
the statement that no term couples $z$ to $\zb$, which says nothing when the two agree,
and there is no $\nu$ in which to have poles. The knowledge along $\nu=0$ therefore fixes
the eigenvalue at one point of each trajectory, and the content below lies off it.

The ladder is what changes between two loops and three. Within the holomorphic block
only the endpoint $z$ occurs at leading order, and Eq.~\eqref{eq:intro:nlo} adds the
conjugate point together with the one bilinear that couples the two. At three loops the
block is spread over the whole segment of integer-shifted arguments $z+k$ running
from $-\zb$ up to $z$ and one step past it, and the coefficient of an atom at rung
$k$ depends on the two distances $\ap=k+n-1$ and $\am=-k$ separating that rung from
the ends. Written out one atom at a time the block is an inventory of forty-seven coefficient
rules, and forty of them close in the two distances. Summing over the atoms before
summing along the ladder carries thirty-seven of those forty into five functions of one
variable, each under a single ladder weight. Two of the five contain all the nested
content of the thirty-seven and the other three reduce to polygammas at halved
argument. Three rules stay outside that count: the rational ladder sum, and the two at
the point beyond the ladder, which merge into a single Hurwitz zeta. In the same step
three separate harmonic weights merge into $S_1$ of the larger of the two distances,
and the exchange $\ap\leftrightarrow\am$, which reflects the ladder about its
midpoint and swaps its two ends, becomes a symmetry one can see directly in the
coefficients.

Seven slots lie outside that description, and the formula is not complete until they are
closed. Three of them are attached to rational atoms, and for one of the three the
failure to close in harmonic sums of the two distances is forced by arithmetic. Its
denominators contain primes as large as $n-2$, set by the ladder as a
whole, while a nested harmonic sum at $\ap$, $\am$ or $\min(\ap,\am)$ is built from primes
only up to $\max(\ap,\am)$, and rational coefficients fixed once and for all can add
only finitely many primes to that. The three are closed by one enlargement, so that
argument settles the sector they close together. It is a proof for one of the three
only; the other two show primes above the bound over the computed range with two
exceptions, and neither is settled on its own. For the remaining four we have no
such argument, only a complete search over bounded bases. The missing ingredient is
the full ladder length, and the functions that supply it are sums along the ladder,
\begin{equation}
\sum_{N=1}^{\ap}\frac{\Sh{1}{\max(\am+N,\,\ap-N)}}{N^{j}} ,
\label{eq:intro:ladder}
\end{equation}
in which the argument of the harmonic sum moves with the summation index. The two
branches $\am+N$ and $\ap-N$ meet at the value $(n-1)/2$, and their maximum reaches the
full ladder length $n-1$ at the endpoint $N=\ap$ of the sum. In this class all seven
rules close exactly.
Three of them come for free from a further property of the block: at each of the
internal points $z=0,1,\dots,n-2$ several of its pieces have a pole and the block does
not, and requiring the poles to cancel there fixes the three rational coefficients once
the others are known.
At odd $n\ge3$ the definition needs no per-spin tabulated coefficient, and the boundary
spin $n=1$ has its own block (App.~\ref{app:construction}), so the eigenvalue is defined
at every odd $n$. The rules themselves are a reconstruction from the computed spins: fixed
at low spin, with no derivation from the integrand at symbolic $n$ behind them.
Section~\ref{sec:checks} fixes the range over which they are verified and the sense in
which they are unproved at arbitrary odd $n$.

We fixed the rules at low spin and then tested them where they had no freedom left.
Four spins, $n=93,95,97,99$, were computed after the rules for the forty closing slots
had been written down, and those rules reproduce all $14580$ of their closing-slot
coefficients exactly. Along $\nu=0$ the intercepts agree spin by spin with the
Quantum Spectral Curve values through $n=91$. The seven rules of
Sec.~\ref{sec:residual} were fixed at low spin in the same way, later than
the four tables, and reproduce every remaining coefficient out to $n=99$, the further
$2672$ at those four spins among them. The two fits behind them are already determined
by the cells with $n\le9$, so a few dozen coefficients fix all seven rules and
everything above is a test. A fifth spin, $n=101$, was generated later still,
and all $4538$ of its coefficients follow from the rules exactly, over all forty-seven
slots. The atom table rebuilt from the rules, with nothing tabulated anywhere, is
the stored table entry for entry at the six spins of Sec.~\ref{sec:chk:closed}, and the
equations as printed reproduce every
coefficient of every computed spin in exact rational arithmetic.

Section~\ref{sec:setup} fixes the variables and draws the ladder.
Section~\ref{sec:closedform} states the block, lists the functions that occur in it,
and shows how to evaluate it. Section~\ref{sec:structure} explains why the nested content of a
forty-rule table is carried by two functions of one variable, and derives the exchange
symmetry from the same reordering. Section~\ref{sec:residual} takes up the last seven
rules: the arithmetic that keeps the rational kernels out of the two ladder distances,
the columns that close on their own, the regularity of the block on the ladder and what
it fixes, and the ladder sums in which all seven close.
Section~\ref{sec:checks} collects the checks and says of each what it does and does not
test, and Sec.~\ref{sec:discussion} what the closed form is for. The appendices contain
the conventions and the continuation in the spin, the extraction of the atom table from
the three-loop integrand, the reduction identities, the coefficient catalogue, the
collapse in detail, the non-closing slots and the validation tables; the per-spin tables
and the evaluator are given in the ancillary files. The result is announced in
Ref.~\cite{Letter}.

\section{Variables, continuation, and the ladder}\label{sec:setup}

\subsection{Variables}
\label{sec:set:vars}
The Regge trajectory is expanded as
\begin{equation}
\omega(n,\nu)=4\cpl\bigl[\chiord{0}+\cpl\,\chiord{1}+\cpl^{2}\,\chiord{2}
   +\cdots\bigr],\qquad
\cpl=g^{2}=\frac{g_{YM}^{2}N_c}{16\pi^{2}},
\label{eq:set:omega}
\end{equation}
so that $\chiord{2}$ is the three-loop eigenvalue. The coupling here is the DRED one of
Kotikov and Lipatov, reached from Eq.~(19) of Ref.~\cite{Kotikov:2002ab} by the finite
renormalization of its Eq.~(24), which removes the term $\tfrac13\chi(n,\gamma)$ that the
DREG scheme leaves behind; Eq.~\eqref{eq:intro:nlo} is their $\delta(n,\gamma)$ at odd
$n$, in that normalization. Everything below is written in the
half-shifted variables
\begin{equation}
z=\frac{|n|-1}{2}+i\nu,\qquad \zb=\frac{|n|-1}{2}-i\nu,\qquad z+\zb=|n|-1 .
\label{eq:set:zdef}
\end{equation}
The eigenvalue depends on $|n|$ only, and we write $n$ for the positive odd spin
throughout.
For real $\nu$ these are complex conjugates, so a combination $F(z)+F(\zb)$ is real
and even in $\nu$ without further work. The spectral variable of
Refs.~\cite{Caron-Huot:2016tzz,Gromov:2015vua} is twice the one used here; lines
quoted from those papers are evaluated at $\nu_{\mathrm{CHH}}=2\nu$ before
comparison.

Nested harmonic sums are taken in the non-strict
convention~\cite{Vermaseren:1998uu,Gromov:2015vua},
\begin{equation}
\Sh{a_1,\dots,a_d}{N}=\sum_{N\ge y_1\ge\cdots\ge y_d\ge1}\;
   \prod_{j=1}^{d}\frac{\operatorname{sgn}(a_j)^{\,y_j}}{y_j^{\,|a_j|}},
\label{eq:set:sdef}
\end{equation}
so a negative letter brings an alternating sign and $\Sh{1}{N}=H_N$. The value at
zero argument is part of the formula and is fixed by it, since the rules
below are evaluated at $\ap=0$ and at $\am=0$ at the two ends of the ladder. A sum over
an empty range is zero, so $\Sh{a_1,\dots,a_d}{0}=0$ for every nonempty word, while the
empty word is an empty product and gives $S_{\emptyset}(0)=1$. Finite sums with no term
in them are zero for the same reason, the ladder sums of
Eq.~\eqref{eq:res:laddersums} at $\ap=0$ among them. The
continuation to complex argument, and the branch used throughout, are in
App.~\ref{app:conventions}; the same appendix gives the
checks that pin Eq.~\eqref{eq:set:omega} against the leading and next-to-leading
orders.

\subsection{Atoms}
\label{sec:set:atoms}
The three-loop block is built from two kinds of one-variable function. The first is
rational,
\begin{equation}
\atom{}^{\Ratom}(w)=\frac{1}{w^{q}},\qquad q=1,\dots,5,
\label{eq:set:Ratom}
\end{equation}
and the second is a nested Mellin sum,
\begin{equation}
\atom{}^{\BAatom}(w)=\sum_{N\ge1}\frac{\sigma^{N}\,\Sh{v}{N}}
{\Bigl[\prod_{(d,p)}(N+d)^{p}\Bigr]\,(N+w)^{q}},
\qquad \sigma=\pm1 ,
\label{eq:set:BAatom}
\end{equation}
with $\Sh{v}{N}$ a nested harmonic sum and $(d,p)$ a short list of non-negative integer
shifts $d$ and powers $p$ fixed by the shape.
We call both kinds \emph{atoms}, and one entry of the list an atom \emph{shape}. There
are $41$ shapes at three loops, five rational and thirty-six nested, and that list does
not grow with the conformal spin.

\subsection{The ladder}
\label{sec:set:ladder}
\begin{figure}[t]
\centering
\begin{tikzpicture}[x=1.05cm,y=1cm]
  \draw[-{Latex}] (-6.4,0) -- (1.9,0);
  \foreach \x in {-6,-5,-4,-3,-2,-1,0,1}
    \draw[fill=black] (\x,0) circle (2.1pt);
  \node[below=5pt] at (-6,0) {$-\zb$};
  \node[below=5pt] at (0,0)  {$z$};
  \node[below=5pt] at (1,0)  {$z\!+\!1$};
  \node[below=5pt] at (-2,0) {$z\!+\!k$};
  \draw[decorate,decoration={brace,amplitude=5pt},thick]
       (-6,0.28) -- (-2,0.28) node[midway,above=5pt] {$\ap=k+n-1$};
  \draw[decorate,decoration={brace,amplitude=5pt},thick]
       (-2,0.28) -- (0,0.28) node[midway,above=5pt] {$\am=-k$};
  \draw[<->,gray] (-6,-0.95) -- (0,-0.95)
       node[midway,below] {$\ap+\am=n-1$};
\end{tikzpicture}
\caption{The ladder at odd conformal spin $n$. The atoms of
Eqs.~\eqref{eq:set:Ratom} and~\eqref{eq:set:BAatom} are evaluated at the marked
points $z+k$. The two distances $\ap$ and $\am$ from a rung to the ends of the
segment sum to $n-1$, and the exchange $\ap\leftrightarrow\am$ is the reflection of
the ladder about its midpoint $i\nu$, which swaps its two ends
(Sec.~\ref{sec:str:reflection}).}
\label{fig:ladder}
\end{figure}
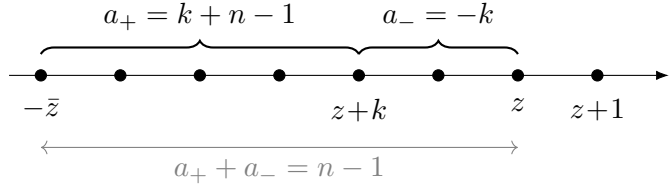

The set of points at which the shapes are evaluated grows with the spin. One shape
placed at one
rung is a shifted \emph{instance}, and it is their number that grows. The arguments
that occur are the shifted points $w=z+k$ with
\begin{equation}
k=-(n-1),\,-(n-2),\,\dots,\,0,\,1 ,
\label{eq:set:krange}
\end{equation}
a segment of integer-spaced points running from the reflected endpoint
$-\zb=z-(n-1)$ up to $z$, and one step beyond. Individual slots do not all reach every
one of these points. Table~\ref{tab:cat:rules} lists the rung range of each closing
slot and App.~\ref{app:grd:seven} that of the seven that do not close. We
call the range
$-(n-1)\le k\le0$ the \emph{ladder} and a point $z+k$ on it a rung; the single point
$k=1$ lies outside it and holds two slots of its own. The two distances from a rung to the ends,
\begin{equation}
\ap=k+n-1,\qquad \am=-k,\qquad \ap+\am=n-1 ,
\label{eq:set:distances}
\end{equation}
are the variables in which the coefficients are naturally written; their sum is
fixed by the spin, so a rung is specified by either distance on its own. Figure~\ref{fig:ladder}
shows the arrangement.

At leading order only the rung $k=0$ occurs. At next-to-leading order the two conjugate
points are still the only arguments, coupled by the single bilinear
$2\,\Del{1}\Del{-2}$ of Eq.~\eqref{eq:intro:nlo}. At three loops every rung is
populated.

\subsection{Vocabulary}
\label{sec:set:vocab}
The words below have fixed senses. App.~\ref{app:conv:map} collects the symbols and the
two senses of \emph{kernel}; the three words that grade a statement, \emph{exact},
\emph{verified} and \emph{proved}, are fixed at the head of Sec.~\ref{sec:checks}.

\medskip
\noindent
\begin{tabular}{@{}p{2.7cm}p{12.6cm}@{}}
\emph{atom} & a function of one variable, Eq.~\eqref{eq:set:Ratom}
or~\eqref{eq:set:BAatom}.\\
\emph{shape} & one entry of that list, carrying neither an argument nor a constant in
front of it. There are $41$ shapes, listed in App.~\ref{app:cat:shapes}.\\
\emph{slot} & a shape together with the constant $\mu\in\{1,\pi^2,\zeta_3\}$
multiplying it. A slot has one coefficient rule $\coeff{}(n,k)$. There are $47$
non-empty slots (the coefficient families of the accompanying Letter).\\
\emph{instance} & one shape placed at one rung, that is one entry of the atom table at
a given $k$. Their number grows with the spin, $39n-3$ at every odd $n\ge3$ of the
computed range, while the numbers of shapes and of slots do not.\\
\emph{ladder} & the rungs $z+k$ with $-(n-1)\le k\le0$; the point $z+1$ of
Eq.~\eqref{eq:set:krange} lies outside it.\\
\emph{rung range} & the rungs at which a given slot is non-zero. It is the
\texttt{supp.} column of Table~\ref{tab:cat:rules} and the field
\texttt{k\_support} of the machine-readable catalogue in the ancillary files.\\
\emph{weight} & the function of $\ap$ and $\am$ that dresses a slot along the
ladder, Eq.~\eqref{eq:str:weights}. A weight quoted as a number, as in weight five, is
the transcendental grading of App.~\ref{app:cat:grading} instead.\\
\emph{kernel} & a coefficient function that is not a combination of harmonic sums in
$\ap$ and $\am$, Sec.~\ref{sec:residual}. Every kernel printed here is raw, in the sense
App.~\ref{app:conv:map} fixes.\\
\end{tabular}
\medskip

\noindent
Of the $47$ slots, $40$ have coefficient rules that close in the two ladder
distances and $7$ do not. Two spans occur below and they are not the same. The
\emph{weight span} is the seven functions of Eq.~\eqref{eq:str:weights} that dress the
slots along the ladder. The \emph{elementary algebra} is the larger set of finite
$\mathbb{Q}$-linear combinations of products of nested harmonic sums in $\ap$, $\am$ and
$\min(\ap,\am)$, at any weight, whose rational coefficients are the same numbers at
every spin. The last clause is the operative one:
Sec.~\ref{sec:res:obstruction} rests on it, and a coefficient that is a rational
function of $n$ or of the two distances is outside the set. A rule \emph{closes} when it lies in the weight span, which is the sense in
which $40$ of the $47$ close; lying in the elementary algebra is the weaker
condition, and Sec.~\ref{sec:res:what} is where the difference matters.

\section{The closed form}\label{sec:closedform}

\subsection{The block}
\label{sec:cf:block}
In the variables of Sec.~\ref{sec:setup} the eigenvalue splits, at every odd $n$, into one block and its conjugate,
\begin{equation}
\chiord{2}(n,\nu)=\tfrac14\bigl[\Fn(z)+\Fn(\zb)\bigr],
\label{eq:cf:separable}
\end{equation}
and at a single fixed odd $n$ this is merely a way of writing the eigenvalue:
$\zb=n-1-z$, so any function even in $\nu$ can be put in this form. The content sits in
the atom class. At fixed odd $n$ the block is product-free in the fixed class of
Eqs.~\eqref{eq:set:Ratom} and~\eqref{eq:set:BAatom}, whose members have weight at most
five: over that class no term couples $z$ to $\zb$, in contrast
with the bilinear $2\,\Del{1}\Del{-2}$ of the previous order. Functions involving both
points are available at these weights, eight of them at weight two against six
one-variable sums~\cite{Joubat:2021pww}. Within the class the property is stable, because
the operations that take the atom table to Eq.~\eqref{eq:cf:ladderform} (partial fractions
in the Mellin variable $N$, the stuffle relations of App.~\ref{app:red:stuffle}, and
rational regrouping of the coefficients over the ladder weights) all act inside it, and
none of them produces a term carrying $z$ and $\zb$ together. This is a property of the
present representation, and it is not a separability theorem. Hermitian separability in
the sense of Refs.~\cite{Bondarenko:2015tba,Joubat:2020hvc}, a property asserted before
any atom class is fixed, is a different statement, and we do not claim it at this order.
Nor do we claim the product-free form under transformations that leave the class, the
reflection identities rewriting one endpoint
against the other~\cite{Prygarin:2018tng,Prygarin:2018cog,Joubat:2019esj} among them.
Beyond that, a single
$\Fn$ serves at every odd spin, with $n$
entering only through the range of the ladder sum and the two distances along it. The
holomorphic block is a prefactor plus a sum over that ladder and reads
\begin{equation}
\Fn(z)=-88\,\zeta_4\,\Sh{1}{z}-16\,\zzt-80\,\zeta_5
       +\sum_{a\in\Aset_n}\coeff{a}\,\atom{a}(z),
\qquad
\coeff{a}=r_0+r_1\,\pi^{2}+r_2\,\zeta_3,\quad r_i\in\mathbb{Q}.
\label{eq:cf:master}
\end{equation}
The prefactor is carried over from the three-loop input at every spin. It is the part of
Eq.~(5.12) of Ref.~\cite{Caron-Huot:2016tzz} that sits outside the Mellin integral,
taken at three loops: $\gamma_K^{(3)}F^{(1)}-C^{(3)}$, where
$\gamma_K=\tfrac14\Gamma_{\mathrm{cusp}}$ has three-loop coefficient
$\gamma_K^{(3)}=\tfrac{11\pi^{4}}{45}=22\,\zeta_4$, where $F^{(1)}=-4\,\Sh{1}{z}$ is the
one-loop eigenvalue of that work in the present variables, and where
$C^{(3)}=16(\zzt+5\,\zeta_5)$. The spin enters those three objects only through $z$, so
the same expression serves at every odd $n$. The constant $C^{(3)}$ is not free there: it
is fixed by the condition $j=1$ at $m=1$ and $\nu=0$, Eq.~(5.8) of
Ref.~\cite{Caron-Huot:2016tzz},
which holds at every loop order and sits at the boundary spin of the present study.
No constant of weight above three occurs at the level of an individual slot; the
weight-five constants $\zeta_5$ and $\zzt$, and the $\zeta_4$ multiplying
$\Sh{1}{z}$, sit in the prefactor. The $\zeta_5$ that appears explicitly in
Eq.~\eqref{eq:cf:ladderform} below is the constant part of a
Hurwitz zeta, exposed by rewriting, and it is no exception to the statement just made; the
slot coefficients there are both equal to $32$. Coefficient and atom together reach uniform
weight five once the Mellin power of the summation variable is counted, in the
grading of App.~\ref{app:cat:grading}.

\subsection{The ladder sum}
\label{sec:cf:ladder}
The sum in Eq.~\eqref{eq:cf:master} collapses to a short
expression. Two slots sit at the single rung $k=1$ and never enter a ladder sum;
the rest organize into four ladder sums and the residual contribution,
\begin{align}
\sum_{a\in\Aset_n}\coeff{a}\,\atom{a}(z)
&=\underbrace{32\,\zeta_5-32\,\Sh{5}{z}}_{\text{the two single-rung slots, merged}}
\nonumber\\[4pt]
&\quad+\sum_{k=-(n-1)}^{0}(-1)^{k}\,\Gladder_{(-1)^k}(z+k)
\;+\;\sum_{k=-(n-1)}^{0}(-1)^{k}\,\Sh{1}{\amax}\,\Phi(z+k)
\nonumber\\[4pt]
&\quad+\sum_{k=-(n-1)}^{0}(-1)^{k}
\Bigl[\Sh{1}{\ap}\,D_{+}+\Sh{1}{\am}\,D_{-}+\Sh{1}{\amin}\,D_{\min}\Bigr](z+k)
\nonumber\\[4pt]
&\quad+\frac{16\pi^{2}}{3}\sum_{k=-(n-1)}^{0}(-1)^{k}\,
\frac{\Sh{2}{\ap}+\Sh{-2}{\ap}}{z+k}
\;+\;\Kres(n,z),
\label{eq:cf:ladderform}
\end{align}
with $\amax=\max(\ap,\am)$ and $\amin=\min(\ap,\am)$. The two slots at $k=1$ sit outside
the ladder range, but they are not two independent objects. They are one Hurwitz zeta
split by peeling off its first term,
$32\,\zeta(5,z+2)+32/(z+1)^{5}=32\,\zeta(5,z+1)$, and written that way they contribute
$16\,\zeta_5-8\,\Sigma_5$ to $\chiord{2}$, with $\Sigma_a=\Sh{a}{z}+\Sh{a}{\zb}$. The
$\Sh{5}{}$ part of that is the whole $\Sh{5}{}$ content of the block, for a reason visible in the atom list: an $\Sh{5}{}$ needs a pole of order five, and every shape of that order sits at $k=1$, at
every spin computed. Nothing on the ladder holds one. The
bookkeeping is fixed from outside: the Caron-Huot--Herranen block at $n=1$ contains
$\Sh{5}{}$ with coefficient $-32$, together with $-16\,\zzt$, $-88\,\zeta_4\Sh{1}{}$ and
$-48\,\zeta_5$, and the prefactor of Eq.~\eqref{eq:cf:master} supplies $-80\,\zeta_5$
against the $+32\,\zeta_5$ exposed here. The same coefficient occurs at $n=0$: the block
of Ref.~\cite{Gromov:2015vua} contains $-32\,\Sh{5}{}$ too, and that determination comes
from the Quantum Spectral Curve, not from the input used here. The same object appears one order at a time,
$-\Sigma_1$ at leading order, $+2\,\Sigma_3$ in Eq.~\eqref{eq:intro:nlo} and
$-8\,\Sigma_5$ at three loops, a polygamma of order $2L-2$ at each order. The rational
multiples $-1$, $+2$, $-8$ depend on the normalization of Eq.~\eqref{eq:set:omega}, so we
draw no continuation from them; what is normalization-free is that the top-weight
non-alternating sum at $L$ loops is $\Sigma_{2L-1}$ at the unhalved argument. Everything in
Eq.~\eqref{eq:cf:ladderform} depends on the spin only through the range of the sum and
through the two distances. The five functions $\Gladder_{(-1)^k}$, $\Phi$, $D_{+}$,
$D_{-}$, $D_{\min}$ are fixed functions of one variable, built once from the atom
list and reused at every spin; they are given explicitly in
App.~\ref{app:cmp:explicit}. Of these, $D_{+}$, $D_{-}$ and $D_{\min}$ have empty nesting
word and reduce to polygammas at halved argument. The nested content sits in
$\Gladder_{(-1)^k}$, of twenty-six atoms, and in $\Phi$, of four. Two alternating sums and
the odd harmonic sum express the latter,
\begin{equation}
A(x)=\sum_{N\ge1}\frac{(-1)^{N}O_2(N)}{N+x},
\qquad
B(x)=\sum_{N\ge1}\frac{(-1)^{N}\Sh{1}{N}}{(N+x)^{2}},
\qquad
O_2(N)=\!\!\sum_{\substack{j\le N\\ j\ \mathrm{odd}}}\!\frac{1}{j^{2}} ,
\label{eq:cf:AB}
\end{equation}
and in terms of them the four atoms collapse to
\begin{equation}
\Phi(x)=\frac{64}{x}\Bigl[\,\frac{\pi^{2}\lntwo-13\,\zeta_3}{4}-2A(x)-B(x)\,\Bigr],
\label{eq:cf:phi}
\end{equation}
where the two constants are known in closed form,
\begin{equation}
A(0)=\frac{\pi^{2}\lntwo}{8}-\frac{21\,\zeta_3}{16},
\qquad
B(0)=-\frac58\,\zeta_3,
\qquad
2A(0)+B(0)=\frac{\pi^{2}\lntwo-13\,\zeta_3}{4}.
\label{eq:cf:AB0}
\end{equation}
Both are weight three, as the grading requires. Since
$O_2(N)=\tfrac12[\Sh{2}{N}-\Sh{-2}{N}]$, the constant $A(0)$ is half the difference of
two alternating Euler sums of weight three, a class treated in
Refs.~\cite{Borwein:1995,FlajoletSalvy:1998}; the value quoted above was fixed here by an
integer-relation search against $\{\zeta_3,\pi^{2}\lntwo,\lntwo^{3}\}$ that returns
$16A(0)+21\zeta_3-2\pi^{2}\lntwo=0$ with a residual of $1.89911\times10^{-65}$ at
sixty-five digits, no $\lntwo^{3}$ and no rational part.
The same $O_2$ governs the continuation between even and odd spin in
App.~\ref{app:conventions}, so the object that flattens $\Phi$ is the one that
mediates between the two parities.

The last term, $\Kres(n,z)$, is the residual contribution of Sec.~\ref{sec:residual}. It
collects the seven slots whose coefficients do not close in $\ap$ and $\am$ alone. They too follow rules
uniform in $n$: Sec.~\ref{sec:residual} gives them, in one further class of function.

Every ladder sum in Eq.~\eqref{eq:cf:ladderform} is performed once and for all in
App.~\ref{app:red:ladder}, which contains the summation-by-parts formulas; the five
functions the sums run against are assembled separately, in
App.~\ref{app:cmp:explicit}. All four ladder sums of Eq.~\eqref{eq:cf:ladderform} are
dressed by $(-1)^{k}$ and none by $1$; the only two slots with that
dressing are the single-rung pair merged above. The
alternation is what turns a ladder into digammas at halved argument. The ladder weight
fixes how far a given sum reduces, and the function it runs against does not: only
the $\Gladder_{(-1)^k}$ sum has a weight independent of $k$ beyond the sign, and it is the only
one resummed outright. The other three are dressed by $\Sh{1}{}$, $\Sh{2}{}$ or
$\Sh{-2}{}$ at $\ap$, $\am$, $\amin$ or $\amax$, the four arguments and three sums of
Eq.~\eqref{eq:str:weights}, which move from rung to rung, and each reduces to a single
nested sum with an elementary summand.

\subsection{The function space}
\label{sec:cf:space}
Reducing every atom of Eqs.~\eqref{eq:set:Ratom} and~\eqref{eq:set:BAatom} to a
canonical form, as in App.~\ref{app:reduction}, leaves a short alphabet. Each atom
denominator is $N^{p}(N+x)^{q}$, so partial fractions apply directly, and the
resulting objects read
\begin{equation}
\begin{gathered}
\Delta_{\sigma,v}(x)=\sum_{N\ge1}\sigma^{N}\Sh{v}{N}
\Bigl[\frac1N-\frac1{N+x}\Bigr],
\qquad
M_{\sigma,v,j}(x)=\sum_{N\ge1}\frac{\sigma^{N}\Sh{v}{N}}{(N+x)^{j}},\\[2pt]
E_{\sigma,v,i}=\sum_{N\ge1}\frac{\sigma^{N}\Sh{v}{N}}{N^{i}} .
\end{gathered}
\label{eq:cf:canonical}
\end{equation}
The first two are functions of $x$ and the third is a constant, an Euler
sum~\cite{Lewin:1981,DeDoelder:1991,Borwein:1995,FlajoletSalvy:1998}. The
$1/N$ and $1/(N+x)$ pieces have to be kept together in $\Delta_{\sigma,v}$: taken
separately they diverge for $\sigma=+1$, and it is the cancellation of their
leading coefficients, which follows from the $1/N^{p+q}$ decay of the atom, that
makes the split legitimate.

Carried out across the $57$ atom entries of the block, which draw
with multiplicity on $39$ of the $41$ shapes, this yields $24$ functions of $x$ and
$12$ constants. The two shapes that do not appear here occur only inside
$\Kres$: the rational atom $1/(z+k)^{3}$ and the nested atom
$\atomlab{\BAatom}{(0,(),((0,1)),1)}$. Eight of the functions have empty nesting word and are polygammas,
\begin{equation}
\begin{gathered}
\Delta_{+,\,()}=\Sh{1}{x},\qquad
\Delta_{-,\,()}=-\lntwo-\betaf(x),\qquad
M_{+,\,(),\,j}=\zeta(j,1+x)\quad(j\ge2),\\[4pt]
M_{-,\,(),\,j}=\frac{\zeta\bigl(j,\tfrac{x+2}{2}\bigr)-\zeta\bigl(j,\tfrac{x+1}{2}\bigr)}{2^{\,j}}
\quad(j\ge2),
\end{gathered}
\label{eq:cf:elementary}
\end{equation}
with halved arguments of the kind App.~\ref{app:red:ladder} accounts for, produced by
splitting the atom's own alternating sum over even and odd $N$ and not by any sum along
the ladder, and with
$\betaf(x)=\sum_{N\ge1}(-1)^{N}/(N+x)$, minus Nielsen's beta function at shifted
argument. After the stuffle relations of App.~\ref{app:reduction}
are applied to the depth-two atoms, the remaining non-elementary content sits in the
fifteen alternating Mellin kernels of App.~\ref{app:red:nested}, and that appendix
writes each of them over standard nested harmonic sums of one variable, twenty-seven
distinct sums in all and none above weight four. Nothing outside
that class occurs in the atoms at odd spin. In general the Mellin transforms of harmonic
polylogarithms reduce to harmonic sums, which is the content of Sec.~9 of
Ref.~\cite{Remiddi:1999ew}; the count above says which of them occur here and at what
weight. Every one of the fifteen carries a nesting word with at most one
negative index, so does every one of the twenty-seven sums, and so does every nesting
word in App.~\ref{app:cat:shapes}, so the
alternation-one restriction proposed for the colour-singlet eigenvalue in
Ref.~\cite{Prygarin:2019ruv} holds of the whole atom class. The closest published
statement of this kind is for the adjoint eigenvalue, conjectured to be a combination of
uniform weight, at every order, of multiple zeta values with the leading-order
eigenvalue, its $\nu$ derivatives and two rational functions of $\nu$ and
$n$~\cite{Dixon:2012yy}. The identities that bring the sums with a
non-empty nesting word into the class of twenty-seven are printed in
App.~\ref{app:red:nested}, sixteen of them, each checked against its defining series
at complex argument. The coefficients need one thing more, the
window sums of Sec.~\ref{sec:res:ladder}, and they enter only through the seven slots of
the residual contribution.

The combinations $D_{+}$ and $D_{\min}$ require the trigamma $\beta'$. The third,
$D_{-}$, does not. Its three alternating atoms add up over the common denominator
$N^{3}(N+x)^{3}$ to the numerator $3N^{2}+3Nx+x^{2}=\bigl[(N+x)^{3}-N^{3}\bigr]/x$, so
the whole alternating part of $D_{-}$ is
$\tfrac{32}{x}\sum_{N\ge1}(-1)^{N}\bigl[(N+x)^{-3}-N^{-3}\bigr]$, in which no
denominator of second order survives. A second-order polygamma does stand in all three,
at $32/x$, $32/x$ and $-32/x$ in $D_{+}$, $D_{-}$ and $D_{\min}$, so it cancels in
$D_{-}+D_{\min}$, the combination Eq.~\eqref{eq:str:R2} forms, and in $D_{+}+D_{\min}$
as well; only $D_{+}+D_{-}$ keeps it, at $64/x$. The single-rung
atom $\atomlab{\BAatom}{(0,(),(),5)}=\zeta(5,1+x)$ involves a fourth-order one.
Catalan's constant is what $\beta'$ gives at half-integer argument; at odd spin the
ladder sits on integer shifts of a single point, so it does not arise here, and it would
only appear in the even sector.

\subsection{Evaluating the eigenvalue}
\label{sec:cf:eval}
The atom table $\Aset_n$ is produced directly from the Caron-Huot--Herranen
integrand at that spin by an exact Mellin extraction in rational arithmetic, with no
floating point at any step: the transform of Eq.~\eqref{eq:cons:mellin} term by term,
each term landing on one atom through Eq.~\eqref{eq:cons:power}
(App.~\ref{app:construction}). A single command
in the ancillary files reconstructs it for any odd $n\ge3$ and reproduces the
tables entry for entry. The boundary spin $n=1$ is supplied as a boundary table of
twenty-eight shifted instances representing the Caron-Huot--Herranen $n=1$ function,
with no reassembly involved.

The evaluator in \texttt{anc/eval/} takes $(n,\nu)$ and returns $\chiord{2}$ on a
stated domain. The infinite sum in each nested atom is truncated
at a cutoff $J$ and closed by a tail, so $J$ must exceed every shift magnitude
$|z+k|$ on the ladder by a margin, and, as App.~\ref{app:val:precision} shows, the
margin needed for a given number of digits is larger than the condition that merely
keeps the tail convergent.

At $\nu=0$ the two arguments of
Eq.~\eqref{eq:cf:separable} coincide, $z=\zb=M$ with $M=(n-1)/2$, so the two halves are
equal and add,
\begin{equation}
\chiord{2}(n,0)=\tfrac12\,\Fn(M).
\label{eq:cf:intercept}
\end{equation}
Nothing cancels between the halves. What does cancel sits inside a single one of them:
$M$ is an interior point of the ladder for every odd $n\ge3$, since $0\le M\le n-2$
there, and at such a point the atom contributions have poles that cancel among
themselves within the block, by Eq.~\eqref{eq:res:laurent} at $m=M$. Term by term
the representation therefore has a removable singularity at $\nu=0$, and the intercept
is taken through a separate path (a mean-value average of $\Fn$ over a small circle
about $M$), not by evaluating the atoms at that point. For example,
\begin{equation}
\chiord{2}(7,\,0.3)=-139.14750268807367049579759648360604\ldots,
\label{eq:cf:worked}
\end{equation}
obtained from Eq.~\eqref{eq:cf:master} with the stored table at $n=7$. The digits are
stable against the truncation and the working precision, provided $\nu$ is taken at
that precision, not at machine precision (App.~\ref{app:val:precision}). The digits shown are truncated, not rounded, so a
reader checking against an independent implementation can compare them one at a time. A reader who
wants the eigenvalue at a spin and a value of $\nu$ needs Eq.~\eqref{eq:cf:master}, the
reductions of App.~\ref{app:reduction}, the forty closing rules of
Table~\ref{tab:cat:rules}, and the seven of Sec.~\ref{sec:residual}:
Eqs.~\eqref{eq:res:c112} to~\eqref{eq:res:f011} for the four nested slots,
Eq.~\eqref{eq:res:KR3} for $\KRthree$, and the regularity identity
Eq.~\eqref{eq:res:laurent} for the two rational kernels left. At odd $n\ge3$ the tables
are a convenience and not the definition; the boundary spin is the exception noted
above.

The extraction of the atom table is exact and works at any odd $n$, since it is a
rewriting of the Caron-Huot--Herranen integrand at that spin with no fitting in it.
The coefficient rules of Sec.~\ref{sec:structure} were instead determined by exact
linear algebra on the spins computed, and hold on every coefficient of every one of
them, out to $n=101$. Section~\ref{sec:chk:oos} gives the spins that were computed after
the rules were written down and what the rules predicted there.

\section{The structure of the ladder block}\label{sec:structure}

\subsection{Atoms first, ladder second}
\label{sec:str:collapse}
The compact form of Eq.~\eqref{eq:cf:ladderform} rests on one fact: the two
summations, over slots and over rungs, can be interchanged. The
coefficient of a given atom depends on the spin and on the rung, but for forty of the
forty-seven slots not freely: each of those forty is built from the same seven
dressings, one of them the constant $1$ and the other six functions of the two ladder
distances,
\begin{equation}
\begin{aligned}
w_f(n,k)\in\bigl\{\,&1,\ (-1)^{k},\ (-1)^{k}\Sh{1}{\ap},\ (-1)^{k}\Sh{1}{\am},\\
&(-1)^{k}\Sh{1}{\amin},\ (-1)^{k}\Sh{2}{\ap},\ (-1)^{k}\Sh{-2}{\ap}\,\bigr\},
\end{aligned}
\label{eq:str:weights}
\end{equation}
and one of them differs from its neighbours only in the rational multiples it assigns to
them, $\coeff{s}(n,k)=\sum_f c_{s,f}\,w_f(n,k)$. These forty are the closing slots of
Sec.~\ref{sec:set:vocab}. The remaining seven do not lie in the span of
Eq.~\eqref{eq:str:weights}: they are collected by $\Kres$ and taken up in
Sec.~\ref{sec:residual}, and the present section is about the forty. The weights are
the same seven functions in every one of the forty slots, so the slot label and the
rung label decouple and
the two sums can be exchanged,
\begin{equation}
\sum_{s\,\in\,\mathrm{closing}}\mu_s\sum_{k\in K_s}\coeff{s}(n,k)\,\atom{s}(z+k)
=\sum_{f}\sum_{k\in K^{f}}w_f(n,k)\,\Gladder_{\!f}(z+k),
\quad
\Gladder_{\!f}(x)=\sum_{s\,\in\,\mathrm{closing}}\mu_s\,c_{s,f}\,\atom{s}(x).
\label{eq:str:regroup}
\end{equation}
The two ranges are not the same one. On the left $K_s$ is the rung range of the slot
$s$, the \texttt{supp.} column of Table~\ref{tab:cat:rules}. On the right $K^{f}$ is
the range the weight $w_f$ runs over: the ladder $-(n-1)\le k\le0$ for the six weights
dressed by $(-1)^{k}$, and the single point $k=1$ for the constant weight, whose two
slots sit there and at no other rung. Where a slot stops short of the ladder its own
weight has already vanished at the rungs left out, since those weights are
$\Sh{1}{\ap}$ at $\ap=0$ and $\Sh{2}{\ap}+\Sh{-2}{\ap}$ at $\ap=0$ and $\ap=1$, so the
wider range on the right adds nothing.
On the right one function of one variable survives per weight, carrying no
spin and no rung index, with all of that moved outside into the range of the ladder
sum. Seven weights survive with nonzero content, six of them on the ladder, and the
sizes of those six are uneven,
\begin{equation}
\begin{array}{lclcl}
\Gladder_{(-1)^k}&:&26\ \text{atoms},\qquad&
\Gladder_{(-1)^k S_1(\amin)}&:\ 8\ \text{atoms},\\[2pt]
\Gladder_{(-1)^k S_1(\ap)}&:&11\ \text{atoms},&
\Gladder_{(-1)^k S_2(\ap)}&:\ 1\ \text{atom},\\[2pt]
\Gladder_{(-1)^k S_1(\am)}&:&8\ \text{atoms},&
\Gladder_{(-1)^k S_{-2}(\ap)}&:\ 1\ \text{atom}.
\end{array}
\label{eq:str:sizes}
\end{equation}
The last two are single rational atoms, both equal to $\tfrac{16}{3}\pi^2/x$, which
is why they appear in Eq.~\eqref{eq:cf:ladderform} as one term. The function
$\Gladder_{(-1)^kS_1(\ap)}$ includes three rational atoms alongside its nested ones,
so rational content enters these objects as well.

The seventh weight is the constant $1$, and it holds two atoms, both of coefficient
$32$, at the rung $k=1$ alone. They merge, since
$\Gladder_{1}(z+1)=32\,\zeta(5,z+2)+32/(z+1)^{5}=32\,\zeta(5,z+1)$, and it is that
single Hurwitz zeta which stands under the underbrace of
Eq.~\eqref{eq:cf:ladderform}. The symbol $\Gladder_{1}$ therefore means the function
at $f=1$ in the indexing of Eq.~\eqref{eq:str:regroup} and nothing else; the
twenty-six-atom function is $\Gladder_{(-1)^k}$ throughout.

\subsection{Three weights become one}
\label{sec:str:threetoone}
The three weights carrying an $S_1$ are not independent, and the way they fail to be
independent is the sharpest statement in the block. Their nested parts
are the same four atoms up to an overall sign,
\begin{equation}
\Bigl[\Gladder_{(-1)^kS_1(\ap)}\Bigr]_{\text{nested}}
=\Bigl[\Gladder_{(-1)^kS_1(\am)}\Bigr]_{\text{nested}}
=-\Bigl[\Gladder_{(-1)^kS_1(\amin)}\Bigr]_{\text{nested}}
=:\ \Phi ,
\label{eq:str:cores}
\end{equation}
so the entire nested content of the $S_1$ sector enters with a single combined
weight $(-1)^k[\Sh{1}{\ap}+\Sh{1}{\am}-\Sh{1}{\amin}]$. Since
$\{\ap,\am\}=\{\amin,\amax\}$ as a pair, $\Sh{1}{\ap}+\Sh{1}{\am}=\Sh{1}{\amin}+\Sh{1}{\amax}$
at every rung, and the three weights merge into
\begin{equation}
(-1)^{k}\,\Sh{1}{\amax},\qquad \amax=\max(\ap,\am),
\label{eq:str:max}
\end{equation}
which is the term that appears in Eq.~\eqref{eq:cf:ladderform}. The merged weight is
built from $\amax$ only, and $\amax$ is invariant under exchange of the two
distances, so this step is also what makes the symmetry of
Sec.~\ref{sec:str:reflection} visible.

The remainder beyond $\Phi$ in the three $S_1$ functions has empty nesting word.
The purely rational combination shows it directly,
\begin{equation}
\Gladder_{(-1)^kS_1(\ap)}+\Gladder_{(-1)^kS_1(\am)}
+2\,\Gladder_{(-1)^kS_1(\amin)}
=\frac{48}{x^{4}}+\frac{16\pi^{2}}{3x^{2}}+\frac{32\zeta_3}{x},
\label{eq:str:R1}
\end{equation}
with no nested sum on the right, and a second combination leaves only atoms of
empty word,
\begin{equation}
\Gladder_{(-1)^kS_1(\am)}+\Gladder_{(-1)^kS_1(\amin)}
=64\!\sum_{N\ge1}\frac{1}{N^{3}(N+x)}
+32\!\sum_{N\ge1}\frac{(-1)^{N}}{N^{2}(N+x)^{2}}
+64\!\sum_{N\ge1}\frac{(-1)^{N}}{N^{3}(N+x)} ,
\label{eq:str:R2}
\end{equation}
which partial fractions turn into zeta values and polygammas at halved argument.
The remainders $D_{+}$, $D_{-}$, $D_{\min}$ of Eq.~\eqref{eq:cf:ladderform}, of
seven, four and four atoms, are what Eqs.~\eqref{eq:str:R1} and~\eqref{eq:str:R2}
leave behind.

The same reduction applied to $\Gladder_{(-1)^k}$ turns its six depth-two atoms into
products of depth-one sums together with a single irreducible nested object,
$\Sh{-2,1}{N}-\Sh{1,-2}{N}$, through the stuffle relations of
App.~\ref{app:reduction}. Three product-type objects survive alongside it, so
$\Gladder_{(-1)^k}$ collects several distinct objects under one ladder weight.

\subsection{Exchanging the two ladder distances}
\label{sec:str:reflection}
The reflection of the ladder about its midpoint sends a rung to
\begin{equation}
k\ \longmapsto\ k^{*}=-(n-1)-k ,
\label{eq:str:refl}
\end{equation}
which exchanges the two distances, $\ap\leftrightarrow\am$. In the centred index
$s=k+(n-1)/2$ the ladder points are $z+k=i\nu+s$ and the exchange is $s\to-s$, a
reflection about the midpoint $i\nu$; equivalently $z+k\to-\overline{(z+k)}$, which
swaps the two ends $z$ and $-\zb$ of the segment. At odd $n$ the alternation is
unaffected, $(-1)^{k^{*}}=(-1)^{k}$, so the exchange grades the coefficients and does
not mix them with the dressing.

The exchange acts on the coefficients cleanly. Of the forty slots that close, thirty-one
satisfy $\coeff{}(n,k)=\coeff{}(n,k^{*})$ exactly, at every rung of every spin
computed. Two more are the single-rung slots at $k=1$, on which the exchange does
not act. Six of the seven closing slots that fail the exchange use $\Sh{1}{\ap}$ and
$\Sh{1}{\am}$ with unequal
coefficients, and $\Sh{1}{\ap}\leftrightarrow\Sh{1}{\am}$ is what the exchange does.
The seventh of them contains no $\Sh{1}{}$: it is the $\pi^2$ slot of $1/(z+k)$, with
rule $\tfrac{16}{3}(-1)^{k}[\Sh{2}{\ap}+\Sh{-2}{\ap}]$, and it fails because
$\Sh{2}{\ap}$ and $\Sh{-2}{\ap}$ are not themselves exchange-invariant. Among the seven
slots that do not close in the two distances, four have a definite sign,
\begin{equation}
\coeff{}(n,k)=-\coeff{}(n,k^{*})\ \ \text{for one},
\qquad
\coeff{}(n,k)=+\coeff{}(n,k^{*})\ \ \text{for the other three},
\label{eq:str:parity}
\end{equation}
on all $8415$ coefficients of those four slots, and on a further $1532$ at the four
spins $n=93,95,97,99$. The first
count is $3\times2115+2070$ and the second $3\times384+380$: the antisymmetric slot
has one rung fewer per spin than the other three, and that rung is always the
midpoint $k=-(n-1)/2$, the fixed point of the exchange, where antisymmetry forces the
coefficient to vanish. The three rational slots among the seven
have no definite sign under the exchange at any centring.

Two of the seven slots that do not close share one coefficient rule. The rules of $\atomlab{\BAatom}{(1,(),((0,1)),2)}$ and
$\atomlab{\BAatom}{(1,(),((0,2)),1)}$ agree on every coefficient, at every spin,
and the two atoms, written $\Om{A}$ and $\Om{B}$ in Sec.~\ref{sec:res:what}, occur
only in the combination
\begin{equation}
\Om{A}(x)+\Om{B}(x)=
\sum_{N\ge1}\frac{(-1)^{N}}{N(N+x)^{2}}+\sum_{N\ge1}\frac{(-1)^{N}}{N^{2}(N+x)}
=\frac1x\Bigl[-\frac{\pi^{2}}{12}
-\frac{\zeta\bigl(2,\tfrac{x+2}{2}\bigr)-\zeta\bigl(2,\tfrac{x+1}{2}\bigr)}{4}\Bigr],
\label{eq:str:merge}
\end{equation}
which is elementary. There are therefore seven rules, one to a slot, and six of them are
distinct, one of the six attached to an atom combination with no nested content.
Section~\ref{sec:res:what} counts the six against the two spans of
Sec.~\ref{sec:set:vocab} and says what those counts settle.

\section{The residual contribution and its arithmetic obstruction}\label{sec:residual}

\subsection{The remaining seven slots}
\label{sec:res:what}
Seven of the forty-seven slots have coefficient rules that do not close in the two
ladder distances. By the shared rule of Sec.~\ref{sec:str:reflection} they follow six
distinct rules. Two spans
are needed to count them and both were fixed in Sec.~\ref{sec:set:vocab}. None of the six
lies in the weight span of Eq.~\eqref{eq:str:weights}, and modulo that span their images
are independent, so counted there they are six. Modulo the wider elementary algebra they
span five directions, because one combination of them is elementary:
\begin{equation}
\begin{aligned}
&\coeff{}^{\,\atomlab{\BAatom}{(0,(),((0,2)),1)}}(n,k)
+\coeff{}^{\,\atomlab{\BAatom}{(1,(),((0,1)),2)}}(n,k)\\
&\qquad=(-1)^{k}\Bigl[\,192\,\Sh{1}{\amin}\Sh{1}{\amax}-64\,\Sh{1}{\amin}^{2}\\
&\qquad\qquad\quad-32\,\Sh{1}{\amax}^{2}+96\bigl(\Sh{2}{\ap}+\Sh{2}{\am}\bigr)\Bigr].
\end{aligned}
\label{eq:res:sixtofive}
\end{equation}
This holds on all $2499$ coefficients of the two arrays at odd $n\le99$, exactly, and
the four rational multiples are the only solution in the weight-two basis up to the
trivial relabelings that exchange $\min$ and $\max$; interchanging them in the
$-64$ and $-32$ terms fails on $2450$ of the $2499$ cells. It is the one closed form we
have for a residual coefficient away from the ends of the ladder that calls on no sum
along the ladder: its right-hand side uses $\Sh{1}{\amin}$, $\Sh{1}{\amax}$,
$\Sh{2}{\ap}$ and $\Sh{2}{\am}$ and nothing else, where
Eqs.~\eqref{eq:res:KR3}, \eqref{eq:res:c112}, \eqref{eq:res:c021},
\eqref{eq:res:c011} and~\eqref{eq:res:f011} all need the sums of
Eq.~\eqref{eq:res:laddersums}.

The step from six to five counts directions and not atoms. It is worth emphasizing that
eliminating a rule removes nothing from the block: the seven atoms stay where they are,
each with a coefficient in front of it, and an evaluable display needs a coefficient on
every atom, not one per direction. The residual contribution to
Eq.~\eqref{eq:cf:ladderform} therefore reads
\begin{equation}
\begin{aligned}
\Kres(n,z)&=\sum_{k=-(n-2)}^{0}\Bigl[\frac{\KRone(n,k)}{z+k}+\frac{\KRtwo(n,k)}{(z+k)^{2}}
+\frac{\KRthree(n,k)}{(z+k)^{3}}\Bigr]\\
&\quad+\sum_{k=-(n-1)}^{0}\Bigl[\cOm(n,k)\,\Om{3}(z+k)+\cCm(n,k)\,\Om{C}(z+k)\\
&\qquad\qquad\quad+\dAB(n,k)\bigl(\Om{A}(z+k)+\Om{B}(z+k)\bigr)\Bigr],
\end{aligned}
\label{eq:res:Kdef}
\end{equation}
six coefficient functions on seven atoms, over the two rung ranges of
App.~\ref{app:grd:seven}: the rational atoms reach $k=-(n-2)$ and the nested ones one
rung further, to $k=-(n-1)$. The three coefficient functions on the nested atoms are
$\cOm=\coeff{}^{\,\atomlab{\BAatom}{(0,(),((0,1)),1)}}$,
$\cCm=\coeff{}^{\,\atomlab{\BAatom}{(0,(),((0,2)),1)}}$ and
$\dAB=\coeff{}^{\,\atomlab{\BAatom}{(1,(),((0,1)),2)}}
     =\coeff{}^{\,\atomlab{\BAatom}{(1,(),((0,2)),1)}}$, the last equality being the
shared rule of Sec.~\ref{sec:str:reflection}; they are given by
Eqs.~\eqref{eq:res:c011}, \eqref{eq:res:c021} and~\eqref{eq:res:c112}. The other three
multiply the rational atoms $1/(z+k)^{q}$ and we call them the rational kernels. The
four nested atoms are
\begin{equation*}
\begin{aligned}
\Om{3}(x)&=\sum_{N\ge1}\frac{1}{N(N+x)}=\frac{\Sh{1}{x}}{x},\\[2pt]
\Om{C}(x)&=\sum_{N\ge1}\frac{1}{N^{2}(N+x)}
 =\frac{\pi^{2}}{6x}-\frac{\Sh{1}{x}}{x^{2}},\\[2pt]
\Om{A}(x)&=\sum_{N\ge1}\frac{(-1)^{N}}{N(N+x)^{2}}
 =-\frac{\lntwo+\betaf(x)}{x^{2}}
  -\frac{\zeta\bigl(2,\tfrac{x+2}{2}\bigr)-\zeta\bigl(2,\tfrac{x+1}{2}\bigr)}{4x},\\[2pt]
\Om{B}(x)&=\sum_{N\ge1}\frac{(-1)^{N}}{N^{2}(N+x)}
 =\frac{\lntwo+\betaf(x)}{x^{2}}-\frac{\pi^{2}}{12x},
\end{aligned}
\end{equation*}
with $\betaf$ as in Sec.~\ref{sec:cf:space}. The last two are the pair of
Eq.~\eqref{eq:str:merge}, and the display shows why they occur only in the sum: the
$\lntwo$ and the $\betaf$ cancel between them. That cancellation does not reach
$\Om{C}$. Folding $\Om{C}$ into the same combination is possible, at the price of an
elementary coefficient left behind on it,
\begin{equation*}
\cCm\,\Om{C}+\dAB\bigl(\Om{A}+\Om{B}\bigr)
=\dAB\bigl(\Om{A}+\Om{B}-\Om{C}\bigr)+\bigl(\cCm+\dAB\bigr)\Om{C},
\end{equation*}
with $\cCm+\dAB$ the elementary right-hand side of Eq.~\eqref{eq:res:sixtofive}. An
evaluable sum has to include that remainder, and $\Om{C}$ therefore keeps a coefficient of
its own above.

None of the six survives this section as an unknown. They are given as exact rational
numbers for every odd $n\le99$, and that tabulation now serves only as a record to
check the rules against; the definition of the coefficients lies in the rules.
The word kernel names a different object in the ancillary files, and
App.~\ref{app:conv:map} keeps the two apart: every coefficient function displayed here
is raw, in the sense fixed there, and the orthogonal residuals the ancillary collapse
works with are not interchangeable with them.

Fitting the seven slots over ladder bases graded by weight, in exact rational
arithmetic and using the exchange grading of Sec.~\ref{sec:str:reflection} to halve
the basis, puts the targets outside the span at weight five, six, seven and, for the
three slots of positive exchange sign, eight. Each of those is a statement about one
finite list of functions, of $486$, $982$, $1616$ and $2450$ members respectively, and
none of them says that no closed form exists. A search for a linear recurrence in $n$
with polynomial coefficients, for $\Kres(n,z)$ at fixed $z$, also comes up empty, over
the $76$ order-degree pairs that
$49$ spins leave overdetermined, at each of three rational values of $z$, which is $228$
stencils per kernel, each run with and without a polynomial right-hand side, and is
again a bound on the search, not on the object. Appendix~\ref{app:grd:open} gives the
basis dimensions and the bounds on both searches. We have no explanation for either
result at this stage. The first is settled below for
three of the six, and the second by the functions the searches were missing.

\subsection{Why the rational kernels cannot close in the two distances}
\label{sec:res:obstruction}
For the rational kernels the failure is not a matter of how far the search
went. It follows from the denominators, and the argument below is a proof for one of the
three, which is enough to settle the sector they close together.

A nested harmonic sum at integer argument $m$ has a denominator dividing
$\mathrm{lcm}(1,\dots,m)^{w}$, with $w$ its weight, and the denominator of a
rational combination divides the least common multiple of the denominators of its
parts. Cancellation can remove a prime; it cannot create one. So any expression
built from nested harmonic sums evaluated at $\ap$, at $\am$ or at $\min(\ap,\am)$, with
rational coefficients of common denominator $D$, has denominators whose primes lie among
those up to $\max(\ap,\am)$ together with the primes dividing $D$, whatever its weight or
depth.
The second set has to be there: a coefficient is a rational number of its own, and
$\tfrac{1}{67}\Sh{1}{1}$ has the prime $67$ in its denominator at $\max=1$.

That is why a bound at one spin proves nothing on its own, and why the argument has to be
asymptotic. A rule uniform in $n$ has one fixed $D$ for all spins, so it can supply only
finitely many primes beyond the ladder bound. Eq.~\eqref{eq:res:primefamily} exhibits
infinitely many spins at which $\KRthree$ needs a prime above $\max(\ap,\am)$,
and those primes are unbounded. No fixed $D$ can cover them, and the exclusion follows.

We write the hypothesis out in full, since what the argument excludes and what it leaves
open both sit inside it. A candidate expression is a finite $\mathbb{Q}$-linear
combination of products of nested harmonic sums evaluated at $\ap$, at $\am$ or at
$\min(\ap,\am)$, whose rational coefficients are the same numbers at every spin. That is
the elementary algebra of Sec.~\ref{sec:set:vocab} and it is the only thing excluded.
Coefficients that are themselves rational functions of $n$ or of the two distances lie
outside the hypothesis and are not excluded by it: the rational tails $1/(\ap+i)$ of
Eq.~\eqref{eq:res:col1} are of that kind, and they are what closes the first four
columns. A denominator that moves with the spin defeats the argument, and we do not
claim otherwise.

The object the argument applies to is the coefficient rule. The
estimate below is computed from Eq.~\eqref{eq:res:KR3}. At the spins that were
generated the rule and the stored coefficient are the same rational number cell by cell,
so there the two statements agree. Past that range the statement is about the rule, and
carrying it to the coefficient of the eigenvalue itself needs the rule to be that
coefficient at every odd spin, which is verified out to the end of the computed range
and is not proved.

The residue of
Eq.~\eqref{eq:res:primeres} is computed for $\KRthree$, so the exclusion is proved for
that kernel only. The other two show the same signature over the computed range
without a proof: on the twenty-three family spins with $p+2\le99$, $\KRone$ shows the
prime $p$ in its denominator at every one of the twenty-three and $\KRtwo$ at twenty-one
of them, the exceptions being $p=5$ and
$p=47$. Since the three are closed together by one enlargement, excluding $\KRthree$
excludes closing the sector as a whole in the elementary algebra, and that is the
statement used below. For $\KRone$ and $\KRtwo$ separately the evidence is the prime
count just quoted, over a finite range, and nothing in the present study excludes them.

The rational kernels violate that bound, and the four nested slots do not.
Table~\ref{tab:res:primes} compares the largest denominator prime of each
coefficient with $\max(\ap,\am)$, over every coefficient with $n\ge21$.

\begin{table}[t]
\centering
\begin{tabular}{@{}lrrr@{}}
\toprule
slot & coefficients & with a prime $>\max(\ap,\am)$ & largest ratio\\
\midrule
$\atomlab{\BAatom}{(0,(),((0,1)),1)}$ & $2360$ & $0$    & n/a\\
$\atomlab{\BAatom}{(0,(),((0,2)),1)}$ & $2400$ & $0$    & n/a\\
$\atomlab{\BAatom}{(1,(),((0,1)),2)}$ & $2400$ & $0$    & n/a\\
$\atomlab{\BAatom}{(1,(),((0,2)),1)}$ & $2400$ & $0$    & n/a\\
\midrule
$1/(z+k)$        & $2360$ & $2112$ & $1.980$\\
$1/(z+k)^{2}$    & $2360$ & $2108$ & $1.980$\\
$1/(z+k)^{3}$    & $2360$ & $2072$ & $1.940$\\
\bottomrule
\end{tabular}
\caption{Largest denominator prime against $\max(\ap,\am)$, over all coefficients
with $n\ge21$. The four nested slots never exceed the bound. The three rational
slots exceed it on almost every coefficient, by a factor approaching two.}
\label{tab:res:primes}
\end{table}

The bound on the other side is sharp. Over the same coefficients no denominator
prime of the rational kernels exceeds $n-2$, and the largest observed ratio to $n-1$
is $0.9898$. Since $\ap+\am=n-1$, the denominators of the rational kernels are built
from $\mathrm{lcm}(1,\dots,n-2)$, which is set by the ladder as a whole, while
those of the four nested slots are built from $\mathrm{lcm}(1,\dots,\max(\ap,\am))$,
set by one distance along it. The two growth rates are incompatible once $n$ is large enough,
so no rewriting of $\KRthree$ inside the elementary algebra can succeed, at any weight
and any depth, and with it no rewriting of the sector the three kernels close together.
That exclusion rests on the hypothesis stated above: the rational coefficients have to be
the same numbers at every spin, whatever the weight and the depth. This does not say the kernels have no closed
form; it says the closed form cannot lie in that algebra, and
Sec.~\ref{sec:res:ladder} exhibits the one it does lie in.

The obstruction is not confined to the computed range, and Eq.~\eqref{eq:res:KR3} makes
that precise. For every prime $p\ge5$ take
\begin{equation}
n=p+2,\qquad \am=\tfrac{p-1}{2},\qquad \ap=\tfrac{p+3}{2},
\label{eq:res:primefamily}
\end{equation}
so that $\ap+\am=n-1$ and $\max(\ap,\am)=\tfrac{p+3}{2}<p$. The $p$-adic valuation of
$\KRthree(n,-\am)$ then follows. The shifted argument never switches, since
$\am+N\ge n-1-\am-N$ for every $N\ge1$ with equality at $N=1$, so the summand contains
$\Sh{1}{\am+N}$ throughout and the argument runs up to $p+1$. The remaining terms
$2/N^{2}$, $(-1)^{N}/N^{2}$ and $\Sh{1}{N}/N$ have $N\le\tfrac{p+3}{2}<p$ and are
$p$-integral. And $\Sh{1}{\am+N}$ contains $1/p$ only for $\am+N\ge p$, that is at
$N=\tfrac{p+1}{2}$ and $N=\tfrac{p+3}{2}$, so the whole $1/p$ content is
\begin{equation}
32\,(-1)^{\am}\,\frac{1}{p}\Bigl(\frac{2}{p+1}+\frac{2}{p+3}\Bigr)
=32\,(-1)^{\am}\,\frac{4(p+2)}{p\,(p+1)(p+3)} ,
\label{eq:res:primeres}
\end{equation}
whose numerator is prime to $p$. Hence $p$ divides the reduced denominator of
$\KRthree$ exactly once, at every prime $p\ge5$. At infinitely many spins the
denominator therefore contains a prime above $\max(\ap,\am)$, and no fixed set of
rational coefficients
can absorb them all. The ratio $p/\max(\ap,\am)=2p/(p+3)$ tends to two, which is the
limit approached in Table~\ref{tab:res:primes}.

The signature is not an artefact of the reduction, and the table above is what shows it.
One and the same extraction produces the four nested slots and the three rational ones,
and it separates them cleanly: over the same coefficients the nested slots never show a
denominator prime above $\max(\ap,\am)$ and the rational ones do so on almost every
cell. A step of the reduction that manufactured large primes would plant them in nested
and rational slots alike. The primes are set by the argument the coefficient needs, and
Eq.~\eqref{eq:res:KR3} shows that directly: the harmonic sum inside it runs up to the
full ladder length, and the two terms that supply $1/p$ in Eq.~\eqref{eq:res:primeres}
are the ones whose argument has passed $p$. This fact is evidence that the
obstruction belongs to the ladder and not to the reduction performed here. It is
still not a proof for $\KRone$ and $\KRtwo$.

The natural argument of these
kernels is the ladder as a whole, not either distance along it. Admitting nested
harmonic sums at the argument $n-2$ alongside $\ap$, $\am$ and $\min(\ap,\am)$ removes the
obstruction but does not close the gap on its own: the fits remain inconsistent at
weight three ($166$ functions), four ($478$) and five ($1126$). The gap is closed by an
identity, not by a larger basis, and Sec.~\ref{sec:res:midpoint} gives it: the
kernels are ladder sums whose summand contains the shifted argument
$\max(m+N,\,n-1-m-N)$, reaching the full ladder length as the denominators
require.

\subsection{Exact columns}
\label{sec:res:columns}
Before the general rules, three columns of the residual contribution close on their
own. They were found first, they are the boundary data everything later had to match,
and each is now a special case of the rules of Secs.~\ref{sec:res:midpoint}
and~\ref{sec:res:ladder}.

At the near end of the ladder every one of the six coefficient functions closes in nested
harmonic sums at argument $n-1$, that is on $\mathrm{lcm}(1,\dots,n-1)$, which is the
argument Sec.~\ref{sec:res:obstruction} says the natural one should be. The two
arguments named across these sections, $n-1$ here and $n-2$ there, carry the same
primes at every odd $n\ge5$, since $n-1$ is even and therefore composite, so nothing
enters $\mathrm{lcm}(1,\dots,n-1)$ that $\mathrm{lcm}(1,\dots,n-2)$ does not already
have. Writing all sums at $n-1$,
\begin{equation}
\begin{aligned}
\KRthree(n,0)&=32\,\Sh{1,1}{}, \\
\KRtwo(n,0)&=-32\,\Sh{1,2}{}+64\,\Sh{1,-2}{}-32\,\Sh{2,1}{}+32\,\Sh{-2,1}{}
             +144\,\Sh{3}{}+16\,\Sh{-3}{}, \\
\KRone(n,0)&=288\,\Sh{4}{}+288\,\Sh{-4}{}+128\,\Sh{-3,1}{}-32\,\Sh{-2,-2}{}
             +32\,\Sh{-2,2}{}-96\,\Sh{1,-3}{}\\
&\quad\ -32\,\Sh{1,3}{}-96\,\Sh{2,-2}{}-32\,\Sh{2,2}{}
             -128\,\Sh{1,-2,1}{}+64\,\Sh{1,1,-2}{} ,
\end{aligned}
\label{eq:res:col0}
\end{equation}
together with, for the three coefficient functions on the nested atoms,
\begin{equation}
\begin{aligned}
\cOm(n,0)&=-128\,\Sh{1,-2}{}+128\,\Sh{-3}{},\\[2pt]
\cCm(n,0)&=32\,\Sh{2}{}-32\,\Sh{-2}{}=64\,O_2(n-1),\\[2pt]
\dAB(n,0)&=-64\,\Sh{1,1}{}+96\,\Sh{2}{}+32\,\Sh{-2}{},
\end{aligned}
\label{eq:res:col0b}
\end{equation}
the second of these the odd harmonic sum of Eq.~\eqref{eq:cf:AB} again, and equal to
$64\,O_2(n-2)$ because $n-1$ is even. Each identity holds at every one of the $49$
computed spins. The last two are not independent of each other, since
Eq.~\eqref{eq:res:sixtofive} at $\am=0$ returns either one from the other.

The weights in Eq.~\eqref{eq:res:col0} follow a pattern. The kernel of the
simple pole closes at weight four, that of the double pole at weight three, that of
the triple pole at weight two. Added to the weight of the atom each multiplies, one,
two and three, every entry reaches five. Equation~\eqref{eq:res:col0b} does the same,
at weights three, two and two against atoms of weight two, three and three. The uniform
weight of Sec.~\ref{sec:dis:weight} therefore survives into the residual contribution, at
least along this column, which is the first quantitative statement we have about it.

At the far end the three rational kernels close as well,
\begin{equation}
\KRthree(n,-(n-2))=-32\,\Sh{1}{},\qquad
\KRtwo(n,-(n-2))=32\,\Sh{1}{}-64\,\Sh{1,1}{}+96\,\Sh{2}{}+32\,\Sh{-2}{},
\label{eq:res:coln}
\end{equation}
with $\KRone(n,-(n-2))=64\,\Sh{1}{}-64\,\Sh{1,1}{}-128\,\Sh{1,-2}{}+64\,\Sh{2}{}
+64\,\Sh{-2}{}+128\,\Sh{-3}{}$, again at argument $n-1$ and again at every computed
spin.

The near end does not stop at one column. Writing the coefficients as functions of
$\ap$ at fixed $\am$, the first four columns close, and they close in the same way:
\begin{equation}
\KRthree\big|_{\am}=32\,(-1)^{\am}\,\Sh{1,1}{\ap}+r_{\am}(\ap),
\qquad
\begin{aligned}
r_0&=0,\\
r_1&=-\frac{32}{\ap+1},\\
r_2&=\frac{96}{\ap}-\frac{48}{\ap+1}+\frac{16}{\ap+2},\\
r_3&=-\frac{160}{3(\ap-1)}-\frac{176}{3(\ap+1)}\\
   &\quad+\frac{80}{3(\ap+2)}-\frac{32}{3(\ap+3)},
\end{aligned}
\label{eq:res:col1}
\end{equation}
each exactly at every computed spin, $193$ coefficients in all. The one cell not
covered, at $n=5$ in the $\am=3$ column, is discussed below. The
transcendental part does not move from column to column: it is the same
$\Sh{1,1}{\ap}$ throughout, alternating in sign with $\am$, and only the rational tail
changes.

How far that tail reaches does change, and the way the
$\am=3$ column was found makes the point, because a function list fixed in advance misses it. Take the
nested harmonic sums of weight at most three in the words used elsewhere in this
paper, at both $\ap$ and $\ap+2$, together with the constant, the tails $1/(\ap+i)$ and
$1/(\ap+i)^{2}$ for $0\le i\le3$, and the two products $\Sh{1}{\ap}/\ap$ and
$\Sh{1}{\ap}^{2}$: thirty-five functions, of rank thirty, against the forty-eight
available coefficients in each column, so each test is overdetermined by eighteen. At
$\am=2$ the system is consistent and the solution of Eq.~\eqref{eq:res:col1} was fitted
on thirty-four spins and reproduces the remaining fourteen. At $\am=3$ that same list
gives $\mathrm{rank}[X|y]=31$ against $\mathrm{rank}[X]=30$ and nothing in it fits. The
deficiency belongs to the list and not to the coefficients: admitting the one further
shift $1/(\ap-1)$ removes it, and $r_3$ above then holds at every spin of the column
with $\ap\ge2$, fitted on five and reproducing the other forty-two. The one spin $n=5$,
where $\ap=1$, sits on the pole of that shift and is not covered by the formula; its
coefficient is $-200/3$.

So the column tests say where the rational tail widens, and no more than that. They are
not where the obstruction of Sec.~\ref{sec:res:obstruction} becomes visible: that
obstruction is arithmetic, it sits in the denominators, and no rank test against a
fixed list of functions can exhibit it.

\subsection{Regularity on the ladder, and what it determines}
\label{sec:res:midpoint}
The block has no singularity at an internal point of the ladder, $z=0,1,\dots,n-2$, and
no shape is singular at every rung. Where the poles sit is decided by the shape. A
rational shape $1/(z+k)^{q}$ has its only pole at its own rung $z=-k$, and the rational
shapes reach no further than $-(n-2)\le k\le0$, the one of pole order five sitting at
$k=1$ instead. A nested shape is regular at its own rung and has poles at $z=-k-N$ for
every $N\ge1$, which is to say at the rungs between its own and the near end, and past
that at the collinear points. The pole of order $q$ at an internal point $z=m$ is
therefore fed by the rational shape at the rung $k=-m$ together with the nested shapes
of the same pole order at the rungs $k=-m-N$, and by nothing else.

The sizes are those of a cancellation. At $n=3$ and $\nu=0.002$ the two contributions
to the pole at $z=1$, which come from neighbouring rungs, are each of order
$3\times10^{12}$, the order of $48/\nu^{4}$, while the sum over atoms is of order $42$.
The cancellation behind that number
is exact and holds at every internal point. For every $m=0,\dots,n-2$ and every pole
order $q$, the coefficient of $(z-m)^{-q}$ vanishes,
\begin{equation}
\coeff{}^{(\Ratom,q)}(n,-m)
+\sum_{N\ge1}\ \sum_{\text{slots of pole order }q}
\coeff{}^{(\BAatom,\sigma,v,\{(d,p)\},q)}(n,-m-N)\;
\frac{\sigma^{N}\Sh{v}{N}}{\prod_{(d,p)}(N+d)^{p}}\;=\;0 ,
\label{eq:res:laurent}
\end{equation}
separately for each of the constants $1$, $\pi^2$ and $\zeta_3$: $2450$ rungs per
pole order over the $49$ computed spins, every one an exact rational zero. This is not
automatic. A Mellin transform is not pole-free where its integrand is regular
($\int_0^1 x^{z-1}dx=1/z$ has one), and nothing in the extraction imposes a
condition at the rungs; repeating the identity in the strict nesting convention breaks
it at every spin. Equation~\eqref{eq:res:laurent} expresses a property of the
eigenvalue itself: at fixed odd $n$ its poles in $\gamma$ sit only on the collinear
and anticollinear towers, and the ladder is the analyticity strip between them. The pole
decomposition of Refs.~\cite{Joubat:2020vrw,Joubat:2023ftd} rests on the same
separation, the sums at $z$ holding one tower and the sums at $\zb$ the other; the
regularity of the block between the towers at three loops is new here.

The point $k=1$ lies outside that strip and behaves in the opposite way. There the
shape $1/(z+1)^{5}$ stands alone at pole order five, with coefficient $32$ and nothing
to cancel it, so the block is singular at $z=-1$. That singularity is the leading
collinear pole of Eq.~\eqref{eq:chk:collinear}, and Eq.~\eqref{eq:res:laurent} is not
asserted there.

At each pole order the only unknowns in Eq.~\eqref{eq:res:laurent} are residual
coefficients, and the count differs with $q$. At $q=3$ no non-closing nested
slot enters, so the identity solves for $\KRthree$ outright, at every cell:
\begin{equation}
\KRthree(n,-m)=32\,(-1)^{m}\sum_{N=1}^{\,n-1-m}
\Bigl[\frac{2}{N^{2}}+\frac{(-1)^{N}}{N^{2}}
-\frac{\Sh{1}{N}}{N}
+\frac{\Sh{1}{\max(m+N,\,n-1-m-N)}}{N}\Bigr].
\label{eq:res:KR3}
\end{equation}
The four closing rules of pole order three assemble into this single finite sum, the
window rule entering through $\Sh{1}{\ap}+\Sh{1}{\am}-\Sh{1}{\amin}=\Sh{1}{\amax}$
evaluated at the shifted rung. Equation~\eqref{eq:res:KR3} reproduces the stored
array on all $2450$ cells, and every column previously obtained by fitting (both ends
of the ladder, the first four columns in $\am$, and the midpoint) follows from it by
specialisation. Its form also closes the loop with Sec.~\ref{sec:res:obstruction}: the
harmonic sum inside runs up to the full ladder length $n-1$, the argument the
denominators require and the two-distance algebra cannot reach.

At $q=2$ the only non-closing nested entrant is $\Om{A}$, so the same identity
determines $\KRtwo$ from that one coefficient; at $q=1$ the entrants are the three
remaining nested slots, $\Om{3}$, $\Om{B}$ and $\Om{C}$, determining $\KRone$ from
them, so all three coefficients of Eq.~\eqref{eq:res:Kdef} are needed here. Both
hold on all $2450$ cells. The cases $q=4$ and $q=5$ impose nothing new: $q=4$ collapses
to a free check on the closing rule $48(-1)^{k}\Sh{1}{\ap}$ of Eq.~\eqref{eq:cat:q4},
and the only pole-order-five atoms sit at $k=1$, off the ladder.

Equation~\eqref{eq:res:laurent} is also a warning for anyone evaluating the block
near $\nu=0$: the cancellation it expresses costs roughly eleven digits even at
$n=3$, $\nu=0.002$, and more at larger spin. Appendix~\ref{app:val:precision} says what
that means in practice.

\subsection{The ladder sums, and the last three rules}
\label{sec:res:ladder}
Equation~\eqref{eq:res:KR3} is more than an answer for one kernel, since it names the
class of function this sector is built from. Every term in it is a sum along the ladder whose
summand is a harmonic sum evaluated at an argument that moves with the summation index
and depends on both distances. Writing
\begin{equation}
\begin{aligned}
\Lam(\ap,\am)&=\sum_{N=1}^{\ap}\frac{\Sh{1}{\max(\am+N,\ap-N)}}{N},\\
\Lax{j}(\ap,\am)&=\sum_{N=1}^{\ap}\frac{\Sh{1}{\max(\am+N,\ap-N)}}{N^{j}},
\end{aligned}
\label{eq:res:laddersums}
\end{equation}
and similarly $\Lin{j}$ with $\min$ in place of $\max$ and $\Lbb{j}(\ap)$ with the
argument $\ap-N$ alone, the three remaining rules close. The first line of
Eq.~\eqref{eq:res:laddersums} is the second at $j=1$, so $\Lam=\Lax{1}$. There is independent reason to
try these functions in place of a larger basis of ordinary harmonic sums: the boundary
coefficient of Eq.~\eqref{eq:res:col0b} is one of them, since at $\am=0$ the sum
$\sum_{N}\Sh{-2}{N}/N$ is $\Sh{1,-2}{n-1}$.

The $\max$ restricts the sum less than the notation suggests. Taken in the order
$\Lam(\amin,\amax)$ it never switches, since $\amax+N$ exceeds $\amin-N$ for every term,
and the sum is the ordinary shifted object $\sum_{N=1}^{\amin}\Sh{1}{\amax+N}/N$. Taken
the other way it switches exactly once, at $N=(\amax-\amin)/2$, which is an integer
because $\ap+\am=n-1$ is even, so
\begin{equation}
\Lam(\amax,\amin)=\sum_{N=1}^{(\amax-\amin)/2}\frac{\Sh{1}{\amax-N}}{N}
+\sum_{N=(\amax-\amin)/2+1}^{\amax}\frac{\Sh{1}{\amin+N}}{N} .
\label{eq:res:split}
\end{equation}
Both pieces are ordinary sums with a shifted argument; the $\max$ only marks where
the larger of the two arguments switches from one branch to the other. The combination that actually occurs in the
rules is the symmetric one, and it collapses further than a first fit suggests. Writing
\begin{equation}
\Psiw_{j,q}(\ap,\am)=-\sum_{N=1}^{(\amax-\amin)/2}\frac{1}{N^{j}}
\sum_{i=\amin+N+1}^{\amax-N}\frac{1}{i^{q}}\,,
\qquad\Psiw\equiv\Psiw_{1,1},
\label{eq:res:window}
\end{equation}
a sum whose two limits move towards each other as $N$ grows and which vanishes whenever
$\amax-\amin\le2$, one has
\begin{equation}
\Lam(\ap,\am)+\Lam(\am,\ap)=\Sh{1,1}{\ap}+\Sh{1,1}{\am}+\Sh{1}{\ap}\Sh{1}{\am}
-\Psiw(\ap,\am),
\label{eq:res:collapse}
\end{equation}
exactly at every cell. The single-argument object reduces as well,
$\Lbb{1}(\ap)=\Sh{1}{\ap}^{2}-\Sh{2}{\ap}$ and
$\Lbb{2}(\ap)=2\,\Sh{2,1}{\ap}+\Sh{1,2}{\ap}-3\,\Sh{3}{\ap}$. The $\max$ and $\min$
orderings are not independent either, since
$\Lax{j}+\Lin{j}=\Lbb{j}(\ap)+\sum_{N=1}^{\ap}\Sh{1}{\am+N}/N^{j}$, though that shifted
sum still reaches the argument $n-1$ and is not elementary; it is the
exchange-symmetrized combination at $j=1$, Eq.~\eqref{eq:res:collapse}, that closes.
The new content therefore sits in the window sums, not in the ladder sums as a class:
$\Psiw$ at weight two together with a second member of the family at
weight three, which the antisymmetric slot of Eq.~\eqref{eq:res:f011} needs. A fit against
products of nested sums at $\ap$, $\am$, $\min$, $\max$ and $n-1$ misses
Eq.~\eqref{eq:res:collapse} and is inconsistent, but that is a statement about the
basis, which does not contain $\Lin{}$; the combination is not excluded.

The window sums do not reduce in turn. Adjoined to a weight-two basis of eighty-five
functions of $\ap$, $\am$, $\min$, $\max$ and $n-1$, of rank thirty across $495$ cells,
$\Psiw$ raises the rank by one. At weight three the two members $\Psiw_{2,1}$ and
$\Psiw_{1,2}$ each raise it by one again, even when $\Psiw$ and its products with
weight-one sums are already in the basis. All are exact rational rank computations, so
they exclude those bases and nothing wider.

Which of the two the antisymmetric slot takes is settled by the same splitting. Writing
$T_j(a,b)=\sum_{N=1}^{a}\Sh{1}{b+N}/N^{j}$, the pair of orderings separates as
$\Lin{j}(a,b)=\Lbb{j}(a)+\theta(a>b)\,\Psiw_{j,1}(a,b)$ and
$\Lax{j}(a,b)=T_j(a,b)-\theta(a>b)\,\Psiw_{j,1}(a,b)$, exactly. Substituted into
Eq.~\eqref{eq:res:f011}, the window content is $(-96)(+1)+(-128)(-1)=+32$ times
$\theta(\ap>\am)\,\Psiw_{2,1}$, and antisymmetrizing turns the step function into a sign:
\begin{equation}
\begin{aligned}
\fa(\ap,\am)-\fa(\am,\ap)&=g(\ap)-g(\am)
-128\bigl[\Sh{2}{\ap}\Sh{1}{\am}-\Sh{2}{\am}\Sh{1}{\ap}\bigr]\\
&\quad+32\,\mathrm{sgn}(\ap-\am)\,\Psiw_{2,1}(\ap,\am),
\end{aligned}
\label{eq:res:windowsel}
\end{equation}
with $g(a)=64\,\Sh{2,1}{a}+96\,\Sh{1,2}{a}-160\,\Sh{3}{a}-64\,\Sh{1,-2}{a}
-32\,\Sh{1}{a}\Sh{2}{a}+32\,\Sh{-3}{a}$, exact at every cell. So the slot takes
$\Psiw_{2,1}$ with coefficient $32$ and $\Psiw_{1,2}$ with coefficient zero. That
$\Psiw_{1,2}$ cannot occur is visible without computing: Eq.~\eqref{eq:res:f011} is built
from ladder sums of outer index two carrying $\Sh{1}{}$, so any window sum it generates
has outer exponent two and inner exponent one. The object the slot needs is the
exchange-odd $\mathrm{sgn}(\ap-\am)\,\Psiw_{2,1}$; the rank tests above are run on one
representative per exchange orbit, so they see $\Psiw_{2,1}$ itself.

The two slots that share a rule, and the one paired with them by
Eq.~\eqref{eq:res:sixtofive}, are
\begin{align}
\coeff{}^{\,\atomlab{\BAatom}{(1,(),((0,1)),2)}}
&=(-1)^{k}\Bigl[-32\bigl(\Lam(\ap,\am)+\Lam(\am,\ap)\bigr)-32\,\Sh{1}{\amin}^{2}\notag\\
&\hspace{3.2em}+96\,\Sh{1}{\ap}\Sh{1}{\am}+32\bigl(\Sh{2}{\ap}+\Sh{2}{\am}\bigr)\Bigr],
\label{eq:res:c112}\\[2pt]
\coeff{}^{\,\atomlab{\BAatom}{(0,(),((0,2)),1)}}
&=(-1)^{k}\bigl[32(\Lam(\ap,\am)+\Lam(\am,\ap))+192\,\Sh{1}{\amin}\Sh{1}{\amax}
 -32\,\Sh{1}{\amin}^{2}\notag\\
&\hspace{3.2em}-32\,\Sh{1}{\amax}^{2}-96\,\Sh{1}{\ap}\Sh{1}{\am}
 +64(\Sh{2}{\ap}+\Sh{2}{\am})\bigr],
\label{eq:res:c021}
\end{align}
each exact on all $2499$ coefficients, and their sum reproduces
Eq.~\eqref{eq:res:sixtofive} identically. The antisymmetric slot needs weight three and
the exchange-odd combination
\begin{equation}
\coeff{}^{\,\atomlab{\BAatom}{(0,(),((0,1)),1)}}(n,k)
=(-1)^{k}\bigl[\,\fa(\ap,\am)-\fa(\am,\ap)\,\bigr],
\label{eq:res:c011}
\end{equation}
\begin{equation}
\begin{aligned}
\fa(\ap,\am)&=192\,\Lbb{2}(\ap)-96\,\Lin{2}(\ap,\am)-128\,\Lax{2}(\ap,\am)
 -64\,\Sh{1,-2}{\ap}\\
&\quad-32\,\Sh{1}{\ap}\Sh{2}{\ap}+128\,\Sh{3}{\ap}+32\,\Sh{-3}{\ap},
\end{aligned}
\label{eq:res:f011}
\end{equation}
exact on all $2450$ coefficients. Both were fixed at low spin and hold
everywhere else without adjustment, and the class is narrow enough for that to mean
something: the span of Eq.~\eqref{eq:res:c112} has rank four across the $2499$ cells
and that of Eq.~\eqref{eq:res:f011} rank seven across $2450$, so the agreement is
heavily overdetermined. Both reach that rank at the lowest spins,
Eq.~\eqref{eq:res:c112} on the eight cells with $n\le5$ and Eq.~\eqref{eq:res:f011} on
the twenty cells with $n\le9$, so the coefficients are fixed there and every cell above
them is a test. Changing a single coefficient by one, flipping the exchange
sign, or offering the other array in its place all make the system inconsistent.

\subsection{The eigenvalue in closed form}
\label{sec:res:remaining}
At odd $n\ge3$ the definition of the eigenvalue needs no tabulated coefficient.
Equations~\eqref{eq:res:c112}, \eqref{eq:res:c021} and~\eqref{eq:res:c011} give the four
nested slots, with the first covering the two that share a rule. Once those
coefficients are known the three rational kernels follow from
Eq.~\eqref{eq:res:laurent}, and $\KRthree$ is explicit through
Eq.~\eqref{eq:res:KR3}. All forty-seven slots therefore follow rules uniform in the
conformal spin at every odd $n\ge3$, and the boundary block of Sec.~\ref{sec:cf:eval}
supplies $n=1$, so Eq.~\eqref{eq:cf:master} is defined at every odd $n$. Reconstructed
from the computed spins, those rules hold exactly wherever a spin was generated and are
unproved at arbitrary odd $n$, with the range and the sense of that fixed in
Sec.~\ref{sec:checks}. Forty-four
of the rules are fixed by exact linear algebra in a finite-dimensional class and hold on
every computed coefficient; the three rational kernels are not fitted but forced
by the regularity of the block on the ladder.

The atom table assembled from the rules is the stored table entry for entry.
Section~\ref{sec:chk:closed} gives the spins at which that rebuild was run, and the comparison
from the printed equations over the whole range.

The alphabet grows by one kind of object only. Beyond the one-variable content that
Sec.~\ref{sec:cf:space} counts, the coefficients need the ladder sums of
Eq.~\eqref{eq:res:laddersums}, in which a harmonic sum of weight at most two is
evaluated at $\max(\am+N,\ap-N)$ or its $\min$ and summed against $1/N^{j}$. They
reduce to ordinary nested sums at either end of the ladder, where a single argument is
the larger one throughout, and the $\max$ is where the ladder length enters in between.

For the rational kernels that enlargement is not a convenience, and the same arithmetic
that obstructed them shows why. A single ladder sum contains denominator primes above
$\max(\ap,\am)$: at $n=99$ and $\ap=65$, for instance, $\Lam$ has the prime $97$ in its
denominator, and so does $\KRthree$ at the same spin and rung, $\KRthree(99,-33)$. By the argument of Sec.~\ref{sec:res:obstruction}, with the hypothesis
stated there, such an object cannot be rewritten in nested harmonic sums at $\ap$,
$\am$ or $\min(\ap,\am)$ with rational coefficients fixed once for all spins, at any
weight or depth.

The same arithmetic does not reach the four nested slots, and we do not claim it does.
There the primes above $\max(\ap,\am)$ cancel in the symmetric combination that actually
occurs, and no coefficient of those four exceeds the bound at any computed cell, as
Table~\ref{tab:res:primes} shows. That the window sums are needed there rests on the
rank statements above and not on the arithmetic. They are sufficient and they lie
outside the bases we searched. Whether a smaller enlargement would serve is beyond the
scope of the present study.

\section{Checks}\label{sec:checks}

The statements made so far come in different strengths, and three words keep those
strengths apart in what follows. A statement is \emph{exact} when it holds in rational
arithmetic at a spin that was generated, with nothing rounded on either side of the
comparison. It is \emph{verified} when it has been tested over a stated finite range
of spins and holds there. It is \emph{proved} when an argument covers every odd spin,
not a range of them: the algebraic reductions of App.~\ref{app:reduction} are of that
kind, those of App.~\ref{app:red:nested} through the weight-four reduction cited
there, and so is the exclusion of Sec.~\ref{sec:res:obstruction}, which states the
hypothesis it needs alongside it. The coefficient rules are not. A figure is quoted
here only for an equation that a program in the ancillary files takes, as printed, as
its input. Where such a program tests something narrower than the display beside it,
the narrower statement is the one made and the restriction is named.

\subsection{The printed equation against the extraction}
\label{sec:chk:closed}
The tightest check is that the equation as printed reproduces the extraction it is a
rewriting of. Equation~\eqref{eq:cf:ladderform} together with
Eq.~\eqref{eq:res:Kdef} is expanded back into the atom basis, with the term inventory,
the coefficient rules, the definitions of the four nested atoms and both summation
ranges read out of the printed text, not supplied from outside, and compared
against the stored per-spin atom tables. Over the forty-nine spins $n=3,\dots,99$ that
is $112112$ coefficients, and a further $4538$ at the holdout spin $n=101$ of
Sec.~\ref{sec:chk:oos}. Every one of them is exact, the two sides hold the same nonzero
cells and nothing besides, and the comparison runs over the whole atom basis with no
part of it left out. The two evaluations share no table and no program at run time: the
tables are the output of the Mellin extraction of App.~\ref{app:construction} applied to
the three-loop integrand, with no closed form anywhere in that extraction, and the other
side is the equation as printed, with no coefficient catalogue behind it. The rules
themselves were inferred from tables at lower spins produced
from the same integrand, so the comparison tests the reconstruction and its
extrapolation in the spin, with no independent input entering it. As a control, removing
the $\cCm\,\Om{C}$ term from Eq.~\eqref{eq:res:Kdef} and changing nothing else
leaves $n$ wrong entries at spin $n$ and no others, all of them on
$\Om{C}$ and each short by $\cCm$.

The method also runs backwards. Rebuilding the atom table from the rules of
Secs.~\ref{sec:closedform} and~\ref{sec:residual}, with no tabulated coefficient
used anywhere and the three rational kernels regenerated from the regularity identity
Eq.~\eqref{eq:res:laurent}, returns the stored table entry for entry at
$n=3,5,9,15,21,31$. The two assemblies of $\chiord{2}(n,\nu)$ are therefore one and the
same number at every $\nu$, with no digit tolerance in the statement. Every
coefficient of the seven slots was also compared rule against table individually,
$2450$ or $2499$ per slot depending on the shift range, with no exception. With the
twenty-eight coefficients of the $n=1$ block the tables hold the $112140$ entries counted in
App.~\ref{app:cat:census}.

\subsection{Spins the rules had never seen}
\label{sec:chk:oos}
The coefficient rules of the forty closing slots were fixed on the spins $n\le49$
and checked against the remaining computed spins out to $n=91$. That check reuses
data produced here earlier, so it tests consistency, not
prediction. To test prediction the rules have to meet coefficients that did not
exist when they were written.

Four spins were computed afterwards. The three-loop integrand was already available
at $n=93,95,97,99$; what was missing was its reduction to an atom table, and that
step was carried out only after the rules had been written down. Two controls came first,
both in App.~\ref{app:cons:repro}: the structured integrand regenerated at $n=33$
reproduces the stored file, and its reduction returns the stored atom table. The results
at the four new spins are collected in Table~\ref{tab:chk:oos}.

\begin{table}[t]
\centering
\begin{tabular}{@{}lrrr@{}}
\toprule
spin & coefficients in closing slots & reproduced exactly & mismatches\\
\midrule
$n=93$ & $3531$  & $3531$  & $0$\\
$n=95$ & $3607$  & $3607$  & $0$\\
$n=97$ & $3683$  & $3683$  & $0$\\
$n=99$ & $3759$  & $3759$  & $0$\\
\midrule
total  & $14580$ & $14580$ & $0$\\
\bottomrule
\end{tabular}
\caption{Coefficients at four spins computed after the rules were fixed, in exact
rational arithmetic. The table counts the closing slots. Including the seven that do
not close, the four spins contain $17252$ coefficients, and the rules of
Sec.~\ref{sec:residual} reproduce those as well. Those seven rules were written after
these tables were reduced, so the further $2672$ coefficients test
consistency where the $14580$ above test prediction.}
\label{tab:chk:oos}
\end{table}

The compact form of Eq.~\eqref{eq:cf:ladderform} was assembled before these tables
existed. Evaluated at the four new spins and three shifted arguments each, it agrees
with the direct atom-by-atom sum over the forty closing slots to a worst relative
$6.38554\times10^{-29}$ at thirty digits and cutoff $J=120$. Both sides of that
comparison are restricted to those forty, so what it tests is the regrouping of
Sec.~\ref{sec:str:collapse} and it
says nothing about $\Kres$. The part left out is not a small one. At $n=99$ and
$z=(n-1)/2+1.7$ the forty closing slots sum to about $-2.0\times10^{4}$ and the seven
remaining ones to about $+2.0\times10^{4}$, the two agreeing in magnitude to five
figures and cancelling down to a sum over atoms of order $10^{-1}$. All forty-seven
slots at these spins are covered by the exact comparison of
Sec.~\ref{sec:chk:closed}.

The exchange
symmetry of Sec.~\ref{sec:str:reflection}, stated on the spins through $n=91$, holds at
the four new spins on a further $1532$ coefficients with no exception, and the three
rational slots continue to show no definite sign, as at the lower spins.

A fifth spin was generated later, for the same purpose. The atom table at $n=101$ was
produced by the extraction of App.~\ref{app:construction} from the same integrand. It
holds $3936$ shifted instances and $4538$ coefficients over all forty-seven slots, and
the rules reproduce every one of them exactly, with no entry left uncovered by a rule
and no rule left without an entry. The regularity identity Eq.~\eqref{eq:res:laurent}
holds there as well, on all $1500$ of its Laurent coefficients, with no rule used
anywhere in that part. The rules predate the table, and the evidence for that order is
in the generation record of the ancillary files. It is one spin only. It tests the
rules and leaves them unproved at arbitrary odd $n$.

\subsection{Intercepts at \texorpdfstring{$\nu=0$}{nu=0}}
\label{sec:chk:qsc}
Along $\nu=0$ the intercepts are known in closed form at arbitrary conformal spin from
the Quantum Spectral Curve analysis of Alfimov, Gromov and
Sizov~\cite{Alfimov:2018cms}, the only one of those
results~\cite{Gromov:2013pga,Alfimov:2014bwa,Gromov:2015vua} that reaches $n\neq0$.
The intercepts $\chiord{2}(n,0)$ obtained from Eq.~\eqref{eq:cf:master} agree with
those values spin by spin through $n=91$, the last odd spin their data set reaches;
the forty-six odd spins up to that bound, the boundary spin $n=1$ among them, are the
original range of the present study, and the run log of App.~\ref{app:val:qsc} carries
forty-five rows because it starts at $n=3$. The reference side of that comparison
reaches us by two routes, and they are kept apart. Through $n=91$ it is read from the
data supplied with Ref.~\cite{Alfimov:2018cms}, as the $g^{6}$ coefficient of the
intercept divided by four, with no re-evaluation of their binomial-sum formula here.
That part of the comparison therefore uses their own values, and the ancillary file
that lists them states where each row comes from. Above $n=91$ their data set ends,
and the reference values are produced here by evaluating their closed form, which is
the test of Sec.~\ref{sec:chk:agsext}. The comparison tests something narrower than
the eigenvalue. It puts the composite of the three-loop integrand and the reduction
performed here against the Quantum Spectral Curve, at the single point $\nu=0$ of each
spin, and it reaches no point off that line. The three-loop kernel had itself been
matched to the Quantum Spectral Curve at $n=0$~\cite{Caron-Huot:2016tzz}, and Alfimov,
Gromov and Sizov report in their Sec.~4.2 that their closed form agrees with that
kernel at several of the first non-negative spins as well~\cite{Alfimov:2018cms},
without saying which. Neither comparison reaches the reduction performed here, so each
intercept above tests the integrand and the reduction as one object. For most of the
spins in $3\le n\le91$ no comparison of that kind had been made before the present
study.

In the other direction, the closed
form of Ref.~\cite{Alfimov:2018cms} was obtained by fitting a basis of binomial harmonic
sums to Quantum Spectral Curve data at odd spin. That fit rests on maximal
transcendentality and on a reciprocity conjecture that reduces the basis to the binomial
sums, a property those authors report
broken at the order above. The
present study starts from the three-loop integrand instead, and reproduces those
intercepts at every odd spin through $n=91$. The agreement therefore supports their closed form
from a different starting point. The comparison also fixes the normalization of
Eq.~\eqref{eq:set:omega}: in the intercept expansion the $g^{2}$ coefficient equals
$4\chiord{0}(n,0)$ and one quarter of the $g^{4}$ coefficient equals
$\chiord{1}(n,0)$, so one quarter of the $g^{6}$ coefficient is $\chiord{2}(n,0)$.
Appendix~\ref{app:validation} lists the residuals.

\subsection{The reference closed form past its data range}
\label{sec:chk:agsext}
The data set of Ref.~\cite{Alfimov:2018cms} stops at $n=91$ and the closed form given
there was fitted inside that range, while the tables of the present study reach
$n=101$. Evaluating that closed form above $n=91$ therefore tests its extrapolation in
the spin, and with it the reciprocity conjecture that narrows its basis. We ran it at
$n=93$, $97$ and $101$. Those are the three spins in that region at
which $M=(n-1)/2$ is even, where the finite binomial sum is the continued intercept and
no branch question arises. At $M$ odd the finite sum differs from the intercept in the
second figure, $-266.658$ against $-266.528$ at $n=91$, so those spins are left out.
The intercepts obtained from Eq.~\eqref{eq:cf:master} agree with the reference closed
form to a relative $3.5\times10^{-43}$, $2.4\times10^{-42}$ and $1.5\times10^{-41}$, at
sixty digits of working precision with cutoff $J=300$. The reference side of that
comparison is
a rational combination of $1$, $\pi^{2}$ and $\pi^{4}$ carried in exact arithmetic and
turned into a decimal at the end, so the residual measures the evaluation on our side.
It is quoted as measured and not as a bound, for the reason
App.~\ref{app:val:precision} gives.

\subsection{The collinear tower}
\label{sec:chk:collinear}
The block has a pole structure at the collinear points $z=-j$, with $j\ge1$, that is
uniform in the spin. Writing $u=z+j$, the tables give
\begin{equation}
\Fn=\frac{32}{u^{5}}
+\frac{48\,(-1)^{j+1}\bigl[\Sh{1}{n+j-1}-\Sh{1}{j-1}\bigr]}{u^{4}}+O(u^{-3}),
\label{eq:chk:collinear}
\end{equation}
exactly at every one of the $588$ cells with $3\le n\le99$ and $1\le j\le12$, without
exception, and the coefficient at order three contains no $\pi^{2}$ at any of them.
Equation~\eqref{eq:chk:collinear} holds for $j\ge1$: at $j=0$ the term $\Sh{1}{j-1}$
sits on the pole of $\Sh{1}{}$ at argument $-1$. At
$j=1$ the second numerator is $48\,\Sh{1}{n}$, so
$\chiord{2}=8/u^{5}+12\,\Sh{1}{n}/u^{4}+O(u^{-3})$ at every odd spin.

The leading number is the one collinear resummation predicts: a change of the energy
scale replaces the anomalous-dimension variable by that variable shifted by
$\pm\omega/2$, the sign set by the direction of the scale
change~\cite{Ciafaloni:1998gs}. Iterating the shifted leading-order
pole~\cite{Salam:1998tj} and solving the resulting quadratic gives, in the
normalization of Eq.~\eqref{eq:set:omega}, the coefficients $1$, $-2$, $8$ and $-40$ at
successive orders, of which $8$ is the three-loop one. The polar terms at $L$ loops follow from the
$L$-loop anomalous dimension~\cite{Marzani:2007gk,Costa:2012cb}, and at $n=0$ the
comparison at this order
was made in Eqs.~(5.16) and~(5.17) of Ref.~\cite{Caron-Huot:2016tzz}.
This is not an independent test, and we do not offer it as
one. Both sides of the comparison descend
from the same input; the coefficient $32$ at $k=1$ is among the checks the
tables are required to pass; and the two orders the tower reaches, $u^{-5}$ and $u^{-4}$, are carried by atoms
of pole order five and four, whereas every one of the seven slots of
Sec.~\ref{sec:residual} has pole order three or less. The check touches none of them.

One relation between the two orders is not part of the tower and does not follow from
it. The
$u^{-4}$ coefficient of $\chiord{2}$ is one quarter of the numerator in
Eq.~\eqref{eq:chk:collinear}, so it is $12\,(-1)^{j+1}$ times the bracket printed
there. The $u^{-2}$ coefficient
of $\chiord{1}$, read off Eq.~\eqref{eq:intro:nlo} by contour integration, is that same
bracket with $-2$ in place of $12$, at every one of the same $588$
cells, so the three-loop coefficient is $-6$ times the two-loop one. We report the
relation as an observation. The $\omega/2$ shift does not give it: applied to the
eigenvalue with its regular part kept, the shift ties the subleading coefficient to the
regular value of $\chiord{0}$ at the collinear point, and so predicts
$2\bigl[\Sh{1}{n+j-1}+\Sh{1}{j-1}\bigr]$, which is $+11/3$ at $n=3$, $j=1$ where the
coefficient is $-11/3$, and $+5$ at $n=1$, $j=2$ where the coefficient is $+1$. The
double pole at two loops and the quartic pole at three therefore stand in a relation we
have observed and have not obtained from the resummation.

\subsection{Internal checks}
\label{sec:chk:internal}
Further checks constrain the method from inside. The reduction of every
atom to the canonical objects of Eq.~\eqref{eq:cf:canonical} was verified
atom by atom against direct summation, to a relative $2.71171\times10^{-31}$, the
atom sums cut at $J=200$ and the working precision thirty digits. The collapse of
Sec.~\ref{sec:str:collapse} was verified against the direct per-spin sum over the
forty closing slots at $138$
points spread over all forty-six spins of the original range, to a relative
$2.61862\times10^{-28}$, thirty digits again and $J=120$; both sides of that
comparison contain those forty and nothing else, as App.~\ref{app:cmp:checks} spells out. The
identities behind Eqs.~\eqref{eq:str:R1} and~\eqref{eq:str:R2} hold as exact
cancellations among the atom coefficients, beyond
any comparison of digits, and the stuffle relations used in App.~\ref{app:reduction} were
checked in the non-strict convention of Eq.~\eqref{eq:set:sdef}, not assumed.

The arithmetic is exact where the objects are rational and carried at thirty digits
or more where they are not, the deepest runs of App.~\ref{app:validation} at $120$. The tables, the evaluator, the coefficient rules and
the programs that check them are in the ancillary files
(App.~\ref{app:ancillary}). For the bounded searches of App.~\ref{app:grd:open} those
files carry the dimension each search is a statement about, rebuilt from the list of
functions it counts, while the fit behind it stands on the outcome stated here. The seven
figures this paper quotes as a relative difference are produced there by the
comparison each one measures.

\subsection{Tests not yet done}
\label{sec:chk:open}
Every check above draws on the same three-loop integrand on both sides, the new spins of
Sec.~\ref{sec:chk:oos} among them, and the intercepts alone reach a source outside it.
The tests that would change that stay within odd spin.

One is the trajectories away from $\nu=0$ from the asymptotic Baxter--Bethe
treatment~\cite{Ekhammar:2024neh,Ekhammar:2025vig}. Unlike the intercepts these probe
the eigenvalue off the line $\nu=0$, which is where the content of
Eq.~\eqref{eq:cf:ladderform} actually lies. For the lowest trajectories the test as
stated is out of reach: the asymptotic treatment of a trajectory of length $L$ is
solvable up to the order $L+1$ in weak coupling, with wrapping corrections entering at
order $g^{L+2}$~\cite{Ekhammar:2024neh}, so at $L=2$ the pomeron trajectory is
contaminated from $g^{4}$ on, below the $g^{6}$ at which $\chiord{2}$ enters.
Trajectories of higher $L$ do reach the needed order: with wrapping at $g^{L+2}$ the
order $g^{6}$ stays clean from $L=5$ on, and the conformal spin enters those
trajectories through the function $\chi(\Delta,S_2)$ of Eq.~(5.19) of
Ref.~\cite{Ekhammar:2025vig} and its derivatives, tabulated in Table~1 of that reference
through $L=4$, so the test asks for one length beyond what that reference prints. A
wrapping-complete Quantum
Spectral Curve evaluation at generic $\nu$ would provide the test without that
restriction. Another is the collinear residues at the
orders below the ones reached above, $u^{-2}$ and $u^{-1}$, against the three-loop
universal anomalous dimension, which is external to the input used here. That needs the
$\omega$-shifted resummation carried past the order at which we have it. We have done
neither.

\section{Discussion and outlook}\label{sec:discussion}

The coefficient rules have the status Sec.~\ref{sec:checks} states in full, and they are
unproved at arbitrary odd $n$. It is worth
emphasizing that the arithmetic exclusion of Sec.~\ref{sec:res:obstruction} is the one
statement in this paper that covers every odd spin on its own. What that exclusion covers
is the coefficient rule, and the hypothesis it rests on is stated beside it.

\subsection{Uses of the closed form}
\label{sec:dis:uses}
The first use of a formula at every odd spin is the multi-Regge limit, and it needs a
distinction drawn before it can be used.
Amplitude data in that limit are organized in the Fourier--Mellin representation,
where the sum over conformal spin and the integral over $\nu$ run against a BFKL
eigenvalue and impact factors~\cite{Lipatov:2010ad,Dixon:2012yy}. The eigenvalue that
appears there is the adjoint, colour-octet one. Its next-to-leading kernel was computed
directly~\cite{Fadin:2011we}; the impact factors at that order, and the eigenvalue two
orders beyond the leading one, were extracted from the Regge limit of the six-point
remainder function~\cite{Lipatov:2010ad,Dixon:2012yy}. Equation~\eqref{eq:cf:master} is
the colour-singlet pomeron. The two are different objects
and a naive comparison is a category error. The singlet eigenvalue controls the pomeron
Green's function, and through the optical theorem the total cross sections that function
gives. Equivalently it is the continuation in the spin of the twist-two anomalous
dimension, and that is the route the Quantum Spectral Curve takes to it. A test
of Eq.~\eqref{eq:cf:master} has to run along one of those two, and we have not
carried one out.

Another is analytic continuation in the conformal spin. Reflection identities
among harmonic sums relate the eigenvalue at $z$ to its value at the reflected
argument~\cite{Prygarin:2018tng,Prygarin:2018cog,Joubat:2023ftd}, and at two loops
those identities are what makes the
hermitian-separable structure of Eq.~\eqref{eq:intro:nlo}
visible~\cite{Kotikov:2002ab,Bondarenko:2015tba,Joubat:2019esj,Joubat:2020vrw,%
Joubat:2020hvc}. With
Eq.~\eqref{eq:cf:master} in hand the same question can be asked at three loops, and
Sec.~\ref{sec:str:reflection} is the first part of the answer: the exchange of the
two ladder distances, which swaps the ends of the segment, already acts cleanly on
the coefficients. That line of work left it open whether the
Caron-Huot--Herranen result can be written through functions of one
variable~\cite{Joubat:2020vrw}; at odd spin Eq.~\eqref{eq:cf:master} shows that it can,
with the sums of one variable evaluated along the ladder and their number growing with
the spin. At odd spin the next-to-leading eigenvalue on the line
$\nu=0$ is a rational combination of $1$, $\pi^{2}$ and $\zeta_3$~\cite{Joubat:2021nss},
the three constants the atom-table coefficients of Eq.~\eqref{eq:cf:master} are made
of.

Everything above is odd $n$. The closed form is built at odd spin, we make no claim
about even conformal spin here, and continuing Eq.~\eqref{eq:cf:master} in $n$ is not
offered as a route to it. That sector we have reduced separately by the same method, in
a calculation of our own that no one outside the present authors has checked and that
will be presented elsewhere.
Beyond the two-reggeon problem the multi-reggeon compound states of multicolour QCD
are an integrable spin chain in the generalized leading-logarithmic
approximation~\cite{Lipatov:1993yb,Faddeev:1994zg}, and how the ladder
structure found here appears there is a question we have not touched.

\subsection{The residual contribution}
\label{sec:dis:residual}
The residual contribution is where the present study closes, and the route to its
rules ran through the attempts that failed. The obstruction of
Sec.~\ref{sec:res:obstruction} is arithmetic and pointed at functions whose argument is
the ladder length instead of a distance along it; cyclotomic and generalized harmonic
sums are the standard enlargements of the alphabet, generated by denominators of higher
cyclotomic polynomials and by numerator weights away from
$\pm1$~\cite{Ablinger:2011te,Ablinger:2013cf}, and in the present setting they arise
when an alternating ladder of fixed total length is summed. Admitting sums at argument
$n-2$ removed the
obstruction without closing the gap, so the kernels are not harmonic sums at a longer
argument in any of the bases we searched.

The obvious next candidate does not work either, and the reasoning that suggests it is
sound, so we say why it fails. A bivariate sum whose outer index runs the whole
ladder while the summand depends on the rung,
\begin{equation}
T^{\sigma}_{a,b}(\ap,\am)=\sum_{j=1}^{\ap}\frac{\sigma^{j}}{j^{a}\,(j+\am)^{b}},
\label{eq:dis:bivariate}
\end{equation}
has denominators on $\mathrm{lcm}(1,\dots,n-1)$ as required, reduces to
$\Sh{\pm(a+b)}{\ap}$ at the near end, and is genuinely new: adjoining these functions
raises the rank of the weight-two basis by eight and of the weight-three basis by
fifty-four beyond products of one-variable sums. They also appear in slices of
$\KRthree$. But they do not close the kernels in the bases we built from them. At the
transcendental weight the denominators force, and again one unit above it, the targets
lie outside the span of one finite list of functions, with
$\mathrm{rank}[X|y]=\mathrm{rank}[X]+1$ over all $49$ spins at once. Beyond that
weight the ansatz stops being falsifiable altogether, its rank reaching the number of
available coefficients, so the higher weights add no information and do not
support the proposal. The result is sharper than a negative one usually is, since
$\KRthree$ does lie in this class column by column at the near end, with the closed
forms of Eq.~\eqref{eq:res:col1} through $\am=3$. We first read the rank deficiency at
$\am=3$ as the obstruction switching on; it is not. It is a property of the
thirty-five function list, and one further rational shift removes it. The four closed
columns therefore share one transcendental part and differ only in a rational tail,
which reaches one shift further at $\am=3$ than before.
Equation~\eqref{eq:res:KR3} says why they close: at fixed small $\am$ the $\max$ inside
the ladder sum is dominated by one argument on all but a bounded number of terms, and
the sum telescopes into $\Sh{1,1}{\ap}$ plus a short rational tail. The restriction that
does survive is the arithmetic one of Sec.~\ref{sec:res:obstruction}, which is fixed by
denominators, not by any list of functions.

The functions that close the sector are the ones the obstruction pointed at,
of the type of Eq.~\eqref{eq:dis:bivariate}
(sums along the ladder whose summand depends on the rung), but with the outer index feeding
a harmonic sum in place of a power, and with $\max(\am+N,\ap-N)$ as the argument.
Section~\ref{sec:res:midpoint} closes the triple-pole kernel that way, and
Sec.~\ref{sec:res:ladder} closes the three remaining rules in the same class, leaving
nothing tabulated. The denominators constrain the argument and leave the summand open: a
bivariate sum with a rational summand cannot
produce the $\Sh{1,1}{\ap}$-type transcendental content that Eq.~\eqref{eq:res:col1}
exhibits, and the sum over a harmonic sum at a moving argument does. That it is the
smallest enlargement which would serve is not something we have shown.

\subsection{Weight}
\label{sec:dis:weight}
The uniform weight now extends to every coefficient in the paper. Assign the ladder
sum of Eq.~\eqref{eq:res:laddersums} the weight $j$ plus the weight of the harmonic sum
inside it, so that $\Lam$ has weight two and $\Lax{2}$ weight three. Then the rules of
Sec.~\ref{sec:res:ladder} have weight two for the three slots on weight-three atoms
and weight three for the slot on the weight-two atom, and $\KRthree$ has weight two
against a weight-three rational atom. Every slot reaches five, coefficient and atom
sharing it between them; Sec.~\ref{sec:res:columns} had seen this, but only
along the boundary column.

The transcendental weight of Eq.~\eqref{eq:cf:master} is uniform and equal to five,
which is what the uniform transcendentality of $\mathcal{N}=4$ quantities leads one to
expect~\cite{Kotikov:2002ab,Kotikov:2006ts,Prygarin:2019ruv,Joubat:2021pww} and what
the $n=0$ result showed~\cite{Gromov:2015vua}. Here it enters as a check. The odd-spin
block adds that uniformity survives the ladder: coefficient and atom share the weight
between them, with the coefficient carrying whatever the atom does not, so neither of
the two separately need have weight five while every complete coefficient-atom
contribution does. The alphabet is nearly the standard one.
Twenty-seven nested harmonic sums of a single variable, none above weight four,
polygammas at halved argument and a
handful of Euler constants cover all of it except the residual sector, which needs the
window sums of Sec.~\ref{sec:res:ladder}. The halved arguments are not a signature of the
ladder: they are the generic image of an alternating letter, already present at
next-to-leading order in $\beta'$, tied by a sign and a constant to the continued
$\Sh{-2}{}$ at shifted argument in Eq.~(28) of Ref.~\cite{Velizhanin:2015xsa}, and a
harmonic sum with one
negative index is traded there for sums of positive index at halved arguments, $z/2$
and $z/2-\tfrac12$~\cite{Prygarin:2019ruv}. They arise inside a single atom before any sum
along the ladder is done.

\subsection{What the closed form makes explicit}
\label{sec:dis:explicit}
The closed form given in the present study makes the pole structure in $\nu$, the
transcendental weight and the product-free form of the block in the fixed atom class of
Sec.~\ref{sec:set:atoms} explicit at every odd spin,
where the intercepts along $\nu=0$ leave all three implicit. Through the
maximal-transcendentality principle the closed form is expected to fix the
highest-weight part of the corresponding QCD eigenvalue at this order.

\section*{Acknowledgements}

We thank Sergey Bondarenko for valuable discussions. The symbolic reduction and the
numerical evaluations were carried out by the authors with computer algebra and with
programs written for the present study. All scientific content, derivations
and conclusions were developed and verified by the authors, who take full responsibility
for the manuscript.

\section*{Declaration of generative AI and AI-assisted technologies in the writing process}
During the preparation of this work the authors used Opus 4.8, Opus 5
and Fable 5 (Anthropic), GPT 5.5 and GPT-5.6-sol (OpenAI), and Gemini 3.1 Pro
(Google) in preparing the accompanying programs and in language editing. After using
these tools, the authors reviewed and edited the content as needed and take full
responsibility for the content of the publication.

\appendix

\section{Conventions and the continuation in the conformal spin}\label{app:conventions}

\subsection{Harmonic sums and their continuation}
\label{app:conv:sums}
The nested harmonic sums of Eq.~\eqref{eq:set:sdef} are continued to complex
argument in the standard way~\cite{Blumlein:1998if,Blumlein:2000hw}.
For a single letter,
\begin{equation}
\Sh{a}{w}=\zeta_a-\sum_{N\ge1}\frac{1}{(N+w)^{a}}\quad(a\ge2),
\qquad
\Sh{1}{w}=\dgam(w+1)+\gE ,
\label{eq:conv:cont}
\end{equation}
and for alternating letters,
\begin{equation}
\Sh{-a}{w}=-\etaf{a}-(-1)^{w}\!\sum_{N\ge1}\frac{(-1)^{N}}{(N+w)^{a}},
\qquad \etaf{a}=(1-2^{1-a})\zeta_a .
\label{eq:conv:contalt}
\end{equation}
At $a=1$ the product $(1-2^{1-a})\zeta_a$ is indeterminate, and the value is taken by
continuation in $a$, which gives $\etaf{1}=\lntwo$; that value is the case $q=1$ of
Eq.~\eqref{eq:red:phi}.
The factor $(-1)^{w}$ in Eq.~\eqref{eq:conv:contalt} is the point at which the
continuation splits into branches. Requiring the sum to agree with its defining series
at integer $w$ of one parity fixes an even branch and an odd branch, and the two differ
by twice the sum over $N$ in Eq.~\eqref{eq:conv:contalt}. That difference falls off like
$|w|^{-a}$ at large $|w|$ in the right half plane, and it has a pole of order $a$ at
every negative integer. Carlson's uniqueness
result~\cite{Carlson:1914} removes the remaining freedom once a growth condition is
imposed, and
the condition follows the spacing of the data. Here they sit on the even integers, so
the rescaling $w=2v$ comes first: it takes them to the integers, where the theorem asks
for exponential type below $\pi$, and it doubles the type. Among continuations of
exponential type below $\pi/2$ in $w$ each branch is therefore unique, and that bound
excludes the ambiguity the data leave open. Two continuations agreeing on the even
integers differ by a function vanishing there, and the only such function of
exponential type below $\pi/2$ is zero. The bound is sharp: $\sin(\pi w/2)$
vanishes at every even integer and has exponential type exactly $\pi/2$, so the
unit-spacing bound would still admit it. Ablinger, Bl\"umlein and
Schneider give the branch statement in their Sec.~6~\cite{Ablinger:2013cf}: no
continuation of the factor $(-1)^{w}$ satisfies the bound, while the sum restricted to
one parity has one and it is the only one. Each branch
is therefore the only continuation of its own parity.

\emph{Every continuation in the atom argument is on the even branch}, the continuation
from the non-negative even integers in the prescription of Kotikov and
Velizhanin~\cite{Kotikov:2005gr}, obtained by setting $(-1)^{w}\to+1$ in
Eq.~\eqref{eq:conv:contalt}. That choice is not a matter of convention. The
identity Eq.~\eqref{eq:red:phi} is an even-branch statement and reverses sign on
the odd branch. And the next-to-leading eigenvalue Eq.~\eqref{eq:intro:nlo} reproduces
the independently known intercepts only on the even branch: evaluated in exact rational
arithmetic against the Quantum Spectral Curve intercept series of Alfimov, Gromov and
Sizov~\cite{Alfimov:2018cms} it agrees at every one of the $46$
odd spins $n\le91$ that their data set reaches, on the even branch, and at none of them
on the odd one.

Nothing is continued in the two ladder distances. They are non-negative integers at
every rung and every spin, so the harmonic sums evaluated at them, in
Table~\ref{tab:cat:rules} and in Eq.~\eqref{eq:res:f011} among others, are the finite
sums of Eq.~\eqref{eq:set:sdef} and no branch enters. Only a word with an alternating
letter has two branches at all, by Eq.~\eqref{eq:conv:contalt}, and at an odd distance
the even-branch value of such a word differs from the finite sum.

Odd conformal spin is the natural sector because of the argument, not because of the
branch: $z=(n-1)/2+i\nu$ has $(n-1)/2$ an integer precisely when $n$ is odd, and that is
what puts
the whole ladder on integer shifts of a single point. The parity of that integer varies
with $n$ modulo four and plays no role in that.

The object that mediates between the two branches is the odd harmonic sum
\begin{equation}
O_k(m)=\!\!\sum_{\substack{j\le m\\ j\ \mathrm{odd}}}\!\frac{1}{j^{k}}
=\tfrac12\bigl[\Sh{k}{m}-\Sh{-k}{m}\bigr],
\label{eq:conv:Ok}
\end{equation}
and it appears in $\Phi$ through Eq.~\eqref{eq:cf:AB} and again in the
boundary closure of Eq.~\eqref{eq:res:col0b}. Written at doubled argument,
$O_k$ interpolates the two parities of the ladder, and the constants
$(1-2^{-k})\zeta_k$ and $2^{-k}\zeta_k$ that distinguish them cancel in the
assembled block.

\subsection{Normalization checks}
\label{app:conv:norm}
Before use, the conventions of Eq.~\eqref{eq:set:omega} were pinned against every
independent data set available. The leading line reduces exactly to
Eq.~\eqref{eq:intro:lo}. The next-to-leading eigenvalue reproduces the $n=0$
result~\cite{Gromov:2015vua} and the general-$n$ intercept
data~\cite{Alfimov:2018cms} to about forty-three digits. In the
Quantum Spectral Curve intercept expansion the $g^{2}$ coefficient equals
$4\chiord{0}(n,0)$ and one quarter of the $g^{4}$ coefficient equals
$\chiord{1}(n,0)$, at the representative spin $n=3$ and to all computed digits; one
spin suffices, since the normalization does not depend on $n$. That identification
extends to the next order, so $\chiord{2}(n,0)$ is one quarter of the $g^{6}$
coefficient.

The spectral variable of Refs.~\cite{Caron-Huot:2016tzz,Gromov:2015vua} is twice the
one used here. Every line quoted from those papers is evaluated at
$\nu_{\mathrm{CHH}}=2\nu$ before comparison, and the tables accompanying this paper
are in the convention of Eq.~\eqref{eq:set:zdef}.

\subsection{Notation map}
\label{app:conv:map}
The symbols are collected here, together with the two senses of kernel. The vocabulary of
the block is fixed in Sec.~\ref{sec:set:vocab}.

\medskip
\noindent
\begin{tabular}{@{}p{3.0cm}p{12.3cm}@{}}
\toprule
$n$ & the conformal spin, odd and positive throughout. The eigenvalue depends on $|n|$
only.\\
$\nu$ & the spectral variable of Eq.~\eqref{eq:set:zdef}. That of
Refs.~\cite{Caron-Huot:2016tzz,Gromov:2015vua} is $2\nu$.\\
$\gamma$ & the anomalous-dimension variable, written $\tfrac12+i\nu$ in
Ref.~\cite{Kotikov:2000pm} and $\tfrac12-i\nu$ below Eq.~(19) of
Ref.~\cite{Kotikov:2002ab}. The two agree on every value quoted here, the eigenvalue
being even in $\nu$. It occurs here only where those results are quoted.\\
$z,\ \zb$ & $z=(n-1)/2+i\nu$ and $\zb=(n-1)/2-i\nu$, with $z+\zb=n-1$. At $\nu=0$ the
two coincide at $M=(n-1)/2$.\\
$-\zb$ & the reflected endpoint $z-(n-1)$, the far end of the ladder. A point of the
segment, not a conjugation of anything.\\
$k$ & the rung label, an integer with $-(n-1)\le k\le0$ on the ladder and $k=1$ at the
single point beyond it, Eq.~\eqref{eq:set:krange}.\\
$\ap,\ \am$ & the distances from the rung $k$ to the two ends, $\ap=k+n-1$ and
$\am=-k$, with $\ap+\am=n-1$, Eq.~\eqref{eq:set:distances}. $\amax$ and $\amin$ are the
larger and the smaller of them.\\
$m$ & an internal point of the ladder, $z=m$ with $0\le m\le n-2$, where
Eq.~\eqref{eq:res:laurent} holds.\\
$N$ & the summation variable inside an atom, Eq.~\eqref{eq:set:BAatom}, and the
summation variable of a ladder sum, Eq.~\eqref{eq:res:laddersums}. Never a rung
label.\\
\midrule
$d$ & the shift in the pair $(d,p)$ of Eq.~\eqref{eq:set:BAatom}, and from
Eq.~\eqref{eq:res:Kdef} on the raw coefficient carried by $\Om{A}$ and $\Om{B}$. The
slot label of Eq.~\eqref{eq:res:laurent} holds both senses at once.\\
$M$ & the canonical kernel $M_{\sigma,v,j}$ of Eq.~\eqref{eq:cf:canonical}, and the
midpoint $(n-1)/2$ of Sec.~\ref{sec:cf:eval} and App.~\ref{app:grd:closures}.\\
$w$ & the atom argument $z+k$ of Eqs.~\eqref{eq:set:Ratom}
and~\eqref{eq:set:BAatom}; the ladder weight $w_f$ of Eq.~\eqref{eq:str:weights}; and the
transcendental weight, which is named as such wherever it is meant.\\
$x$ & the argument of the one-variable functions from Eq.~\eqref{eq:cf:phi} on; the
integration variable of App.~\ref{app:construction}; and, carrying four indices, the
recurrence coefficients $x_{\alpha\beta d_Md_j}$ of Table~\ref{tab:grd:rec}.\\
$j$ & the collinear index of Eq.~\eqref{eq:chk:collinear}, where $z=-j$; the outer
exponent of a ladder or window sum, Eqs.~\eqref{eq:res:laddersums}
and~\eqref{eq:res:window}; the pole order in $M_{\sigma,v,j}$; and $j=-k$ in
App.~\ref{app:grd:closures}.\\
$\mu$ & the constant $\mu\in\{1,\pi^{2},\zeta_3\}$ that multiplies a shape in a slot,
Eq.~\eqref{eq:str:regroup} and Table~\ref{tab:cat:rules}; and, where
Eq.~\eqref{eq:intro:nlo} is discussed, the conformal weight $\mu=\gamma+|n|/2$ of
Refs.~\cite{Kotikov:2000pm,Kotikov:2002ab}, with $\tilde\mu=\mu-|n|$ beside it.\\
$\coeff{}^{\,s}$ & the coefficient rule of the slot $s$, in four notations for one
object. The slot is named either by the atom label
$\atomlab{\BAatom}{(\sigma,v,((d,p)),q)}$ of App.~\ref{app:cat:shapes} or, in
Eq.~\eqref{eq:res:laurent}, by the tuple $(\BAatom,\sigma,v,\{(d,p)\},q)$ that spells
out the same data; and the arguments $(n,k)$ are carried where the rung is in play and
left off where the rule is meant as a whole.\\
\bottomrule
\end{tabular}
\medskip

\noindent
\begin{tabular}{@{}p{3.0cm}p{12.3cm}@{}}
\toprule
raw kernel & the exact rational number a non-closing slot carries at a given rung, as
stored in the atom table. The six coefficient functions $\KRone$, $\KRtwo$, $\KRthree$,
$\cOm$, $\cCm$ and $\dAB$ of Eq.~\eqref{eq:res:Kdef} are raw in this sense, and every
rule and every displayed formula of this paper is raw.\\
projected kernel & the orthogonal residual $c^{\perp}$ of such a coefficient against the
thirty-four function elementary spanning set described below, taken spin by
spin. It vanishes identically at every odd $n\le17$, and it occurs only in the collapse
layer of the ancillary files. Nothing displayed here is projected, and a projected
coefficient put in place of a raw one would drop the elementary part of the residual
contribution.\\
\bottomrule
\end{tabular}
\medskip

\noindent
The ancillary files use the word kernel for a different object, and the two entries above
keep them apart. The collapse layer of those files works with the orthogonal residuals
of the six coefficient functions of Eq.~\eqref{eq:res:Kdef}, taken spin by spin by least
squares against a thirty-four function weight-two spanning set in $\ap$, $\am$ and
$\amin$, and we write $c^{\perp}$ for the residual of a coefficient $c$. Every one of the
residuals vanishes identically at every odd $n\le17$, where the raw coefficients do not,
and in the residual layer $\cCm^{\perp}=-\dAB^{\perp}$ exactly, which is
Eq.~\eqref{eq:res:sixtofive} taken modulo the elementary span. That relation is what lets
the ancillary collapse work with the single combination $\Om{A}+\Om{B}-\Om{C}$ of the
regrouping in Sec.~\ref{sec:res:what}, and it is not available to
Eq.~\eqref{eq:res:Kdef}, where the elementary complement stays. The ancillary names for
$\cOm^{\perp}$ and $\dAB^{\perp}$ are $\Kcwthree$ and $\Kcwtwo$; for the three rational
residuals those files use $\KRone$, $\KRtwo$ and $\KRthree$, which in this paper always
denote the raw coefficients of Eq.~\eqref{eq:res:Kdef}.

\section{From the three-loop integrand to the atom table}\label{app:construction}

The external input to the atom table is the three-loop integrand of Caron-Huot and
Herranen~\cite{Caron-Huot:2016tzz}. At a given conformal spin it is a function of a
single variable $x$ built from iterated integrals over the three-letter alphabet
$\{x,\,1-x,\,1+x\}$, multiplied by rational functions and by constants. Those authors
say where the three letters stop: for even $m=2,4,6,\dots$ the integrand needs in
addition the square-root letter $(1+i\sqrt{x})/(1-i\sqrt{x})$, and they leave the
resulting class of sums as an open problem. That at odd conformal spin the three letters
are the whole of the alphabet, and that the constants there are $1$, $\pi^{2}$ and
$\zeta_3$ with no $\pi^{4}$, is an observation of the present study, taken from the
integrand at every odd spin generated here; their $m=0$ and $m=2$ expressions do
contain $\pi^{4}$. The half-integer powers of $x$ that appear at
odd spin sit outside the iterated integral as an overall factor and are not letters. Everything below is a
transformation of that object, carried out in exact rational arithmetic.

\subsection{From iterated integrals to harmonic polylogarithms}
\label{app:cons:hpl}
The iterated integrals are first written as harmonic
polylogarithms~\cite{Remiddi:1999ew}. With the
letters mapped as $x\to0$, $1-x\to1$, $1+x\to-1$ and the word reversed, an iterated
integral of depth $d$ becomes $(-1)^{c}H_{w}(x)$ with $c$ the number of $(1-x)$
letters and $w$ the reversed image of the word. That map is the one supplied with
Ref.~\cite{Caron-Huot:2016tzz}, whose files note that their iterated integrals with
entries $x$, $1+x$ and $1-x$ are harmonic polylogarithms up to signs and a reversal of
indices. Expanding the result and stripping
the factor $(1-x)^{-e}$ that contains the endpoint singularity leaves each term in
the form
\begin{equation}
x^{\,p}\;H_{w}(x)\;(1-x)^{-e},\qquad e\in\{0,1\},
\label{eq:cons:term}
\end{equation}
with a rational or $\zeta$-valued prefactor. At $n=33$, for example, there are
$1113$ such terms; at $n=99$ there are $3357$. The count grows linearly with the
spin.

\subsection{Mellin extraction}
\label{app:cons:mellin}
Each term of the form~\eqref{eq:cons:term} is Mellin transformed in closed form. The
transform is taken in the measure the integrand comes with,
\begin{equation}
\widetilde{I}_n(z)=\int_{0}^{1}x^{\,z-n/2}\,I_n(x)\;dx ,
\label{eq:cons:mellin}
\end{equation}
with $I_n(x)$ the integrand at the spin $n$ and $z$ the variable of
Eq.~\eqref{eq:set:zdef}. Expanding $H_w(x)$ in powers of $x$ and of $\ln x$ takes a
term of Eq.~\eqref{eq:cons:term} to $x^{\,p+N}\ln^{j}x$ with $N\ge0$, and
\begin{equation}
\int_{0}^{1}x^{\,z-n/2+p+N}\ln^{j}x\;dx
=\frac{(-1)^{j}\,j!}{\bigl(N+z+p-n/2+1\bigr)^{\,j+1}} ,
\label{eq:cons:power}
\end{equation}
so the term lands on an atom of pole order $j+1$ at the shift $k=p-n/2+1$. At odd
spin the power $p$ is a half-integer and so is $n/2$; the two combine into an
integer, and that is what puts every instance on a rung of
Eq.~\eqref{eq:set:krange}. The endpoint factor does not move that argument. Since
$(1-x)^{-1}=\sum_{N\ge0}x^{N}$, it replaces the coefficient of $x^{N}\ln^{j}x$ in the
expansion by the sum of the coefficients of $x^{N'}\ln^{j}x$ over $N'\le N$, so a
single power of $x$ turns into every power above it, all on the one rung. The term
$32\,x^{5/2}H_{0,0,0,0}(x)$ of the spin $n=5$ shows what that does: by itself it
integrates to $32/(z+1)^{5}$, and with the factor $(1-x)^{-1}$ in place to
$32\,\zeta(5,1+z)$, which is that rational atom together with the nested atom of pole
order five at the same rung. Carrying this out term by term and collecting like
structures gives the atom table $\Aset_n$ of Eq.~\eqref{eq:cf:master}: a list of
shifted instances in the sense of Sec.~\ref{sec:set:vocab}, each one shape at one
rung and each carrying an exact coefficient of the form $r_0+r_1\pi^2+r_2\zeta_3$.

There is no fitting in this step and no floating point. The extraction is a rewriting
of the integrand, and its output is determined by the input. The number of instances
grows with the spin, from $1284$ at $n=33$ to $3858$ at $n=99$, because the ladder
grows; the number of distinct shapes does not, and stays at $41$.

\subsection{Reproducing the tables}
\label{app:cons:repro}
One command regenerates the structured integrand at a given odd spin from the
Caron-Huot--Herranen data, and a second reduces it to the
atom table. Both were checked against the stored files before being used to produce
anything new: at $n=33$ the regenerated integrand agrees with the stored file term
for term, and the atom table reduced from it agrees with the stored table exactly,
all $1284$ instances with no coefficient different. The four spins of
Sec.~\ref{sec:chk:oos} were produced through the same two commands after those
controls had been run.

The boundary spin $n=1$ is supplied, not reassembled. It contains twenty-eight
instances, on the two rungs $k=0$ and $k=1$, where the uniform count $39n-3$ would ask
for thirty-six, and it stands for the Caron-Huot--Herranen $n=1$ function.

Supplying it is a statement about provenance and not about the extraction. The generic
odd-$n$ route runs at $n=1$ exactly as it runs anywhere else, and it was run: the
structured integrand at that spin has twenty entries, and reducing them gives back the
same twenty-eight instances with the same exact coefficients, byte for byte the stored
table. This fact is what grounds the statement that the boundary block is the same
object that the generic route would produce, and the check is in the ancillary files, where the reduction of
$n=1$ by the generic route and its comparison against the stored table can be repeated
in one command.

\section{Reducing the atoms}\label{app:reduction}

Every atom of Eqs.~\eqref{eq:set:Ratom} and~\eqref{eq:set:BAatom} reduces to the
canonical objects of Eq.~\eqref{eq:cf:canonical}. What follows performs the
reduction and states the identities it rests on.

\subsection{Sums with empty nesting word}
\label{app:red:empty}
The atoms with empty nesting word rest on two identities. The alternating one is a
meromorphic identity valid for all $q\ge1$,
\begin{equation}
\Phibare_q(w)=\sum_{N\ge1}\frac{(-1)^{N}}{(N+w)^{q}}
   =-\bigl(\Sh{-q}{w}+\etaf{q}\bigr),
\label{eq:red:phi}
\end{equation}
which is the even branch of $\Sh{-q}{}$ written out; on the odd branch the right-hand
side changes sign. It is best read as the definition of that branch, not as
something to be derived from a difference equation in $w$: the recursion
$\Sh{-q}{w+1}=\Sh{-q}{w}+(-1)^{w+1}/(w+1)^{q}$ takes the two branches into each
other, so no single branch satisfies a first-order difference equation on its own and
the usual telescoping argument is not available here. The non-alternating one is immediate for $q\ge2$,
\begin{equation}
\zeta(q,w+1)=\sum_{N\ge1}\frac{1}{(N+w)^{q}}=\zeta_q-\Sh{q}{w},
\label{eq:red:psi}
\end{equation}
and diverges at $q=1$, which is why the $q=1$ case is kept in the difference form of
Eq.~\eqref{eq:cf:canonical}. The two are Eqs.~(5) and (6) of
Bl\"umlein~\cite{Blumlein:2000hw} solved for the sum, with $\zeta_q$ the constant
$c_q^{+}$ of that paper and $\etaf{q}$ its $c_q^{-}$; in its Eq.~(6) the branch sits in the same
place, in a factor $(-1)^{N}$ multiplying a difference of polygammas at halved
argument.

\subsection{Partial fractions, and why the split is allowed}
\label{app:red:pf}
Every nested atom in the block has denominator $N^{p}(N+x)^{q}$, with no shift in the
first factor. Partial fractions in $N$ give
\begin{equation}
\frac{1}{N^{p}(N+x)^{q}}=\sum_{i=1}^{p}\frac{a_i}{N^{i}}
+\sum_{j=1}^{q}\frac{b_j}{(N+x)^{j}},
\label{eq:red:pf}
\end{equation}
with $a_i$ and $b_j$ rational in $1/x$. Taken separately the $i=1$ and $j=1$ pieces
diverge when the atom is non-alternating. They are not separate: since the left-hand
side decays as $N^{-(p+q)}$ with $p+q\ge2$, the coefficients of $1/N$ on the right
must cancel,
\begin{equation}
a_1+b_1=0 ,
\label{eq:red:cancel}
\end{equation}
which we have verified symbolically for every $p\le4$ and $q\le5$ occurring in the
block. The convergent combination that survives is $\Delta_{\sigma,v}$ of
Eq.~\eqref{eq:cf:canonical}, and the remaining pieces with $i\ge2$ and $j\ge2$ are
the constants $E_{\sigma,v,i}$ and the functions $M_{\sigma,v,j}$.

Applied across the $57$ atom entries of the block this produces $13$ objects of type
$\Delta$, $11$ of type $M$ and $12$ constants. The reduction was checked atom by
atom against direct summation, to a relative $2.71171\times10^{-31}$, at thirty
digits with $J=200$.

\subsection{The sums with a non-empty nesting word}
\label{app:red:nested}
Of the twenty-four functions of one variable that Sec.~\ref{sec:cf:space} counts,
eight have an empty nesting word and are the polygammas of
Eq.~\eqref{eq:cf:elementary}. The sixteen that remain contain the whole non-elementary
content of the block, and each of them is a finite combination of nested harmonic
sums of the single argument $w$. That fact is what the statement of
Sec.~\ref{sec:cf:space} rests on.

Fifteen of the sixteen alternate. Ten of them are the sums $\Delta_{-,v}$ that
Eq.~\eqref{eq:red:pf} leaves at $i=j=1$, and five are the $M_{-,v,j}$ it leaves at
$j\ge2$; an alternating $\Delta_{-,v}$ is $E_{-,v,1}-M_{-,v,1}$, with both pieces
convergent on their own, so all fifteen are carried by the identities for $M_{-,v,j}$
below, at the pole orders $j\le q$ that occur in the atoms. The sixteenth is
$\Delta_{+,(1)}$, left by $\sum_{N\ge1}\Sh{1}{N}/[N^{3}(N+w)]$, the one shape of
App.~\ref{app:cat:shapes} that has a nesting word and does not alternate. Every harmonic sum
below is at the argument $w$ and on the even branch of App.~\ref{app:conv:sums}, and
the constants are $\zeta_2$, $\zeta_3$, $\zeta_4$, $\lntwo$ and $\Lifour$.

The linear representations the sixteen rest on are those of Bl\"umlein and
Kurth~\cite{Blumlein:1998if}. Their Sec.~4 gives every nested harmonic sum of weight
at most four as a Mellin transform of a Nielsen function together with sums of lower
weight, over the constants $\lntwo$, $\zeta_2$, $\zeta_3$ and $\Lifour$ of their own
list, $\zeta_4$ appearing there as $\zeta_2^{2}$. Equation~(50) of
Ref.~\cite{Blumlein:1998if} at $N=w-1$, solved for that transform, is the identity for
$M_{-,(-2),1}$ below, which we checked against the defining series and against the
integral at $w=4,6,8$ to thirty digits. The substitution is $N=w-1$ and not $N=w$
because the harmonic-sum transforms of that paper are written with an explicit
$x^{N}$, one power above the definition in its Eq.~(2)~\cite{Blumlein:1998if}. The
other fifteen follow in the same way, each from the relation of that section whose
transform it is, and none of the sixteen has weight above four, which is the range
that section covers. Here those representations are continued to complex $w$ on the
even branch, and the argument is shifted by one, since the atom leaves the transform
at $w-1$ where Eq.~\eqref{eq:cf:canonical} asks for a function of $w$. At depth one
the identities
read
\begin{align*}
M_{-,(1),1}(w)&= \Sh{-1,1}{w} - \Sh{-2}{w} - \Sh{-1}{w}\,\Sh{1}{w} + \frac{\Sh{-1}{w}}{w} + \lntwo\,\Sh{-1}{w}\\
&\quad - \lntwo\,\Sh{1}{w} + \tfrac{1}{2}\,\lntwo^{2}\\[4pt]
M_{-,(1),2}(w)&= \Sh{-1,2}{w} + \Sh{-2,1}{w} - 2\,\Sh{-3}{w} - \Sh{-1}{w}\,\Sh{2}{w} - \Sh{-2}{w}\,\Sh{1}{w}\\
&\quad - \tfrac{1}{2}\,\zeta_2\,\Sh{1}{w} + \lntwo\,\Sh{-2}{w} - \lntwo\,\Sh{2}{w} + \frac{\Sh{-2}{w}}{w}\\
&\quad + \frac{\Sh{-1}{w}}{w^{2}} + \frac{\zeta_2}{2\,w} + \tfrac{1}{8}\,\zeta_3\\[4pt]
M_{-,(1),3}(w)&= \Sh{-1,3}{w} + \Sh{-2,2}{w} + \Sh{-3,1}{w} - 3\,\Sh{-4}{w} - \Sh{-3}{w}\,\Sh{1}{w}\\
&\quad - \Sh{-2}{w}\,\Sh{2}{w} - \Sh{-1}{w}\,\Sh{3}{w} - \tfrac{1}{2}\,\zeta_2\,\Sh{2}{w}\\
&\quad - \tfrac{3}{4}\,\zeta_3\,\Sh{1}{w} + \lntwo\,\Sh{-3}{w} - \lntwo\,\Sh{3}{w} + \frac{\Sh{-3}{w}}{w}\\
&\quad + \frac{\Sh{-2}{w}}{w^{2}} + \frac{\Sh{-1}{w}}{w^{3}} + \frac{3\,\zeta_3}{4\,w}\\
&\quad + \frac{\zeta_2}{2\,w^{2}} - \tfrac{15}{8}\,\zeta_4 + \tfrac{7}{4}\,\zeta_3\,\lntwo\\
&\quad - \tfrac{1}{2}\,\zeta_2\,\lntwo^{2} + \tfrac{1}{12}\,\lntwo^{4} + 2\,\Lifour\\[4pt]
M_{-,(2),1}(w)&= - \tfrac{1}{4}\,\zeta_3 + \tfrac{1}{2}\,\zeta_2\,\lntwo - \frac{\zeta_2}{2\,w} - \Sh{-1,2}{w}\\
&\quad + \tfrac{1}{2}\,\zeta_2\,\Sh{-1}{w} + \Sh{-1}{w}\,\Sh{2}{w} - \frac{\Sh{-1}{w}}{w^{2}}\\
&\quad - \lntwo\,\Sh{-2}{w} + \Sh{-3}{w} + \lntwo\,\Sh{2}{w}\\[4pt]
M_{-,(2),2}(w)&= \tfrac{1}{2}\,\zeta_2\,\Sh{-2}{w} + \tfrac{1}{2}\,\zeta_2\,\Sh{2}{w} + \Sh{2,-2}{w} + 2\,\Sh{3,-1}{w}\\
&\quad - 2\,\lntwo\,\Sh{-3}{w} + 2\,\lntwo\,\Sh{3}{w} - \frac{\Sh{-2}{w}}{w^{2}} - \frac{\zeta_2}{w^{2}}\\
&\quad - \frac{2\,\Sh{-1}{w}}{w^{3}} + \tfrac{65}{16}\,\zeta_4 - \tfrac{7}{2}\,\zeta_3\,\lntwo\\
&\quad + \zeta_2\,\lntwo^{2} - \tfrac{1}{6}\,\lntwo^{4} - 4\,\Lifour\\[4pt]
M_{-,(3),1}(w)&= - \Sh{3,-1}{w} - \tfrac{1}{2}\,\zeta_2\,\Sh{-2}{w} + \tfrac{3}{4}\,\zeta_3\,\Sh{-1}{w}\\
&\quad + \lntwo\,\Sh{-3}{w} - \lntwo\,\Sh{3}{w} + \frac{\Sh{-1}{w}}{w^{3}} + \frac{\zeta_2}{2\,w^{2}}\\
&\quad - \frac{3\,\zeta_3}{4\,w} - \tfrac{5}{16}\,\zeta_4 + \tfrac{3}{4}\,\zeta_3\,\lntwo\\[4pt]
M_{-,(-2),1}(w)&= \Sh{-2,1}{w} - \zeta_2\,\Sh{-1}{w} - \frac{\Sh{1}{w}}{w^{2}} + \frac{\zeta_2}{w} + \tfrac{5}{8}\,\zeta_3\\
&\quad - \zeta_2\,\lntwo\\[4pt]
M_{-,(-2),2}(w)&= 2\,\Sh{-3,1}{w} + \Sh{-2,2}{w} - 2\,\zeta_2\,\Sh{-2}{w} - \tfrac{3}{16}\,\zeta_4\\
&\quad + \frac{2\,\zeta_2}{w^{2}} - \frac{\Sh{2}{w}}{w^{2}} - \frac{2\,\Sh{1}{w}}{w^{3}}\\[4pt]
M_{-,(-3),1}(w)&= - \Sh{-3,1}{w} + \zeta_2\,\Sh{-2}{w} - \zeta_3\,\Sh{-1}{w} + \frac{\Sh{1}{w}}{w^{3}}\\
&\quad - \frac{\zeta_2}{w^{2}} + \frac{\zeta_3}{w} - \tfrac{3}{2}\,\zeta_4 + \tfrac{3}{4}\,\zeta_3\,\lntwo\\
&\quad - \tfrac{1}{2}\,\zeta_2\,\lntwo^{2} + \tfrac{1}{12}\,\lntwo^{4} + 2\,\Lifour
\end{align*}
and at depth two, with the non-alternating sum after them,
\begin{align*}
M_{-,(1,1),1}(w)&= \tfrac{1}{2}\,\lntwo^{2}\,\Sh{1}{w} - \tfrac{1}{2}\,\lntwo^{2}\,\Sh{-1}{w}\\
&\quad + \tfrac{1}{2}\,\zeta_2\,\Sh{-1}{w} - \lntwo\,\Sh{1,1}{w} + \lntwo\,\Sh{1,-1}{w} + \lntwo\,\Sh{2}{w}\\
&\quad - \lntwo\,\Sh{-2}{w} - \Sh{1,1,-1}{w} + \Sh{2,-1}{w} + \frac{\lntwo\,\Sh{1}{w}}{w}\\
&\quad - \frac{\lntwo\,\Sh{-1}{w}}{w} + \frac{\Sh{1,-1}{w}}{w} - \frac{\zeta_2}{2\,w}\\
&\quad - \frac{\Sh{-1}{w}}{w^{2}} - \tfrac{1}{4}\,\zeta_3 + \tfrac{1}{2}\,\zeta_2\,\lntwo\\
&\quad - \tfrac{1}{6}\,\lntwo^{3}\\[4pt]
M_{-,(1,1),2}(w)&= - \Sh{1,1,-2}{w} - \Sh{1,2,-1}{w} - \Sh{2,1,-1}{w} + \Sh{2,-2}{w} + 2\,\Sh{3,-1}{w}\\
&\quad - \lntwo\,\Sh{1,2}{w} - \lntwo\,\Sh{2,1}{w} + \lntwo\,\Sh{1,-2}{w} + \lntwo\,\Sh{2,-1}{w}\\
&\quad + 2\,\lntwo\,\Sh{3}{w} - 2\,\lntwo\,\Sh{-3}{w} + \tfrac{1}{2}\,\lntwo^{2}\,\Sh{2}{w}\\
&\quad - \tfrac{1}{2}\,\lntwo^{2}\,\Sh{-2}{w} + \tfrac{1}{2}\,\zeta_2\,\Sh{2}{w}\\
&\quad + \tfrac{1}{2}\,\zeta_2\,\Sh{-2}{w} - \tfrac{1}{2}\,\zeta_2\,\Sh{1,1}{w}\\
&\quad + \tfrac{1}{8}\,\zeta_3\,\Sh{1}{w} + \frac{\Sh{1,-1}{w}}{w^{2}} + \frac{\Sh{1,-2}{w}}{w}\\
&\quad + \frac{\Sh{2,-1}{w}}{w} - \frac{\Sh{-2}{w}}{w^{2}} - \frac{2\,\Sh{-1}{w}}{w^{3}}\\
&\quad + \frac{\lntwo\,\Sh{1}{w}}{w^{2}} + \frac{\lntwo\,\Sh{2}{w}}{w} - \frac{\lntwo\,\Sh{-1}{w}}{w^{2}}\\
&\quad - \frac{\lntwo\,\Sh{-2}{w}}{w} + \frac{\zeta_2\,\Sh{1}{w}}{2\,w} - \frac{\zeta_2}{w^{2}}\\
&\quad - \frac{\zeta_3}{8\,w} + \tfrac{49}{16}\,\zeta_4 - \tfrac{21}{8}\,\zeta_3\,\lntwo\\
&\quad + \tfrac{3}{4}\,\zeta_2\,\lntwo^{2} - \tfrac{1}{8}\,\lntwo^{4} - 3\,\Lifour\\[4pt]
M_{-,(1,2),1}(w)&= \Sh{1,2,-1}{w} - \Sh{3,-1}{w} + \lntwo\,\Sh{1,2}{w} - \lntwo\,\Sh{1,-2}{w} - \lntwo\,\Sh{3}{w}\\
&\quad + \lntwo\,\Sh{-3}{w} - \tfrac{1}{2}\,\zeta_2\,\Sh{-2}{w} + \tfrac{1}{2}\,\zeta_2\,\Sh{1,-1}{w}\\
&\quad - \tfrac{1}{4}\,\zeta_3\,\Sh{1}{w} + \tfrac{1}{2}\,\zeta_2\,\lntwo\,\Sh{1}{w} + \zeta_3\,\Sh{-1}{w}\\
&\quad - \tfrac{1}{2}\,\zeta_2\,\lntwo\,\Sh{-1}{w} - \frac{\Sh{2,-1}{w}}{w} + \frac{\Sh{-1}{w}}{w^{3}}\\
&\quad - \frac{\lntwo\,\Sh{2}{w}}{w} + \frac{\lntwo\,\Sh{-2}{w}}{w} - \frac{\zeta_2\,\Sh{-1}{w}}{2\,w}\\
&\quad - \frac{3\,\zeta_3}{4\,w} + \frac{\zeta_2}{2\,w^{2}} - \tfrac{53}{16}\,\zeta_4\\
&\quad + \tfrac{29}{8}\,\zeta_3\,\lntwo - \zeta_2\,\lntwo^{2} + \tfrac{1}{8}\,\lntwo^{4} + 3\,\Lifour\\[4pt]
M_{-,(2,1),1}(w)&= \Sh{2,1,-1}{w} - \Sh{3,-1}{w} + \lntwo\,\Sh{2,1}{w} - \lntwo\,\Sh{2,-1}{w} - \lntwo\,\Sh{3}{w}\\
&\quad + \lntwo\,\Sh{-3}{w} - \tfrac{1}{2}\,\lntwo^{2}\,\Sh{2}{w} + \tfrac{1}{2}\,\lntwo^{2}\,\Sh{-2}{w}\\
&\quad - \tfrac{1}{2}\,\zeta_2\,\Sh{-2}{w} + \tfrac{5}{8}\,\zeta_3\,\Sh{-1}{w} - \frac{\Sh{1,-1}{w}}{w^{2}}\\
&\quad + \frac{\Sh{-1}{w}}{w^{3}} - \frac{\lntwo\,\Sh{1}{w}}{w^{2}} + \frac{\lntwo\,\Sh{-1}{w}}{w^{2}}\\
&\quad - \frac{5\,\zeta_3}{8\,w} + \frac{\zeta_2}{2\,w^{2}} + \tfrac{43}{16}\,\zeta_4 - 2\,\zeta_3\,\lntwo\\
&\quad + \tfrac{3}{4}\,\zeta_2\,\lntwo^{2} - \tfrac{1}{8}\,\lntwo^{4} - 3\,\Lifour\\[4pt]
M_{-,(-2,1),1}(w)&= \Sh{-2,1,1}{w} - \Sh{-3,1}{w} + \zeta_2\,\Sh{-2}{w} - 2\,\zeta_3\,\Sh{-1}{w} - \frac{\Sh{1,1}{w}}{w^{2}}\\
&\quad + \frac{\Sh{1}{w}}{w^{3}} + \frac{2\,\zeta_3}{w} - \frac{\zeta_2}{w^{2}} - \tfrac{29}{16}\,\zeta_4\\
&\quad + \tfrac{5}{8}\,\zeta_3\,\lntwo - \tfrac{3}{4}\,\zeta_2\,\lntwo^{2} + \tfrac{1}{8}\,\lntwo^{4}\\
&\quad + 3\,\Lifour\\[4pt]
M_{-,(1,-2),1}(w)&= \Sh{1,-2,1}{w} - \Sh{-3,1}{w} + \zeta_2\,\Sh{-2}{w} - \zeta_2\,\Sh{1,-1}{w}\\
&\quad + \tfrac{5}{8}\,\zeta_3\,\Sh{1}{w} - \zeta_2\,\lntwo\,\Sh{1}{w} - \tfrac{13}{8}\,\zeta_3\,\Sh{-1}{w}\\
&\quad + \zeta_2\,\lntwo\,\Sh{-1}{w} - \frac{\Sh{-2,1}{w}}{w} + \frac{\Sh{1}{w}}{w^{3}}\\
&\quad + \frac{\zeta_2\,\Sh{-1}{w}}{w} + \frac{\zeta_3}{w} - \frac{\zeta_2}{w^{2}} - \tfrac{21}{16}\,\zeta_4\\
&\quad + \tfrac{1}{8}\,\zeta_3\,\lntwo + \tfrac{1}{12}\,\lntwo^{4} + 2\,\Lifour\\[4pt]
\Delta_{+,(1)}(w)&= \Sh{1,1}{w} - \frac{\Sh{1}{w}}{w} + \zeta_2
\end{align*}

Each identity is homogeneous in the transcendental weight, which is the weight of $v$
plus $j$ with $1/w$ counted as one unit, so the grading of
App.~\ref{app:cat:grading} survives the reduction term by term. Both sides were
compared at six complex points, two of them with $\mathrm{Re}\,w<0$, at sixty digits
of working precision, the left-hand side summed from its defining series and the
right-hand side evaluated through the continuation of App.~\ref{app:conv:sums}: the
largest difference over the ninety-six evaluations is $8.7\times10^{-50}$. The
program that does it is \texttt{anc/eval/closure\_check.py} of
App.~\ref{app:anc:eval}, and it keeps the residual of every point.

\subsection{Stuffle reduction of the depth-two atoms}
\label{app:red:stuffle}
The function $\Gladder_{(-1)^k}$ contains six atoms of depth two. Three stuffle relations,
in the non-strict convention of Eq.~\eqref{eq:set:sdef}, reduce them:
\begin{equation}
\Sh{1,2}{N}+\Sh{2,1}{N}=\Sh{1}{N}\Sh{2}{N}+\Sh{3}{N},\qquad
2\Sh{1,1}{N}=\Sh{1}{N}^{2}+\Sh{2}{N},
\label{eq:red:stuffle1}
\end{equation}
\begin{equation}
\Sh{-2,1}{N}+\Sh{1,-2}{N}=\Sh{1}{N}\Sh{-2}{N}+\Sh{-3}{N}.
\label{eq:red:stuffle2}
\end{equation}
All three are instances of the product rule of
Bl\"umlein~\cite{Blumlein:2003gb}, whose Eq.~(1.1) is the same non-strict nesting as
Eq.~\eqref{eq:set:sdef} and whose Eq.~(2.2), with $a\wedge b=\mathrm{sgn}(a)\,
\mathrm{sgn}(b)\,(|a|+|b|)$, gives them at the index pairs $(1,2)$, $(1,1)$ and
$(1,-2)$; the same relation is Eq.~(9) of Ref.~\cite{Vermaseren:1998uu}. We checked
all three against the defining sums before use. Splitting the
$(-2,1)$ pair into its symmetric and antisymmetric parts,
\begin{equation}
128\,\Sh{-2,1}{N}-64\,\Sh{1,-2}{N}
=32\bigl[\Sh{1}{N}\Sh{-2}{N}+\Sh{-3}{N}\bigr]
+96\bigl[\Sh{-2,1}{N}-\Sh{1,-2}{N}\bigr],
\label{eq:red:split}
\end{equation}
leaves one irreducible nested object, $\Sh{-2,1}{N}-\Sh{1,-2}{N}$, at coefficient
$96$. The reduction changes three depth-one coefficients: $\Sh{3}{N}$ from $160$ to
$32$, $\Sh{-3}{N}$ from $-96$ to $-64$, and, if the $\Sh{2}{N}$ half generated by the
$(1,1)$ pair is absorbed into the two $\Sh{2}{N}$ atoms already present, those from
$96$ to $32$ and from $160$ to $96$. Both presentations reproduce $\Gladder_{(-1)^k}$; the
reduced form was checked against the original to a relative
$2.74321\times10^{-29}$, thirty digits and $J=200$.

Three product-type objects survive the reduction, $\Sh{1}{N}^{2}$,
$\Sh{1}{N}\Sh{2}{N}$ and $\Sh{1}{N}\Sh{-2}{N}$ under the Mellin sum. They are not
reducible to polygammas; for that reason $\Gladder_{(-1)^k}$ stays several distinct objects
sharing one ladder weight.

\subsection{The ladder sums}
\label{app:red:ladder}
The sums along the ladder in Eq.~\eqref{eq:cf:ladderform} are performed once. With
$\Sigq_q(x,m)=\sum_{b=0}^{m-1}(x-b)^{-q}$, a finite descending sum to be kept apart
from the $\Sig{a}$ fixed under Eq.~\eqref{eq:intro:nlo}, a ladder sum whose weight does not move with
the rung is elementary against a summand of empty nesting word. Only the alternating
weight occurs here. The first identity below defines $\Sigq_q$ and the second is the one
used,
\begin{equation}
\sum_{a=0}^{n-1}\frac{1}{(w-a)^{q}}=\Sigq_q(w,n),
\qquad
\sum_{a=0}^{n-1}\frac{(-1)^{a}}{(w-a)^{q}}
=\frac{1}{2^{q}}\Bigl[\Sigq_q\bigl(\tfrac{w}{2},\lceil\tfrac n2\rceil\bigr)
-\Sigq_q\bigl(\tfrac{w-1}{2},\lfloor\tfrac n2\rfloor\bigr)\Bigr],
\label{eq:red:Lalt}
\end{equation}
and $\Sigq_q$ is a difference of polygammas. The halving of the argument in the
alternating case is what the ladder sum does to an alternating weight. The same
mechanism acts one level down, inside a single atom, and it is there that the halved
arguments of Eq.~\eqref{eq:cf:elementary} come from: the atom brings its own
alternating sum $\sum_{N\ge1}(-1)^{N}/(N+x)^{j}$, and splitting it over even and odd
$N$ produces them directly, with no sum along the ladder involved. The two are the same
duplication at two levels and should not be conflated.

The three sums dressed by $\Sh{1}{}$ or $\Sh{2}{}$ are not elementary. Summation by
parts against the elementary partial sums brings each of them to the same shape, a
single nested sum with an elementary summand,
\begin{equation}
\mathrm{Nest}_q(n,u;f)=\sum_{b=1}^{n-1}P_q(b-1,u)\,f(b),
\qquad
P_q(b,u)=\sum_{c=0}^{b}\frac{(-1)^{c}}{(u+c)^{q}},
\label{eq:red:nest}
\end{equation}
with $f(b)=1/b$ for the $\Sh{1}{}$ weights and $f(b)=1/b^{2}+(-1)^{b}/b^{2}$ for the
$\Sh{2}{}+\Sh{-2}{}$ weight. The upper limit is the only place the spin appears, so
one expression serves at every $n$. It is not
elementary, and Eq.~\eqref{eq:red:nest} is as far as the reduction goes.

\section{The coefficient catalogue}\label{app:catalogue}

\subsection{The slot census}
\label{app:cat:census}
The block contains $41$ atom shapes, five rational and thirty-six nested. Paired
with the constants $1,\pi^2,\zeta_3$ these give $47$ non-empty slots, of which $40$
have coefficient rules that close in the two ladder distances and $7$ do not. Neither
number grows with the spin; the number of shifted instances does, one shape
placed at one rung, $39n-3$ at every odd $n\ge3$ of the computed range, so
$114$ at $n=3$ and $1284$ at $n=33$, against $28$ in the boundary block at $n=1$. Over
the range $n\le99$ the tables hold $112140$ coefficients in total, $94843$ of
them in closing slots and $17297$ in the seven that do not close; the holdout spin
$n=101$ of Sec.~\ref{sec:chk:oos} adds a further $4538$ and is counted apart from
these, since every count in this paper that says $n\le99$ was taken before it existed. That last figure counts table entries, with no measure of
size or of transcendental content in it, and fixes only the scale of the catalogue.

\subsection{The weight grading}
\label{app:cat:grading}
The grading is fixed once here and used throughout the paper. The weight of a nested
atom of Eq.~\eqref{eq:set:BAatom} is the weight of its nesting word $v$, plus the
powers $p$ of the summation variable carried by its extra denominators, plus its pole
order $q$; the weight of a rational atom of Eq.~\eqref{eq:set:Ratom} is $q$. On that
counting the $41$ shapes have weights one through five, twenty-four of them five. The
weight of a coefficient is the weight of its constant (zero for $1$, two for $\pi^{2}$,
three for $\zeta_3$, and none for a rational multiple) plus the weight of the harmonic
sums in $\ap$ and $\am$ that dress it, with the ladder sums of
Eq.~\eqref{eq:res:laddersums} graded as in Sec.~\ref{sec:dis:weight}.

Neither factor separately need have weight five, and the atom weight in particular
moves from slot to slot. Each complete coefficient-atom contribution is homogeneous of
total weight five: on all $47$ slots the two weights add to that number. For the forty
closing slots this is read off Table~\ref{tab:cat:rules} against
App.~\ref{app:cat:shapes}, and for the seven of Sec.~\ref{sec:residual} it is the
content of Sec.~\ref{sec:dis:weight}. No
constant of weight above three appears at slot level. That is a property of the
extraction and not of the grading, since the rational atom $1/w$ contributes only one unit
and would leave room for one: what the integrand of
Ref.~\cite{Caron-Huot:2016tzz} supplies at odd spin is $1$, $\pi^{2}$
and $\zeta_3$, as App.~\ref{app:construction} states, and nothing of higher weight
reaches a slot.

\subsection{The closing rules}
\label{app:cat:rules}
Each of the $40$ closing slots is a rational combination of the seven weights of
Eq.~\eqref{eq:str:weights}. The catalogue in the ancillary files stores those
combinations as exact fractions, keyed by shape and constant. The rational slot of pole
order four is
\begin{equation}
\coeff{}^{\,(\Ratom,\,q=4)}(n,k)=48\,(-1)^{k}\,\Sh{1}{\ap},
\label{eq:cat:q4}
\end{equation}
a single weight with a single rational multiple. The slot of pole order five and the
nested slot of pole order five are constants, $32$ each, non-zero only at the single
rung $k=1$; these are the two that never enter a ladder sum. And a representative
multi-weight slot reads
\begin{equation}
\coeff{}^{\,(\BAatom,\,1,\,(1),\,((0,1)),\,2)}(n,k)
=(-1)^{k}\Bigl[\,c_1\,\Sh{1}{\ap}+c_2\,\Sh{1}{\am}
+c_3\,\Sh{1}{\amin}\,\Bigr],
\label{eq:cat:multi}
\end{equation}
with $c_i\in\mathbb{Q}$ fixed in the catalogue. The remaining thirty-six are of the
same kind, rational combinations of the seven weights with different multiples, and
Table~\ref{tab:cat:rules} gives all forty.

\subsection{The rules}
\label{app:cat:table}
Table~\ref{tab:cat:rules} is the catalogue. Each row gives one slot's coefficient rule as the
rational combination of the seven weights of Eq.~\eqref{eq:str:weights}; blank cells are zero.
Thirty-one of the forty rules use a single weight, and no rule mixes the weight
$(-1)^{k}$ with an $\Sh{1}{}$ weight. In the row labels the nesting word, the extra
denominators and the pole order are given in the order of
Eq.~\eqref{eq:set:BAatom}, so $\BAatom^{1}_{1;\,(0,1);\,2}$ is the atom
$\atomlab{\BAatom}{(1,(1),((0,1)),2)}$ of the body. With this table, the atom list of
App.~\ref{app:cat:shapes}, the reductions of App.~\ref{app:reduction} and the rules of
Sec.~\ref{sec:res:ladder} for the seven remaining slots,
Eq.~\eqref{eq:cf:master} can be evaluated without the ancillary files.

\clearpage
{\small\setlength{\tabcolsep}{2pt}%
\begin{longtable}{@{}llcrrrrrrr@{}}
\caption{The forty closing coefficient rules. Each row gives $\coeff{}(n,k)$ as the
rational combination of the seven weights of Eq.~\eqref{eq:str:weights} shown in the
columns; blank cells are zero. The rung range is the full ladder $-(n-1)\le k\le0$ unless
the third column says otherwise, where $k\!=\!1$ marks the two single-rung slots and an
entry $n{-}j$ means that the rung range ends one or more rungs early, at $k=-(n-j)$.}\label{tab:cat:rules}\\
\toprule
shape & $\mu$ & supp. & $1$ & $\epsk$ & $\epsk\Sh{1}{\ap}$ & $\epsk\Sh{1}{\am}$ & $\epsk\Sh{1}{\amin}$ & $\epsk\Sh{2}{\ap}$ & $\epsk\Sh{-2}{\ap}$\\
\midrule\endfirsthead
\toprule
shape & $\mu$ & supp. & $1$ & $\epsk$ & $\epsk\Sh{1}{\ap}$ & $\epsk\Sh{1}{\am}$ & $\epsk\Sh{1}{\amin}$ & $\epsk\Sh{2}{\ap}$ & $\epsk\Sh{-2}{\ap}$\\
\midrule\endhead
$\BAatom^{0}_{\varnothing;\,(0,2);\,3}$ & $1$ &  &  & $-32$ &  &  &  &  & \\
$\BAatom^{0}_{\varnothing;\,(0,3);\,1}$ & $1$ &  &  &  & $-32$ & $96$ & $-32$ &  & \\
$\BAatom^{0}_{\varnothing;\,(0,3);\,2}$ & $1$ &  &  & $-48$ &  &  &  &  & \\
$\BAatom^{0}_{\varnothing;\,(0,4);\,1}$ & $1$ &  &  & $-96$ &  &  &  &  & \\
$\BAatom^{0}_{\varnothing;\,\varnothing;\,5}$ & $1$ & $k\!=\!1$ & $32$ &  &  &  &  &  & \\
$\BAatom^{0}_{1;\,(0,3);\,1}$ & $1$ &  &  & $-64$ &  &  &  &  & \\
$\BAatom^{1}_{\varnothing;\,(0,1);\,3}$ & $1$ &  &  &  & $-32$ & $-32$ & $32$ &  & \\
$\BAatom^{1}_{\varnothing;\,(0,1);\,4}$ & $1$ &  &  & $-48$ &  &  &  &  & \\
$\BAatom^{1}_{\varnothing;\,(0,2);\,2}$ & $1$ &  &  &  & $-96$ & $-32$ & $64$ &  & \\
$\BAatom^{1}_{\varnothing;\,(0,2);\,3}$ & $1$ &  &  & $-64$ &  &  &  &  & \\
$\BAatom^{1}_{\varnothing;\,(0,3);\,1}$ & $1$ &  &  &  & $-160$ & $-32$ & $96$ &  & \\
$\BAatom^{1}_{\varnothing;\,(0,3);\,2}$ & $1$ &  &  & $-144$ &  &  &  &  & \\
$\BAatom^{1}_{\varnothing;\,(0,4);\,1}$ & $1$ &  &  & $-288$ &  &  &  &  & \\
$\BAatom^{1}_{-2,1;\,(0,1);\,1}$ & $1$ &  &  & $128$ &  &  &  &  & \\
$\BAatom^{1}_{-2;\,(0,1);\,1}$ & $1$ &  &  &  & $-64$ & $-64$ & $64$ &  & \\
$\BAatom^{1}_{-2;\,(0,1);\,2}$ & $1$ &  &  & $-32$ &  &  &  &  & \\
$\BAatom^{1}_{-3;\,(0,1);\,1}$ & $1$ &  &  & $-96$ &  &  &  &  & \\
$\BAatom^{1}_{1,-2;\,(0,1);\,1}$ & $1$ &  &  & $-64$ &  &  &  &  & \\
$\BAatom^{1}_{1,1;\,(0,1);\,2}$ & $1$ &  &  & $-128$ &  &  &  &  & \\
$\BAatom^{1}_{1,1;\,(0,2);\,1}$ & $1$ &  &  & $-128$ &  &  &  &  & \\
$\BAatom^{1}_{1,2;\,(0,1);\,1}$ & $1$ &  &  & $-128$ &  &  &  &  & \\
$\BAatom^{1}_{1;\,(0,1);\,2}$ & $1$ &  &  &  & $64$ & $64$ & $-64$ &  & \\
$\BAatom^{1}_{1;\,(0,1);\,3}$ & $1$ &  &  & $32$ &  &  &  &  & \\
$\BAatom^{1}_{1;\,(0,2);\,1}$ & $1$ &  &  &  & $64$ & $64$ & $-64$ &  & \\
$\BAatom^{1}_{1;\,(0,2);\,2}$ & $1$ &  &  & $96$ &  &  &  &  & \\
$\BAatom^{1}_{1;\,(0,3);\,1}$ & $1$ &  &  & $160$ &  &  &  &  & \\
$\BAatom^{1}_{2,1;\,(0,1);\,1}$ & $1$ &  &  & $-128$ &  &  &  &  & \\
$\BAatom^{1}_{2;\,(0,1);\,1}$ & $1$ &  &  &  & $64$ & $64$ & $-64$ &  & \\
$\BAatom^{1}_{2;\,(0,1);\,2}$ & $1$ &  &  & $96$ &  &  &  &  & \\
$\BAatom^{1}_{2;\,(0,2);\,1}$ & $1$ &  &  & $160$ &  &  &  &  & \\
$\BAatom^{1}_{3;\,(0,1);\,1}$ & $1$ &  &  & $160$ &  &  &  &  & \\
$\Ratom_{4}$ & $1$ & $n{-}2$ &  &  & $48$ &  &  &  & \\
$\Ratom_{5}$ & $1$ & $k\!=\!1$ & $32$ &  &  &  &  &  & \\
$\BAatom^{0}_{\varnothing;\,(0,2);\,1}$ & $\pi^2$ &  &  & $-\tfrac{16}{3}$ &  &  &  &  & \\
$\BAatom^{1}_{\varnothing;\,(0,1);\,2}$ & $\pi^2$ &  &  & $-\tfrac{16}{3}$ &  &  &  &  & \\
$\BAatom^{1}_{\varnothing;\,(0,2);\,1}$ & $\pi^2$ &  &  & $-\tfrac{16}{3}$ &  &  &  &  & \\
$\Ratom_{1}$ & $\pi^2$ & $n{-}3$ &  &  &  &  &  & $\tfrac{16}{3}$ & $\tfrac{16}{3}$\\
$\Ratom_{2}$ & $\pi^2$ & $n{-}2$ &  &  & $\tfrac{16}{3}$ &  &  &  & \\
$\BAatom^{1}_{\varnothing;\,(0,1);\,1}$ & $\zeta_3$ &  &  & $-32$ &  &  &  &  & \\
$\Ratom_{1}$ & $\zeta_3$ & $n{-}2$ &  &  & $32$ &  &  &  & \\
\bottomrule
\end{longtable}}

\subsection{The atom shapes}
\label{app:cat:shapes}
The $41$ shapes, five rational and thirty-six nested, in the notation of
Eqs.~\eqref{eq:set:Ratom} and~\eqref{eq:set:BAatom} with $w$ the argument:

\begin{align*}
&1/w^{1}\\
&1/w^{2}\\
&1/w^{3}\\
&1/w^{4}\\
&1/w^{5}\\[4pt]
&\sum_N 1\,1/[(N+w)^{5}]\\
&\sum_N 1\,1/[(N+0)^{1}(N+w)^{1}]\\
&\sum_N 1\,1/[(N+0)^{2}(N+w)^{1}]\\
&\sum_N 1\,1/[(N+0)^{2}(N+w)^{3}]\\
&\sum_N 1\,1/[(N+0)^{3}(N+w)^{1}]\\
&\sum_N 1\,1/[(N+0)^{3}(N+w)^{2}]\\
&\sum_N 1\,1/[(N+0)^{4}(N+w)^{1}]\\
&\sum_N 1\,S_{1}(N)/[(N+0)^{3}(N+w)^{1}]\\
&\sum_N (-1)^N\,1/[(N+0)^{1}(N+w)^{1}]\\
&\sum_N (-1)^N\,1/[(N+0)^{1}(N+w)^{2}]\\
&\sum_N (-1)^N\,1/[(N+0)^{1}(N+w)^{3}]\\
&\sum_N (-1)^N\,1/[(N+0)^{1}(N+w)^{4}]\\
&\sum_N (-1)^N\,1/[(N+0)^{2}(N+w)^{1}]\\
&\sum_N (-1)^N\,1/[(N+0)^{2}(N+w)^{2}]\\
&\sum_N (-1)^N\,1/[(N+0)^{2}(N+w)^{3}]\\
&\sum_N (-1)^N\,1/[(N+0)^{3}(N+w)^{1}]\\
&\sum_N (-1)^N\,1/[(N+0)^{3}(N+w)^{2}]\\
&\sum_N (-1)^N\,1/[(N+0)^{4}(N+w)^{1}]\\
&\sum_N (-1)^N\,S_{-3}(N)/[(N+0)^{1}(N+w)^{1}]\\
&\sum_N (-1)^N\,S_{-2}(N)/[(N+0)^{1}(N+w)^{1}]\\
&\sum_N (-1)^N\,S_{-2}(N)/[(N+0)^{1}(N+w)^{2}]\\
&\sum_N (-1)^N\,S_{-2,1}(N)/[(N+0)^{1}(N+w)^{1}]\\
&\sum_N (-1)^N\,S_{1}(N)/[(N+0)^{1}(N+w)^{2}]\\
&\sum_N (-1)^N\,S_{1}(N)/[(N+0)^{1}(N+w)^{3}]\\
&\sum_N (-1)^N\,S_{1}(N)/[(N+0)^{2}(N+w)^{1}]\\
&\sum_N (-1)^N\,S_{1}(N)/[(N+0)^{2}(N+w)^{2}]\\
&\sum_N (-1)^N\,S_{1}(N)/[(N+0)^{3}(N+w)^{1}]\\
&\sum_N (-1)^N\,S_{1,-2}(N)/[(N+0)^{1}(N+w)^{1}]\\
&\sum_N (-1)^N\,S_{1,1}(N)/[(N+0)^{1}(N+w)^{2}]\\
&\sum_N (-1)^N\,S_{1,1}(N)/[(N+0)^{2}(N+w)^{1}]\\
&\sum_N (-1)^N\,S_{1,2}(N)/[(N+0)^{1}(N+w)^{1}]\\
&\sum_N (-1)^N\,S_{2}(N)/[(N+0)^{1}(N+w)^{1}]\\
&\sum_N (-1)^N\,S_{2}(N)/[(N+0)^{1}(N+w)^{2}]\\
&\sum_N (-1)^N\,S_{2}(N)/[(N+0)^{2}(N+w)^{1}]\\
&\sum_N (-1)^N\,S_{2,1}(N)/[(N+0)^{1}(N+w)^{1}]\\
&\sum_N (-1)^N\,S_{3}(N)/[(N+0)^{1}(N+w)^{1}]
\end{align*}

\noindent
Thirty-nine of these have closing slots. The two that occur only inside $\Kres$ are the
rational atom $1/w^{3}$ and the nested atom $\atomlab{\BAatom}{(0,(),((0,1)),1)}$.

\subsection{How the rules were fixed}
\label{app:cat:fixing}
The rules were determined by exact rational linear algebra: for each slot, the
coefficients at spins $n\le49$ were fitted over the span of the seven weights, and
the solution was then required to reproduce every coefficient at every remaining
computed spin. An identification restricted to $n\le17$ gives the same
coefficient vectors, so the fit is heavily overdetermined. Section~\ref{sec:chk:oos} is
what makes this more than an interpolation: the rules were subsequently tested against
four spins that had not been computed when they were written, and reproduced
all $14580$ of their closing-slot coefficients exactly; with the seven remaining slots
included the four spins hold $17252$ coefficients, and every one of them follows from a
rule.

The split into closing and non-closing slots is a property of the basis of
Eq.~\eqref{eq:str:weights}: enlarging the basis
moves the line, and Sec.~\ref{sec:res:obstruction} shows how far it can be moved for
three of the seven.

\section{The collapse in detail}\label{app:compact}

\subsection{The five functions}
\label{app:cmp:five}
The functions appearing in Eq.~\eqref{eq:cf:ladderform} are assembled from the atom
list by Eq.~\eqref{eq:str:regroup}. Each is a fixed rational-and-$\zeta$ combination
of atoms, with no dependence on the spin or the rung.

The nested core of the $S_1$ sector is four atoms,
\begin{equation}
\Phi(x)=64\sum_{N\ge1}(-1)^{N}\Bigl[
-\frac{\Sh{-2}{N}}{N(N+x)}
+\frac{\Sh{1}{N}}{N(N+x)^{2}}
+\frac{\Sh{1}{N}}{N^{2}(N+x)}
+\frac{\Sh{2}{N}}{N(N+x)}\Bigr].
\label{eq:cmp:phiraw}
\end{equation}
The identity $\Sh{2}{N}-\Sh{-2}{N}=2O_2(N)$ merges the outer pair
into the odd harmonic sum of Eq.~\eqref{eq:conv:Ok}, and the partial-fraction
identity
\begin{equation}
\frac{1}{N(N+x)^{2}}+\frac{1}{N^{2}(N+x)}=\frac1x\Bigl[\frac{1}{N^{2}}-\frac{1}{(N+x)^{2}}\Bigr]
\label{eq:cmp:pf}
\end{equation}
merges the inner pair. Together they give Eq.~\eqref{eq:cf:phi}, verified against
Eq.~\eqref{eq:cmp:phiraw} to $1.51446\times10^{-41}$ at forty digits of working
precision. The difference is set by that precision and falls with it: at thirty
digits the same five arguments give $6.21147\times10^{-31}$.

The four original atoms of Eq.~\eqref{eq:cmp:phiraw} converge absolutely. After the
split, $B(x)$ still does, while $A(x)$ converges only conditionally, since
$O_2(N)\to\pi^2/8$ and its terms fall off like $1/N$. The split is legitimate because
each of $A(0)$, $A(x)$, $B(0)$, $B(x)$ converges on its own and no rearrangement of
terms is involved.

The conditional convergence is removed, not tolerated, when the constant is
required. Pairing $N=2m-1$ with $N=2m$, and using $O_2(2m)=O_2(2m-1)$ because an even
index contributes no new odd term, gives
\begin{equation}
A(0)=-\sum_{m\ge1}\frac{P(m)}{2m(2m-1)},
\qquad
P(m)=\sum_{i=1}^{m}\frac{1}{(2i-1)^{2}}
    =\frac{\pi^{2}}{8}-\frac{\dgam'\!\left(m+\tfrac12\right)}{4},
\label{eq:cmp:A0}
\end{equation}
whose terms fall off like $m^{-2}$. Equivalently, and more transparently,
$\Sh{2}{N}-\Sh{-2}{N}=2O_2(N)$ splits $A(0)$ into two standard alternating Euler sums,
\begin{equation}
\sum_{N\ge1}\frac{(-1)^{N}\Sh{2}{N}}{N}=\frac{\pi^{2}\lntwo}{12}-\zeta_3,
\qquad
\sum_{N\ge1}\frac{(-1)^{N}\Sh{-2}{N}}{N}=\frac{13\,\zeta_3}{8}-\frac{\pi^{2}\lntwo}{6},
\label{eq:cmp:euler}
\end{equation}
whose half-difference is Eq.~\eqref{eq:cf:AB0}.

The identity is sensitive to the conventions of Eq.~\eqref{eq:set:sdef} and is therefore
a check on them as well as an evaluation. Taking $O_2$ over strictly fewer terms shifts
$A(0)$ by $\tfrac78\zeta_3$, and replacing $\Sh{1}{N}$ by $\Sh{1}{N-1}$ in $B$ changes
its sign and magnitude to $+\tfrac18\zeta_3$. Neither variant reproduces the block.

The function $\Gladder_{(-1)^k}$ is twenty-six atoms, of which fifteen are nested, nine at
depth one and six at depth two. Section~\ref{sec:str:threetoone} and
App.~\ref{app:red:stuffle} reduce the depth-two part. The remainders $D_{+}$,
$D_{-}$ and $D_{\min}$ are seven, four and four atoms, all of empty nesting word, so
each reduces by Eqs.~\eqref{eq:red:phi}, \eqref{eq:red:psi} and~\eqref{eq:red:pf} to
polygammas at halved argument together with $\zeta$ values.

\subsection{The five functions, explicitly}
\label{app:cmp:explicit}
Written against the atoms of App.~\ref{app:cat:shapes}, in the shorthand
\begin{equation*}
\A{\pm}{S_v}{\prod(N{+}d)^{p}}{q}
=\sum_{N\ge1}\frac{(\pm1)^{N}S_v(N)}{\prod(N{+}d)^{p}\,(N{+}w)^{q}}\,,
\end{equation*}
and $w^{-q}$ for the rational atoms, the five functions of Eq.~\eqref{eq:cf:ladderform} are

\begin{align}
\Gladder_{(-1)^k}(w)&= -\,32\,\A{+}{\cdot}{(N{+}0)^{2}}{3} -\,48\,\A{+}{\cdot}{(N{+}0)^{3}}{2} -\,96\,\A{+}{\cdot}{(N{+}0)^{4}}{1}\nonumber\\
&\quad -\,64\,\A{+}{S_{1}}{(N{+}0)^{3}}{1} -\,48\,\A{-}{\cdot}{(N{+}0)^{1}}{4} -\,64\,\A{-}{\cdot}{(N{+}0)^{2}}{3}\nonumber\\
&\quad -\,144\,\A{-}{\cdot}{(N{+}0)^{3}}{2} -\,288\,\A{-}{\cdot}{(N{+}0)^{4}}{1} +\,128\,\A{-}{S_{-2,1}}{(N{+}0)^{1}}{1}\nonumber\\
&\quad -\,32\,\A{-}{S_{-2}}{(N{+}0)^{1}}{2} -\,96\,\A{-}{S_{-3}}{(N{+}0)^{1}}{1} -\,64\,\A{-}{S_{1,-2}}{(N{+}0)^{1}}{1}\nonumber\\
&\quad -\,128\,\A{-}{S_{1,1}}{(N{+}0)^{1}}{2} -\,128\,\A{-}{S_{1,1}}{(N{+}0)^{2}}{1} -\,128\,\A{-}{S_{1,2}}{(N{+}0)^{1}}{1}\nonumber\\
&\quad +\,32\,\A{-}{S_{1}}{(N{+}0)^{1}}{3} +\,96\,\A{-}{S_{1}}{(N{+}0)^{2}}{2} +\,160\,\A{-}{S_{1}}{(N{+}0)^{3}}{1}\nonumber\\
&\quad -\,128\,\A{-}{S_{2,1}}{(N{+}0)^{1}}{1} +\,96\,\A{-}{S_{2}}{(N{+}0)^{1}}{2} +\,160\,\A{-}{S_{2}}{(N{+}0)^{2}}{1}\nonumber\\
&\quad +\,160\,\A{-}{S_{3}}{(N{+}0)^{1}}{1} -\,\tfrac{16}{3}\,\pi^{2}\,\A{+}{\cdot}{(N{+}0)^{2}}{1} -\,\tfrac{16}{3}\,\pi^{2}\,\A{-}{\cdot}{(N{+}0)^{1}}{2}\nonumber\\
&\quad -\,\tfrac{16}{3}\,\pi^{2}\,\A{-}{\cdot}{(N{+}0)^{2}}{1} -\,32\,\zeta_3\,\A{-}{\cdot}{(N{+}0)^{1}}{1}
\label{eq:cmp:G}\\[3pt]
\Phi(w)&= -\,64\,\A{-}{S_{-2}}{(N{+}0)^{1}}{1} +\,64\,\A{-}{S_{1}}{(N{+}0)^{1}}{2} +\,64\,\A{-}{S_{1}}{(N{+}0)^{2}}{1}\nonumber\\
&\quad +\,64\,\A{-}{S_{2}}{(N{+}0)^{1}}{1}
\label{eq:cmp:Phi}\\[3pt]
D_{+}(w)&= -\,32\,\A{+}{\cdot}{(N{+}0)^{3}}{1} -\,32\,\A{-}{\cdot}{(N{+}0)^{1}}{3} -\,96\,\A{-}{\cdot}{(N{+}0)^{2}}{2}\nonumber\\
&\quad -\,160\,\A{-}{\cdot}{(N{+}0)^{3}}{1} +\,48\,w^{-4} +\,\tfrac{16}{3}\,\pi^{2}\,w^{-2} +\,32\,\zeta_3\,w^{-1}
\label{eq:cmp:Dp}\\[3pt]
D_{-}(w)&= 96\,\A{+}{\cdot}{(N{+}0)^{3}}{1} -\,32\,\A{-}{\cdot}{(N{+}0)^{1}}{3} -\,32\,\A{-}{\cdot}{(N{+}0)^{2}}{2}\nonumber\\
&\quad -\,32\,\A{-}{\cdot}{(N{+}0)^{3}}{1}
\label{eq:cmp:Dm}\\[3pt]
D_{\min}(w)&= -\,32\,\A{+}{\cdot}{(N{+}0)^{3}}{1} +\,32\,\A{-}{\cdot}{(N{+}0)^{1}}{3} +\,64\,\A{-}{\cdot}{(N{+}0)^{2}}{2}\nonumber\\
&\quad +\,96\,\A{-}{\cdot}{(N{+}0)^{3}}{1}
\label{eq:cmp:Dmin}
\end{align}

\noindent
Equations~\eqref{eq:cmp:G} to~\eqref{eq:cmp:Dmin} are the five in the order they stand
in Eq.~\eqref{eq:cf:ladderform}: $\Gladder_{(-1)^k}$, $\Phi$, $D_{+}$, $D_{-}$ and
$D_{\min}$.
With these, Table~\ref{tab:cat:rules}, the reductions of App.~\ref{app:reduction} and
the rules of Sec.~\ref{sec:res:ladder} for the seven remaining slots,
Eq.~\eqref{eq:cf:master} is fully determined by the text.

\subsection{Checks on the collapse}
\label{app:cmp:checks}
The numbers in this subsection are about the forty closing slots and about nothing
else. Both sides of each comparison below are built from those forty, the seven slots
of Sec.~\ref{sec:residual} appearing on neither, so what is measured is the exchange of
the two summations, with the block common to both sides. Section~\ref{sec:chk:closed} is where the two are
compared over all forty-seven, in rational arithmetic with nothing rounded.

With that said, the regrouping of Eq.~\eqref{eq:str:regroup} was verified against the
direct per-spin sum over the closing part of the atom table at $138$ points spread over
all forty-six spins of the original range, at thirty digits and cutoff $J=120$,
with a worst relative difference of $2.61862\times10^{-28}$. The assembled form
of Eq.~\eqref{eq:cf:ladderform}, with the three $S_1$ weights merged into
$\Sh{1}{\amax}$, was verified the same way at $n=3,5,9,15,21$ and reproduces that
same sum to $7.0155\times10^{-30}$ at the same precision and cutoff.

The part set aside is not a remainder. At $n=99$ the closing part of the atom sum and the
seven remaining slots agree in magnitude to five figures and cancel, with the numbers in
Sec.~\ref{sec:chk:oos}. A comparison
that drops the second of them is therefore a comparison of a large quantity with itself,
and it constrains only the regrouping.

The relations of Eqs.~\eqref{eq:str:R1} and~\eqref{eq:str:R2} are exact cancellations
among atom coefficients, not numerical coincidences: in
Eq.~\eqref{eq:str:R1} all eight $\BAatom$-atoms of the three $S_1$ functions cancel
identically, leaving only the three rational ones, and in Eq.~\eqref{eq:str:R2} the
nested atoms cancel and the depth-zero coefficients come out $64$, $0$, $32$, $64$.

\subsection{Two counts}
\label{app:cmp:counts}
Two functions in Eq.~\eqref{eq:cf:ladderform} hold nested content, $\Gladder_{(-1)^k}$ and
$\Phi$, and that is the reason the block can be written in a few lines. Expanded over the
canonical objects of Eq.~\eqref{eq:cf:canonical}, and after the stuffles of
App.~\ref{app:red:stuffle}, the same block spreads over the fifteen alternating Mellin
kernels that App.~\ref{app:red:nested} resolves into nested harmonic sums of one
variable. The
difference is one of grouping. $\Phi$, $D_{+}$,
$D_{-}$ and $D_{\min}$ are separate because they sit under different ladder weights,
whereas $\Gladder_{(-1)^k}$ collects several distinct objects that happen to
share the weight $(-1)^k$.

\section{The non-closing slots}\label{app:grids}

\subsection{The seven slots}
\label{app:grd:seven}
The slots whose coefficient rules do not close in the two ladder distances are
\begin{equation}
\begin{array}{llll}
\atomlab{\Ratom}{q=1},&
\atomlab{\Ratom}{q=2},&
\atomlab{\Ratom}{q=3},&\\[3pt]
\atomlab{\BAatom}{(0,\,(),\,((0,1)),\,1)},&
\atomlab{\BAatom}{(0,\,(),\,((0,2)),\,1)},&
\atomlab{\BAatom}{(1,\,(),\,((0,1)),\,2)},&
\atomlab{\BAatom}{(1,\,(),\,((0,2)),\,1)},
\end{array}
\label{eq:grd:seven}
\end{equation}
all with constant $1$. The four nested ones are the atoms $\Om{3}$, $\Om{C}$, $\Om{A}$
and $\Om{B}$ of Sec.~\ref{sec:res:what}, in that order. The last two follow identical
coefficient rules, on every coefficient of every spin computed, and their atoms enter
only through the combination of Eq.~\eqref{eq:str:merge}, which is elementary. Seven
slots therefore follow six distinct rules, counted as in Sec.~\ref{sec:res:what}.

The rules are given exactly, as rational numbers, for every odd $n\le99$. Their
rung ranges differ: the three rational slots occupy $k\in[-(n-2),0]$ and the four
nested ones $k\in[-(n-1),0]$, and these are the two ranges carried by the two sums of
Eq.~\eqref{eq:res:Kdef}. The slot of negative exchange sign, $\Om{3}$, vanishes at the
midpoint rung and is not stored there, which is why its counts run one lower per spin
than those of the other three.

\subsection{Exchange signs}
\label{app:grd:parity}
Under the exchange $k\to-(n-1)-k$ of Eq.~\eqref{eq:str:refl} the four nested slots
have definite sign,
\begin{equation}
\begin{aligned}
\atomlab{\BAatom}{(0,(),((0,1)),1)}&:\ \coeff{}(n,k)=-\coeff{}(n,k^{*}),\\
\atomlab{\BAatom}{(0,(),((0,2)),1)},\ \
\atomlab{\BAatom}{(1,(),((0,1)),2)},\ \
\atomlab{\BAatom}{(1,(),((0,2)),1)}&:\ \coeff{}(n,k)=+\coeff{}(n,k^{*}),
\end{aligned}
\label{eq:grd:parity}
\end{equation}
on all $8415$ coefficients of the original range and on a further $1532$ at the four
spins $n=93,95,97,99$. The three rational slots have no
definite sign at any of the four possible centrings, which is consistent with their
different rung ranges and with the arithmetic of Sec.~\ref{sec:res:obstruction}.

\subsection{The determined arrays}
\label{app:grd:determined}
None of the seven arrays is independent data any longer. The four nested ones are
given by Eqs.~\eqref{eq:res:c112}, \eqref{eq:res:c021} and~\eqref{eq:res:c011},
$\KRthree$ in closed form
by Eq.~\eqref{eq:res:KR3}, and $\KRtwo$ and $\KRone$ by the regularity identity
Eq.~\eqref{eq:res:laurent}; every one was checked against the stored array on all of
its cells with no mismatch. The arrays remain in the ancillary files as the record the
rules were verified against, and as a convenience for anyone who would rather read a
coefficient than evaluate a ladder sum.

\subsection{The closures}
\label{app:grd:closures}
The closures at the two ends of the ladder and along the first four columns are
collected in Sec.~\ref{sec:res:columns}, and Eq.~\eqref{eq:res:KR3} of
Sec.~\ref{sec:res:midpoint} returns them, the midpoint among them, by specialisation.
Each was fitted on a subset of the spins and validated on the remainder. A column at
$\am$ needs $\am\le n-2$, so the four columns of Eq.~\eqref{eq:res:col1} exist at $49$,
$49$, $48$ and $48$ spins. The closures hold on all of those cells but one, $n=5$ at
$\am=3$, named in Sec.~\ref{sec:res:columns}, so on $193$ in all. The
programs that reproduce the closures are in the ancillary files.

One of the nested slots also satisfies an explicit two-dimensional recurrence. With
$M=(n-1)/2$ and $j=-k$ the coefficient array $c(M,j)$ obeys
\begin{equation}
\sum_{\alpha=0}^{1}\sum_{\beta=0}^{3}\sum_{d_M=0}^{3}\sum_{d_j=0}^{3}
x_{\alpha\beta d_Md_j}\;M^{d_M}\,j^{d_j}\;c(M+\alpha,\,j+\beta)=0,
\label{eq:grd:recurrence}
\end{equation}
with $68$ of the $128$ coefficients $x$ nonzero, all integers of content one, the
largest of modulus $28$ and the smallest $-26$; they are listed in
Table~\ref{tab:grd:rec}. The solution is unique up to normalisation: the stencil
matrix has rank $127$ and nullity one, at every fitting window tested. It belongs to the rule shared by
$\atomlab{\BAatom}{(1,(),((0,1)),2)}$ and $\atomlab{\BAatom}{(1,(),((0,2)),1)}$, and
annihilates neither of the other five arrays, failing on essentially every cell of each.
The recurrence was fixed on the low spins and annihilates every interior coefficient of that slot out to
$n=99$, including the four spins of Sec.~\ref{sec:chk:oos} that postdate it. A
recurrence of this kind determines the slot from finitely many seeds even in the
absence of a closed form.

\begin{longtable}{@{}cccr@{\qquad}cccr@{\qquad}cccr@{}}
\caption{The $68$ nonzero integers of the recurrence, Eq.~\eqref{eq:grd:recurrence}.
Columns give $(\alpha,\beta,d_M,d_j)$ and the coefficient.}\label{tab:grd:rec}\\
\toprule
$\alpha$&$\beta$&$d_Md_j$&$x$ & $\alpha$&$\beta$&$d_Md_j$&$x$ & $\alpha$&$\beta$&$d_Md_j$&$x$\\
\midrule\endfirsthead\toprule
$\alpha$&$\beta$&$d_Md_j$&$x$ & $\alpha$&$\beta$&$d_Md_j$&$x$ & $\alpha$&$\beta$&$d_Md_j$&$x$\\
\midrule\endhead
0 & 0 & 01 & 3 & 0 & 0 & 02 & 6 & 0 & 0 & 03 & 3\\
0 & 0 & 10 & -6 & 0 & 0 & 11 & -10 & 0 & 0 & 12 & -2\\
0 & 0 & 13 & 2 & 0 & 0 & 20 & -4 & 0 & 0 & 21 & -8\\
0 & 0 & 22 & -4 & 0 & 1 & 00 & 12 & 0 & 1 & 01 & 24\\
0 & 1 & 02 & 18 & 0 & 1 & 03 & 6 & 0 & 1 & 10 & -16\\
0 & 1 & 11 & -20 & 0 & 1 & 12 & -6 & 0 & 1 & 13 & 4\\
0 & 1 & 20 & -4 & 0 & 1 & 21 & -12 & 0 & 1 & 22 & -12\\
0 & 1 & 30 & 8 & 0 & 1 & 31 & 8 & 0 & 2 & 00 & 6\\
0 & 2 & 01 & 15 & 0 & 2 & 02 & 12 & 0 & 2 & 03 & 3\\
0 & 2 & 10 & -20 & 0 & 2 & 11 & -26 & 0 & 2 & 12 & -4\\
0 & 2 & 13 & 2 & 0 & 2 & 20 & 8 & 0 & 2 & 21 & -12\\
0 & 2 & 22 & -8 & 0 & 2 & 30 & 16 & 0 & 2 & 31 & 8\\
1 & 1 & 01 & 1 & 1 & 1 & 03 & -1 & 1 & 1 & 10 & -2\\
1 & 1 & 11 & 4 & 1 & 1 & 12 & 4 & 1 & 1 & 13 & -2\\
1 & 1 & 20 & -8 & 1 & 1 & 22 & 8 & 1 & 1 & 30 & -8\\
1 & 1 & 31 & -8 & 1 & 2 & 01 & -4 & 1 & 2 & 02 & -6\\
1 & 2 & 03 & -2 & 1 & 2 & 10 & 8 & 1 & 2 & 11 & 8\\
1 & 2 & 12 & -6 & 1 & 2 & 13 & -4 & 1 & 2 & 20 & 8\\
1 & 2 & 21 & 28 & 1 & 2 & 22 & 12 & 1 & 2 & 30 & -16\\
1 & 2 & 31 & -8 & 1 & 3 & 00 & -6 & 1 & 3 & 01 & -11\\
1 & 3 & 02 & -6 & 1 & 3 & 03 & -1 & 1 & 3 & 11 & -12\\
1 & 3 & 12 & -10 & 1 & 3 & 13 & -2 & 1 & 3 & 20 & 24\\
1 & 3 & 21 & 20 & 1 & 3 & 22 & 4 &  &  &  & \\
\bottomrule
\end{longtable}

\subsection{The searches that failed}
\label{app:grd:open}
Every negative result in this paper is of one kind. A finite list of functions, or a
finite set of stencils, is written down and the target is shown not to lie in its span,
in exact rational arithmetic. The list is what the statement is about. None of these
results says that no closed form exists, and the one statement of that kind we do make
is the arithmetic exclusion of Sec.~\ref{sec:res:obstruction}, which rests on
denominators, not on a list, and comes with the hypothesis given there.

The bases used for these fits are built as follows. Take all monomials of total weight
at most $w$ in the nested harmonic sums $\Sh{a}{\ap}$, $\Sh{a}{\am}$ and
$\Sh{a}{\amin}$, with and without the alternation $(-1)^k$. Where the target has a
definite exchange sign $\sigma$, replace each monomial $m$ by the projection
$m+\sigma\,m^{*}$, with $m^{*}$ the monomial with $\ap$ and $\am$ interchanged, and
keep one representative of each orbit. That halves the column count without losing
anything the target can use, since a target of definite sign is orthogonal to the
opposite sector. With the bases built that way the fits are inconsistent at every
weight tested: five ($486$ functions), six ($982$), seven ($1616$), and eight ($2450$
functions against $2499$ coefficients) for the three slots of positive sign. The
exchange-odd count is the smaller one, because there the projection $m-m^{*}$
annihilates every monomial already symmetric in the two distances: there are $55$,
$98$, $138$ and $195$ such monomials at the four weights, each counted twice by the
alternation, which leaves $376$, $786$, $1340$ and $2060$ functions. The slot of
negative sign holds $2450$ coefficients and was searched through weight seven only.

No linear recurrence in $n$ with polynomial coefficients was found for
$\Kres(n,z)$. The search covered three rational values of $z$ and every order up to
fourteen and degree up to fourteen for which the $49$ available spins give an
overdetermined system. A stencil of order $R$ and degree $D$ has $(R+1)(D+1)$ unknowns and
$49-R$ usable rows, and it is admitted when the rows exceed the unknowns by at least
six, $49-R\ge(R+1)(D+1)+6$. Only $76$ of the nominal order-degree pairs meet that at
$49$ spins, and $76$ pairs at three values of $z$ is the $228$ stencils per kernel, each
run in two variants, with and without a polynomial right-hand side; so these numbers
bound the search and say nothing about the object. Each dimension quoted in this subsection
is rebuilt in the ancillary files from the list of functions it counts and compared
there against the number printed here; the fits themselves are not rerun. For this
sector those files add the recurrence of Eq.~\eqref{eq:grd:recurrence}, checked
there on a holdout, and the denominator-prime audit behind
Sec.~\ref{sec:res:obstruction}.

Following the prescription of Sec.~\ref{sec:res:obstruction} and admitting nested
harmonic sums at the argument $n-2$ alongside $\ap$, $\am$ and
$\min(\ap,\am)$ removes the arithmetic obstruction but does not produce a fit: the
three rational slots remain inconsistent at weight three ($166$ functions), four
($478$) and five ($1126$).

\section{Validation tables}\label{app:validation}

\subsection{Intercepts against the Quantum Spectral Curve}
\label{app:val:qsc}
Along $\nu=0$ the eigenvalue obtained from Eq.~\eqref{eq:cf:master} is compared with the
closed-form intercepts of the Quantum Spectral Curve analysis at non-zero conformal
spin~\cite{Alfimov:2018cms}, which is the only one of the Quantum Spectral Curve
results~\cite{Gromov:2013pga,Alfimov:2014bwa,Gromov:2015vua} that reaches $n\neq0$. The
comparison runs spin by spin through $n=91$. Table~\ref{tab:val:qsc} lists a
representative sample of the measured residuals; all $45$ are in
\texttt{anc/eval/reproduce\_tableI\_all.runlog.csv}, one row per spin. The file
\texttt{anc/tables/intercepts\_check.csv} is a different record: it lists every odd
spin through $n=91$ and gives an evaluated intercept at the twenty of them for which
one was computed for the comparison of Table~\ref{tab:val:qsc}.

\begin{table}[t]
\centering
\begin{tabular}{@{}rll@{}}
\toprule
$n$ & absolute residual & relative residual\\
\midrule
$3$   & $2.0\times10^{-23}$ & $2.4\times10^{-25}$\\
$9$   & $2.0\times10^{-34}$ & $1.3\times10^{-36}$\\
$17$  & $3.5\times10^{-37}$ & $1.9\times10^{-39}$\\
$41$  & $1.5\times10^{-37}$ & $6.4\times10^{-40}$\\
$85$  & $1.5\times10^{-23}$ & $5.5\times10^{-26}$\\
$91$  & $3.3\times10^{-22}$ & $1.2\times10^{-24}$\\
\bottomrule
\end{tabular}
\caption{Intercepts at $\nu=0$ against the Quantum Spectral Curve, at $40$ digits of
working precision, with cutoff $J=120$ and $K=24$ nodes in the circle average. The relative
residual is smallest near the middle of the range, at $n=41$, and grows towards both
ends: by fifteen orders of magnitude at $n=3$ and by fifteen again at $n=91$. That
two-sided growth is a property of the settings of this run and not of the method or
of the reference. At deeper settings both ends improve, for different reasons. At large
$n$ the limit is the cutoff margin $J-\max_k|z+k|=J-(n+1)/2$, the maximum running over
the ladder rungs and the rung at $k=1$, so it is $J$ that matters there: at the
reference configuration of $120$ digits and $J=300$, run at $n=51$, $71$ and $91$, the
$n=91$ residual is $3.5\times10^{-41}$. At small $n$ the limit is the node count $K$ of
the circle average used for the removable singularity at $\nu=0$: the deepest run at
$n=3$, a circle average of radius $r=0.15$ with $K=48$ nodes at $60$ digits, leaves a
residual of $2.0\times10^{-49}$. The configuration behind each reference residual is
listed in \texttt{anc/tables/intercepts\_check.csv}, and both limits are listed spin by
spin in \texttt{anc/eval/reproduce\_tableI\_all.runlog.csv}, which covers all $45$
spins together with the programs that produce them.}
\label{tab:val:qsc}
\end{table}

\subsection{The spins computed after the rules were fixed}
\label{app:val:oos}
Table~\ref{tab:chk:oos} of the main text gives the coefficient counts at the four
spins computed after the rules were fixed, and Sec.~\ref{sec:chk:oos} the order in which
the integrand, its reduction and the rules were produced; the controls that reduction was
run against are those of App.~\ref{app:cons:repro}. The comparison of the compact form of
Eq.~\eqref{eq:cf:ladderform} with the direct atom-by-atom sum at those four spins is
given there too. That comparison's worst relative difference is a measure of cancellation
and moves with the
cutoff: at the same thirty digits and $J=80$ it becomes $5.81791\times10^{-29}$, against
the $6.38554\times10^{-29}$ quoted at $J=120$.

\subsection{Precision and domain}
\label{app:val:precision}
Wherever the objects are rational the arithmetic is exact, with no floating point
anywhere from the integrand to the coefficients. The floating-point side needs more care than a single tolerance.

The evaluator closes each nested atom's infinite sum by a direct sum to a cutoff $J$
plus a tail, and the evaluation requires $J$ to exceed every shift magnitude $|z+k|$ on
the ladder before it returns an answer. Across working precisions of $40$, $60$ and $80$
digits and cutoffs $J=120$, $240$ and $400$, Eq.~\eqref{eq:cf:worked} is stable in
every one of those nine combinations, so neither limit is what bounds the result here.

The one thing that does bound it is the spectral variable itself. If $\nu$ is given
at machine precision, or as a multiprecision number built before the working precision is
raised, it brings a representation error of its own: $0.3$ at machine precision is
$0.3-1.11\times10^{-17}$. The block is smooth in $\nu$, so the error is carried through
linearly, and with $\partial\chiord{2}/\partial\nu\simeq-1.95$ at this point it displaces the
answer by $2.16\times10^{-17}$, which is to say it corrupts everything past the
nineteenth digit while leaving the first nineteen intact. The nineteen belongs to this
point, since the derivative does, and where the block is steeper fewer digits survive.
The remedy is to form $\nu$
at the working precision, as a string or otherwise; the evaluator accepts a string for
that reason. We state this explicitly because the failure is silent: the domain
check passes, the tail converges, and the answer is wrong in a way that looks like
noise, not like an error.

The same caution applies at $\nu=0$, where the mechanism is a different one. The two
halves of Eq.~\eqref{eq:cf:separable} coincide at that point and add,
Eq.~\eqref{eq:cf:intercept}. The cancellation is among the atom contributions inside a
single block, at the interior point $z=M=(n-1)/2$, where poles of order up to four
meet. That
cancellation deepens with the spin, and the intercept is taken through a separate
removable-singularity path, not as a limit. The residuals of
App.~\ref{app:val:qsc} are quoted as measured, not as a bound, for that reason.

\subsection{The holdout spin}
\label{app:val:holdout}
The atom table at $n=101$ was generated after the complete rule set had been written
down, by the extraction of App.~\ref{app:construction} applied to the same integrand. The table
lists $3936$ shifted instances and $4538$ coefficients, and the census law $39n-3$ and
the controls of App.~\ref{app:cons:repro} hold on it. The rules reproduce all
$4538$ exactly, over all forty-seven slots, with no entry uncovered; ladder regularity
holds on all $1500$ of its Laurent coefficients with no rule used; and the printed
Eqs.~\eqref{eq:cf:ladderform} and~\eqref{eq:res:Kdef}, expanded, reproduce the same
$4538$. The run record and the software versions are in the generation record of the
ancillary files. One spin tests the rules and does no more than that: it does not extend the proof of
Sec.~\ref{sec:res:obstruction}, which stands on its own, and it does not make the rules
proved at arbitrary odd $n$.

\section{Ancillary files}\label{app:ancillary}

Most of what this paper quotes is reproduced by the ancillary files. The rest is
listed here. The fits behind the bounded searches of App.~\ref{app:grd:open} are not
rerun there, although the dimension each of them is a statement about is rebuilt and
checked against the number printed here. The rank statements of
Sec.~\ref{sec:res:columns}, Sec.~\ref{sec:res:ladder} and Sec.~\ref{sec:dis:residual}
are given with their provenance and not with a program, since the function list each
of them rests on is not printed here. Four checks have no program either: the
exchange-symmetry totals of Sec.~\ref{sec:str:reflection}, the collinear tower of
Sec.~\ref{sec:chk:collinear}, the annihilation and the rank of the stencil of
Eq.~\eqref{eq:grd:recurrence}, whose coefficients \texttt{anc/kernels/} stores and no
program there repeats, and the comparisons that need the two lower orders in the
coupling, the normalization of App.~\ref{app:conv:norm} and the branch test of
App.~\ref{app:conv:sums} among them, for which the bundle holds no reference values.

The files run from the raw data through to the evaluation. The bundle contains the
directories \texttt{anc/tables/}, \texttt{anc/catalog/}, \texttt{anc/kernels/},
\texttt{anc/eval/}, \texttt{anc/gate/}, \texttt{anc/source\_chain/},
\texttt{anc/collapse/}, \texttt{anc/obstruction/}, \texttt{anc/searches/} and
\texttt{anc/residuals/}. Beside them are
\texttt{anc/anc\_closed\_form.py}, a \texttt{README.md}, a
\texttt{LICENSE} and a \texttt{THIRD\_PARTY} record. Four of these hold supporting
data and the programs that read it: the collapse of the ladder sum, the
denominator-prime audit of Sec.~\ref{sec:res:obstruction}, the search dimensions of
App.~\ref{app:grd:open} with what is known about the three rank statements, and the
comparison behind each figure this paper quotes as a relative difference.

\subsection{The closed-form rules}
\label{app:anc:closed}
\texttt{anc/anc\_closed\_form.py} implements the rules of Sec.~\ref{sec:res:ladder} and
Eq.~\eqref{eq:res:KR3} in exact rational arithmetic: the ladder sums of
Eq.~\eqref{eq:res:laddersums}, the four nested coefficient rules, and $\KRthree$. Run
against the stored arrays it reproduces $12397$ coefficients exactly, and the two
remaining rational kernels follow from Eq.~\eqref{eq:res:laurent} as described there.
It needs nothing but the Python standard library.

\subsection{Tables}
\label{app:anc:tables}
\texttt{anc/tables/} holds the per-spin atom tables for odd $n\le101$, one file per
spin, keyed by atom and storing exact rational coefficients as strings. These are
the output of App.~\ref{app:construction} and the input to everything else. The
files for $n\le91$ are those of the original range; those for $n=93,95,97,99$ were
produced after the coefficient rules were fixed, and the one for $n=101$ later still,
and they are what Sec.~\ref{sec:chk:oos} tests against. The generation record for the
holdout spin, with the hashes of the three runs and the software versions, is in the
same directory.

\subsection{Catalogue}
\label{app:anc:catalogue}
\texttt{anc/catalog/} holds the coefficient rules of App.~\ref{app:catalogue} in
structured form: for each of the $40$ closing slots, the rational multiples of the
seven weights of Eq.~\eqref{eq:str:weights}, together with the slot's rung range.
The seven non-closing slots are listed there as such, with their coefficients kept
in the next layer, not reduced to a rule.

\subsection{Kernels}
\label{app:anc:kernels}
\texttt{anc/kernels/} holds the coefficients of the seven non-closing slots exactly,
for every odd $n\le99$, together with the integer coefficients of the recurrence of
Eq.~\eqref{eq:grd:recurrence}. These are raw data, not reduced expressions,
and they are the object any future identification of the residual kernels has to
reproduce.

\subsection{Evaluator}
\label{app:anc:eval}
\texttt{anc/eval/} holds a self-contained evaluator that takes $(n,\nu)$ and returns
$\chiord{2}$, assembling the block from the atom table and closing each nested sum
by a direct sum plus a tail. The worked value of Eq.~\eqref{eq:cf:worked} is produced
by it, and the programs that reproduce the validation tables are alongside it. The atom
tables are regenerated from \texttt{anc/source\_chain/} instead, and the catalogue fit is
repeated by \texttt{anc/catalog/certificate.py}.
\texttt{closure\_check.py} is there as well. It evaluates both sides of every identity
of App.~\ref{app:red:nested} at complex argument, along two routes with no summation
in common, and keeps the residual of each point.
\texttt{intercept\_extension.py} is beside them and runs the comparison of
Sec.~\ref{sec:chk:agsext}: it evaluates the binomial closed form of
Ref.~\cite{Alfimov:2018cms} in rational arithmetic at $n=93$, $97$ and $101$, takes the
other side from the atom tables through the wrapper named below, and prints the two
values with their difference. It also prints the finite binomial sum at an odd
$M=(n-1)/2$, where that sum is not the continued intercept, as a control.

The domain condition of App.~\ref{app:val:precision} is enforced at the call, at a
different place for each of the two evaluators. The low-spin one,
\texttt{chi\_nnlo\_eval.py}, rejects out-of-domain calls on its own. The extended-spin
\texttt{chi\_eval\_ext.py}, which is the one that reaches the whole table range, checks the
spin and rejects a non-integer or a spin it has no table for, but leaves the summation
cutoff to the caller, since the reproducer that drives it pins that cutoff. The cutoff
at every spin is enforced by a guarded wrapper alongside them,
\texttt{chi\_ext\_guarded.py}, which reads the rungs of the requested spin out of its
own atom table, takes the floor from them, and raises an error instead of returning a
divergent tail or a value from inside the small-$\nu$ cancellation band. That wrapper is the
entry point, and the directory's own documentation names it as such. The spin range
is nowhere written out as a constant: it is computed from the tables present, so
that a range in a header and a range in a guard cannot come apart.

\subsection{The comparison programs}
\label{app:anc:gate}
\texttt{anc/gate/} holds the programs behind the counts of Sec.~\ref{sec:chk:closed}
and Sec.~\ref{sec:chk:oos}. \texttt{gate\_e2e.py} rebuilds every atom coefficient from
the rules and nothing else and compares against the tables, \texttt{gate\_rules.py} checks the
residual coefficient arrays against the rules that generate them, \texttt{gate\_regularity.py} checks
Eq.~\eqref{eq:res:laurent} from the tables with no rule used,
\texttt{build\_from\_rules.py} writes an atom table carrying no tabulated coefficient
anywhere, and \texttt{gate\_holdout.py} runs all forty-seven rules against a single spin
outside the fitted set, $n=101$ by default.

The last one is \texttt{gate\_compact.py}, and it is the only one that reads the
manuscript. With \texttt{texbind.py} it takes Eqs.~\eqref{eq:cf:ladderform}
and~\eqref{eq:res:Kdef}, the rule displays of Sec.~\ref{sec:residual},
Table~\ref{tab:cat:rules} and the function list of App.~\ref{app:cmp:explicit} out of
the \LaTeX{} sources at run time, expands them into the atom basis and compares against
the tables in exact rational arithmetic. If a display it depends on can no longer be
read it aborts and guesses nothing, and the few interpretations it has to supply are listed in
the file. Each of these programs returns a non-zero status if any comparison fails.

\subsection{Regenerating from scratch}
\label{app:anc:regen}
Two commands take the Caron-Huot--Herranen integrand at an odd spin to a finished
atom table, one producing the structured integrand and one reducing it. Running them
at a spin that is already stored reproduces the stored files exactly, which is the
control described in App.~\ref{app:cons:repro}. Nothing in that route requires the
stored tables; both are included so that a reader can start anywhere.


\end{document}